\documentclass[prd,aps,eqsecnum]{revtex4}
\usepackage{epsfig}
\usepackage{amsfonts}
\usepackage{rotating} 
\usepackage{sparticles}
\usepackage{dcolumn}
\usepackage{bm}
\usepackage{hyperref}
\hypersetup{
    bookmarks=true,         
    unicode=false,          
    pdftoolbar=true,        
    pdfmenubar=true,        
    pdffitwindow=false,     
    pdfstartview={FitH},    
    pdftitle={My title},    
    pdfauthor={Author},     
    pdfsubject={Subject},   
    pdfcreator={Creator},   
    pdfproducer={Producer}, 
    pdfkeywords={keyword1} {key2} {key3}, 
    pdfnewwindow=true,      
    colorlinks=true,       
    linkcolor=red,          
    citecolor=cyan,        
    filecolor=magenta,      
    urlcolor=green,           
    linktocpage=true
}
\usepackage{amsmath,amssymb,amsthm,graphicx,latexsym}
\usepackage{subfigure}
\usepackage{pstricks}
\usepackage{color}

\usepackage{mathrsfs}

\newcommand{\be}{\begin{equation}}
\newcommand{\ee}{\end{equation}}
\newcommand{\bea}{\begin{eqnarray}}
\newcommand{\eea}{\end{eqnarray}}
\newcommand{\bml}{\begin{subequations}}
\newcommand{\eml}{\end{subequations}}
\newcommand{\bfig}{\begin{figure}}
\newcommand{\efig}{\end{figure}}

\begin{document}

\title{Primordial non-Gaussian features from  DBI Galileon inflation}

\author{Sayantan Choudhury \footnote{\bf\textcolor{red}{\bf Electronic address: {sayantan@theory.tifr.res.in, sayanphysicsisi@gmail.com}}}}
\affiliation{Department of Theoretical Physics, Tata Institute of Fundamental Research, Homi Bhabha Road, Colaba, Mumbai - 400005, India} 
\affiliation{Physics and Applied Mathematics Unit, Indian Statistical Institute, 203 B.T. Road, Kolkata 700 108, India}
\author{Supratik Pal\footnote{\bf\textcolor{blue}{\bf Electronic address: {supratik@isical.ac.in}}} ${}^{}$}
\affiliation{Physics and Applied Mathematics Unit, Indian Statistical Institute, 203 B.T. Road, Kolkata 700 108, India}

\begin{abstract}
 We have studied primordial non-Gaussian features 
from a model of  potential driven single field DBI Galileon inflation.  
We have computed the bispectrum from the three point correlation function considering all possible 
cross correlation between scalar and tensor modes from the proposed setup. Further, we have  computed the 
trispectrum from four point correlation function considering the contribution from contact interaction, scalar and graviton exchange diagrams in the in-in picture. 
Finally we have obtained the non-Gaussian consistency conditions from the four point correlator, 
which results in partial violation of the  {\it Suyama-Yamaguchi} four-point consistency relation.
This further leads to the conclusion that sufficient primordial non-Gaussianities can be
obtained from DBI Galileon inflation.

\end{abstract}


\maketitle
\tableofcontents
\newpage
\section{\bf Introduction}

The physics of the early universe is a very rich area of theoretical physics, for there is a plethora of potential models that
solve, at least partially, the well-known problems of the standard cosmological paradigm.
 Inflationary cosmology is the most successful branch which addressed all of these problems meticulously. 
This can however be explained by several class of models originated from a proper field theoretic or particle physics framework.
But from observational point view a big issue may crop up in model discrimination and also in the removal of the degeneracy of cosmological parameters
obtained from Cosmic Microwave Background (CMB) observations \cite{wmap, Ade:2014xna,planck}. 
In this context the study of primordial non-Gaussian feature acts as a powerful computational tool
 to discriminate among inflationary 
models. In the very recent days  the analysis of bispectrum and trispectrum derived from the
study of primordial features of non-Gaussianity \cite{hall, maldacena, shell, lim, wands1, wands2, sherry, esther, zalda, paolo, lyth, riotto, felice, shun, kobayashi, ranu, jerome}
 from different models of inflation has thus become an intriguing aspect in the context of inflationary model building as well as studies of 
CMB physics.


Galileon based inflationary models \cite{claudia, trodden, goon} and DBI inflationary models \cite{gauss}, \cite{edmu} are both in vogue for
quite some time now. Despite its successes, Galileon models generically give rise to  unwanted degrees of freedom like 
ghosts, Laplacian and Tachyonic instabilities. Recently, a natural extension to these class of models has been brought forth by the present authors 
  \cite{sayan1}  in which  DBI was clubbed together with Galileon. The  
framework, called  DBI Galileon, consists of a D3 brane in the background of ${\cal N}$=1,${\cal D}$=4
SUGRA derived from D4 brane in ${\cal N}$=2,${\cal D}$=5
bulk SUGRA background. The interesting feature of this treaty is that
those unwanted debris can be successfully thrown away  keeping
all the good features of Galileon intact.  In the present paper, our prime objective is to investigate for some more 
interesting features of this rich structure of DBI Galileon
\cite{sayan1}, which ultimately results in
sufficient non-Gaussianity in this framework. Specifically, we explicitly calculate  
bispectrum and trispectrum from three and four point correlation functions  by exploiting
third and fourth order actions. The calculations reveal, along with the feature of large non-Gaussianity, 
some other interesting results like partial violation of the  {\it Suyama-Yamaguchi} four-point consistency relation.
Subsequently, we demonstrate that, in this framework, it is  possible to have a parameter space for both non-Gaussianity and tensor-to-scalar ratio ($r$) 
consistent with combined constraint obtained from {\it Planck+WMAP9+high-L+BICEP2} data \cite{Ade:2014xna,planck}.

The plan of the paper is as follows.
First we explore primordial non-Gaussian features from the third order
 action through the non-linear parameter $f_{NL}$ calculated from 
bispectrum (in equilateral limit configuration) including all possible scalar - tensor type of cross correlations in the different polarizing modes.
 Hence from the fourth order action we derive the expression for other two non-linear parameters $g_{NL}$ and $\tau_{NL}$ through trispectrum analysis
considering the contribution from contact interaction, scalar and graviton exchange diagrams in the in-in picture.
 Finally, we explicitly derive the four point consistency relation from scalar and graviton exchange diagrams and also find a partial
 violation of {\it standard Suyama-Yamaguchi relation} \cite{ken,futa}.
We also attempt to give some possible explanations for this violation.
We end up with scanning the parameter space for non-Gaussianity and tensor-to-scalar ratio 
in the light of {\it Planck+WMAP9+high-L+BICEP2} data.


\section{\bf The Background Model}

For systematic development of the formalism, let us
briefly review from our previous paper \cite{sayan1} how one can
construct the effective 4D inflationary potential 
for DBI Galileon starting from ${\cal N}=2, {\cal D}=5$ SUGRA
along with Gauss-Bonnet correction in the bulk geometry and D4 brane setup 
leads to an effective ${\cal N}=1, {\cal D}=4$ SUGRA in the D3 brane. 
 Here the total five dimensional model is described by the following action
\be\label{totac}S^{(5)}_{Total}=S^{(5)}_{EH}+S^{(5)}_{GB}+S^{(5)}_{DBI}+S^{(5)}_{WZ}+S^{(5)}_{BSUG}\ee
where
\be\begin{array}{llll}\label{5eh}S^{(5)}_{EH}=\frac{1}{2\kappa^{2}_{5}}\int d^{5}x\sqrt{-g^{(5)}}\left[R_{(5)}-2\Lambda_{5}\right],~~~
S^{(5)}_{GB}=\frac{\alpha_{(5)}}{2\kappa^{2}_{5}}\int d^{5}x\sqrt{-g^{(5)}}
\left[R^{ABCD(5)}R^{(5)}_{ABCD}-4R^{AB(5)}R^{(5)}_{AB}+R^{2}_{(5)}\right],\\
S^{(5)}_{DBI}=-\frac{T_{4}}{2}\int d^{5}x \exp(-\Phi)\sqrt{-\left(\gamma^{(5)}+B^{(5)}+2\pi\alpha^{'}F^{(5)}\right)},\\
S^{(5)}_{WZ}=-\frac{T_{4}}{2}
\int \sum_{n=0,2,4} \hat{C_{n}}\wedge \exp\left(\hat{B}_{2}+2\pi\alpha^{'}F_{2}\right)|_{4~form}\\~~~~~~
=\frac{1}{2}\int d^{5}x\sqrt{-g^{(5)}}\left\{\epsilon^{ABCD}\left[\partial_{A}\Phi^{I}\partial_{B}\Phi^{J}
\left(\frac{C_{IJ}B_{KL}}{4T_{4}}\partial_{C}\Phi^{K}\partial_{D}\Phi^{L}+\frac{\pi\alpha^{'}C_{IJ}F_{CD}}{2}\right.
\right.\right.\\ \left.\left.\left.~~~~~~
+\frac{C_{0}}{8T_{4}}B_{IJ}B_{KL}\partial_{C}\Phi^{K}\partial_{D}\Phi^{L}+
\frac{\pi\alpha^{'}C_{0}}{2}B_{IJ}F_{CD}\right)+2\pi^{2}\alpha^{'2}T_{4}C_{0}F_{AB}F_{CD}-T_{4}\left(\nu_{0}+\frac{\nu_{4}}{\Phi^{4}}\right)\right]\right\},\\
S^{(5)}_{BSUG}=\frac{1}{2}\int d^{5}x\sqrt{-g_{(5)}}
e_{(5)}\left[-\frac{M^{3}_{5}R^{(5)}}{2}+\frac{i}{2}\bar{\Psi}_{i\tilde{m}}
\Gamma^{\tilde{m}\tilde{n}\tilde{q}}\nabla_{\tilde{n}}
\Psi^{i}_{\tilde{q}}-{S}_{IJ}F^{I}_{\tilde{m}\tilde{n}}F^{I\tilde{m}\tilde{n}}-\frac{1}{2}
g_{\alpha\beta}(D_{\tilde{m}}\phi^{\mu})(D^{\tilde{m}}\phi^{\nu})
\right.\\ \left.~~~~~~~~~~~~~~~~~~~~~~~~~~~~~~~~~~~~~~~~~~~~~ + {\rm Fermionic} + {\rm Chern-Simons} + {\rm Pauli~~ mass}\right]\end{array}\ee
where $T_{(4)}$ is the D4 brane tension, $\alpha^{'}$ is the Regge Slope, $\exp(-\Phi)$
is the closed string dilaton and $C_{0}$ is the Axion. Here $\gamma^{(5)}$, $B^{(5)}$ and $F^{(5)}$ represent the determinant
of the 5D induced metric ($\gamma_{AB}$) and the gauge fields ($B_{AB},F_{AB}$) respectively.
Additionally here $\nu_{0}$ and $\nu_{4}$ represent the constants characterizing the
interaction strength between  D4-$\bar{D4}$ brane.
In the present context 5-dimensional coordinates
$X^{A}=(x^{\alpha},y)$, where $y$ parameterizes the extra
dimension compactified on the closed interval $[-\pi R,+\pi R]$. 

It is 
useful to introduce the 5D metric in conformal form 
\be\label{metric}ds^{2}_{4+1}=g_{AB}dX^{A}dX^{B}
=
\frac{b^{2}_{0}}{R^{2}\left(\exp(\beta y)
+\frac{\Lambda_{(5)}b^{4}_{0}}{24R^{2}}\exp(-\beta y)\right)}\left(ds^{2}_{4}+R^{2}\beta^{2}dy^{2}\right),\ee
and $ds^{2}_{4}=g_{\alpha\beta}dx^{\alpha}dx^{\beta}$ is FLRW counterpart. The parameter $\beta$ determines the slope of the
warp factor and $R$ represents the compactification radius. 
Applying dimensional reduction technique via ${\bf S^{1}/Z_{2}}$ orbifolding symmetry and using the metric stated in equation(\ref{metric}) the total effective model for {\it D3 DBI Galileon} in background ${\cal N}$=1, ${\cal D}$=4 SUGRA is described by the 
following action \cite{sayan1}:
\be\begin{array}{lllllll}\label{model1}
  \displaystyle S= \int d^{4}x \sqrt{-g^{(4)}}
\left[\hat{\tilde{K}}(\phi,X)
-\tilde{G}(\phi,X)\Box^{(4)}\phi+\tilde{l}_{1}R_{(4)}\right.\\ \left.
\displaystyle~~~~~~~~~~~~~~~~~~~~~~~~~~~~~~~~~~~~~~~~~~~~~+\tilde{l}_{4}\left({\cal C}(1)R^{\alpha\beta\gamma\delta(4)}R^{(4)}_{\alpha\beta\gamma\delta}
-4{\cal I}(2)R^{\alpha\beta(4)}R^{(4)}_{\alpha\beta}+{\cal A}(6)R^{2}_{(4)}\right)+\tilde{l}_{3}\right],\end{array}\ee
where
\be\begin{array}{lllll}\label{effcons} \hat{\tilde{K}}(\phi,X)=-\frac{\tilde{D}}{\tilde{f}(\phi)}\left[\sqrt{1-2QX\tilde{f}}-Q_{1}\right]
-\tilde{C}_{5}\tilde{G}(\phi,X)-2X\tilde{M}(T,T^{\dag})-V(\phi),\\ 
\tilde{M}(T,T^{\dag})=\frac{M(T,T^{\dag})}{2\kappa^{2}_{(4)}},~M(T,T^{\dag})=\frac{\sqrt{2}\beta R^{2}}{(T+T^{\dag})},~\tilde{D}=\frac{D}{2\kappa^{2}_{(4)}},\\
\tilde{G}(\phi,X)=\left(\frac{\tilde{g}(\phi)k_{1}\tilde{C}_{4}}{2(1-2\tilde{f}(\phi)Xk_{2}))}\right),~
\tilde{g}(\phi)=\tilde{g}_{0}+\tilde{g}_{2}\phi^{2},~\tilde{f}(\phi)\simeq\frac{1}{(\tilde{f}_{0}+\tilde{f}_{2}\phi^{2}+\tilde{f}_{4}\phi^{4})}\\
\tilde{l}_{1}=\left\{\frac{1}{2\kappa^{2}_{(4)}}
\left[1+\frac{\alpha_{(4)}}{R^{2}\beta^{2}}\left(24{\cal I}(2)-24{\cal A}(9)-16{\cal A}(10)\right)
\right]-\frac{\alpha_{(4)}{\cal C}(2)}{\kappa^{2}_{(4)}R^{2}\beta^{2}}\right\},\tilde{l}_{4}=\frac{\alpha_{(4)}}{2\kappa^{2}_{(4)}}, \\
\tilde{l}_{3}=\frac{1}{2\kappa^{2}_{(4)}}\left[\frac{\alpha_{(4)}}{R^{4}\beta^{4}}
\left(24{\cal C}(4)-144{\cal I}(4)-64{\cal A}(5)+144{\cal A}(7)+64{\cal A}(8)+192{\cal A}(11)\right)-\frac{3M^{3}_{5}\beta b^{6}_{0}}
{2\kappa^{2}_{(4)}M^{2}_{PL}R^{5}}{\cal I}(1)\right]\end{array}.\ee
where $\alpha_{(4)},\tilde{l}_{1},\tilde{l}_{3},\tilde{l}_{4}$ are effective 4D couplings and $\kappa_{(4)}$ be the gravitational coupling strength. Here $X$ represents
the 4D kinetic term after dimensional reduction given by $X:=-\frac{1}{2}g_{\mu\nu}\partial^{\mu}\phi\partial^{\nu}\phi$.
 In this context $(T,T^{\dagger})$ are the four dimensional background SUGRA moduli fields
 which are constant after dimensional reduction. 

The one-loop corrected Coleman Weinberg potential is given by \cite{sayan1}:
\be\begin{array}{lllllllll}\label{modelpot} 
\displaystyle V(\phi)=\sum^{2}_{m=-2,m\neq-1}C_{2m}\left[1+D_{2m}\ln\left(\frac{\phi}{M}\right)\right]\phi^{2m},\end{array}\ee
where $D_{0}=0$ and the other constants are function of 
effective the brane tension for D3 brane and constant moduli in 4D.
Hence using equation(\ref{model1}) the modified {\it Friedman } equation in presence of effective 4D Gauss-Bonnet coupling
can be expressed as \cite{sayan1}:
\be\label{fr1}
H^{4}=\frac{\Lambda_{(4)}+8\pi G_{(4)}\rho_{\phi}}{\tilde{g}_{1}}\approx\frac{\Lambda_{(4)}+8\pi G_{(4)}V(\phi)}{\tilde{g}_{1}},\ee

where $\rho_{\phi}$ plays the role of energy density of the inflation in 4D effective theory, $\tilde{g}_{1}$ represents the effective 4D Gauss-Bonnet coupling dependent function on FLRW background which can be expressed in terms of the brane tension of $D3$ brane 
and $\Lambda_{(4)}$ is the
4D effective cosmological constant. It is important to note that in the 4D effective action as stated in Eq~(\ref{model1}), the contribution of higher curvature effective Gauss-Bonnet like correction 
term is dominant compared to Ricci scalar. More precisely one can interpret this
to be a non-perturbative solution of the effective field theory where the effective coupling parameter $\tilde{l}_{4}>>\tilde{l}_{1}$. Consequently the 
effective Friedmann equation in 4D takes a non-trivial form in the high energy regime, where energy density of the inflaton $\rho_{\phi}\approx V(\phi)>>\tilde{g}_{1}$ 
of $D3-\bar{D3}$ system. Here Eq~(\ref{fr1}) also implies that within our prescribed setup 
the non-perterbative regime of effective field theory cannot able to produce the well known solutions of GR in the low energy 
limiting situation where $\rho_{\phi}\approx V(\phi)<<\tilde{g}_{1}$. But in the perturbative regime of the effective theory the situation is completely 
different compared the non-perturbative case. In the regime where the effective coupling parameter $\tilde{l}_{4}<<\tilde{l}_{1}$,
it is possible to get 
back the known solution of GR. In literature it usually identified to be the low energy regime,
where the inflaton energy density, $\rho_{\phi}\approx V(\phi)<<\tilde{g}_{1}$ in $D3-\bar{D3}$ system. But in the high energy regime, 
where $\rho_{\phi}\approx V(\phi)>>\tilde{g}_{1}$, it is not possible to realize the essence of the higher curvature terms
through Fridemann equations, which will
finally control the cosmological dynamics in a non-trivial manner. For more details see ref.~\cite{sayan1}, where the Friedmann equations are derived in
detail.


\section{\bf Tree Level Bispectrum Analysis}
\subsection{\bf Three scalar correlation}
To calculate the scalar bispectrum for D3 DBI Galileon we consider here the third order action up to total derivatives.
Using the uniform field gauge analysis the third order action for three scalar interaction can be written as:
\be\begin{array}{llllll}\label{opiu}
 \displaystyle S_{\zeta\zeta\zeta}
=\int dt d^{3}x \left\{a^{3}\bar{C}_{1}M^{2}_{PL}\zeta{\dot{\zeta}}^{2}+a\bar{C}_{2}M^{2}_{PL}\zeta(\partial\zeta)^{2}
+a^{3}\bar{C}_{3}M_{PL}{\dot{\zeta}}^{3}+a^{3}\bar{C}_{4}\dot{\zeta}(\partial_{i}\zeta)(\partial_{i}\tilde{\chi})
+a^{3}\left(\frac{\bar{C}_{5}}{M^{2}_{PL}}\right)\partial^{2}\zeta(\partial\tilde{\chi})^{2}\right.\\ \left.
\displaystyle~~~~~~~~~~~~a\bar{C}_{6}{\dot{\zeta}}^{2}\partial^{2}\zeta+\left(\frac{\bar{C}_{7}}{a}\right)
\left[\partial^{2}\zeta(\partial\zeta)^{2}-\zeta\partial_{i}\partial_{j}(\partial_{i}\zeta)(\partial_{j}\zeta)\right]
+a\frac{\bar{C}_{8}}{M_{PL}}\left[\partial^{2}\zeta\partial_{i}\zeta\partial_{i}\tilde{\chi}
-\zeta\partial_{i}\partial_{j}(\partial_{i}\zeta)(\partial_{j}\tilde{\chi})
\right]+{\cal R}\frac{\delta{\cal L}_{2}}{\delta\zeta}|_{1}\right\},  
   \end{array}\ee

where 
\be\label{lagden}\frac{\delta{\cal L}_{2}}{\delta\zeta}|_{1}=-2\left[\frac{d}{dt}\left(a^{3}Y_{S}\dot{\zeta}\right)-aY_{S}c^{2}_{s}\partial^{2}\zeta\right]\ee
 can be calculated from 
the second order action \cite{sayan1} 
\be\label{sect}\left(S^{(4)}\right)_{\zeta\zeta}=\int dt d^{3}xa^{3}Y_{S}\left[\dot{\zeta}^{2}-\frac{c^{2}_{s}}{a^{2}}(\partial\zeta)^{2}\right].\ee

Here $\bar{C}_{i}(i=1,2,3,......,8)$ are dimensionless co-efficients defined as:\\
\be\begin{array}{lllll}\label{jhg}\displaystyle\bar{C}_{1}=\frac{Y_{S}}{M^{2}_{PL}}\left[3-\frac{L_{1}H}{c^{2}_{s}}\left(3+\frac{\dot{Y_{S}}}{HY_{S}}\right)
+\frac{d}{dt}\left(\frac{L_{1}}{c^{2}_{s}}\right)\right], \displaystyle  \bar{C}_{2}=
\left[1+\frac{1}{a}\frac{d}{dt}\left(aL_{1}\{Y_{S}-t_{1}\}\right)\right],\\ \displaystyle
\bar{C}_{3}=\frac{L_{1}}{M_{PL}}\left[L_{1}(L_{1}a_{1}+a_{3})+a_{12}+(a_{9}+L_{1}a_{4})
\frac{Y_{S}}{t_{1}}+\frac{Y_{S}}{c^{2}_{s}}\right], \\ \displaystyle 
\bar{C}_{4}=-\frac{Y_{S}}{2t_{1}}\left\{1+2t_{1}\left[\frac{d}{dt}\left(\frac{A_{5}}{t^{2}_{1}}\right)
-\frac{3HA_{5}}{t^{2}_{1}}\right]\right\}, \displaystyle\bar{C}_{5}=\frac{M^{2}_{PL}}{2t^{2}_{1}}\left[\frac{3M^{2}_{PL}}{2}\left(1-HL_{1}\right)
\right]-\frac{M^{2}_{PL}}{2}\frac{d}{dt}\left(\frac{A_{5}}{t^{2}_{1}}\right), \\ \displaystyle
\bar{C}_{6}=L^{2}_{1}\left[2M^{2}_{PL}-L_{1}a_{4}\right], \displaystyle
\bar{C}_{7}=\frac{L^{2}_{1}M^{2}_{PL}(1-HL_{1})}{6}-\frac{c^{2}_{s}Y_{S}L^{2}_{1}M^{2}_{PL}}{2t_{1}}
+\frac{M^{2}_{PL}}{6}\frac{d}{dt}\left(L^{3}_{1}\right), \\ \displaystyle
\bar{C}_{8}=M_{PL}\left\{\frac{L_{1}M^{2}_{PL}}{t_{1}}\left(HL_{1}-1\right)
+\frac{c^{2}_{s}Y_{S}L_{1}M^{2}_{PL}}{t^{2}_{1}}\right\} \end{array}\ee
and the co-efficient of $\frac{\delta{\cal L}_{2}}{\delta\zeta}|_{1}$ involves spatial and time derivative in equation(\ref{opiu}) is defined by
the following expression: 
\be\begin{array}{lllll}\label{corr}\displaystyle
{\cal R}=\frac{A_{5}}{t^{2}_{1}}\left\{(\partial_{k}\zeta)(\partial_{k}\tilde{\chi})
-\partial^{-2}\partial_{i}\partial_{j}\left[(\partial_{i}\zeta)(\partial_{j}\tilde{\chi})\right]\right\}
+p_{1}\zeta\dot{\zeta}-\frac{A_{5}L_{1}}{2t_{1}a^{2}}\left\{(\partial\zeta)^{2}-
\partial^{-2}\partial_{i}\partial_{j}\left[(\partial_{i}\zeta)(\partial_{j}\zeta)\right]\right\}.\end{array}\ee

In this context ${\cal R}\rightarrow 0$ as $k\rightarrow 0$ at large scale.
 Additionally 
\be \begin{array}{lllll}\label{newconty}\displaystyle L_{1}=\left(\frac{M^{2}_{PL}}{HM^{2}_{PL}-\dot{\phi}X\tilde{G}_{X}}\right),
\displaystyle 
 \tilde{\chi}=\partial^{-2}(Y_{S}\dot{\zeta}),~~~ A_{3}=2Y_{S},~~~ A_{5}=-\frac{L_{1}M^{2}_{PL}}{2},\\
\displaystyle Y_{S}=\frac{t_{1}\left(4t_{1}t_{3}+9t^{2}_{2}\right)}{3t^{2}_{2}}, ~~~~~~~
c^{2}_{s}=\frac{3\left(2Ht_{2}t^{2}_{1}-t_{4}t^{2}_{2}-2t^{2}_{1}\dot{t}_{2}\right)}{t_{1}\left(4t_{1}t_{3}+9t^{2}_{2}\right)},
\displaystyle t_{1}=\tilde{l}_{1},~~~ t_{2}=\left(2H\tilde{l}_{1}-2\dot{\phi}X\tilde{G}_{X}\right),\\
\displaystyle t_{3}=-9\tilde{l}_{1}H^{2}+3\left(X\hat{\tilde{K}}_{X}+2X^{2}\hat{\tilde{K}}_{XX}\right)+18H\dot{\phi}
\left(2X\tilde{G}_{X}+X^{2}\tilde{G}_{XX}\right),\\
\displaystyle a_{1}=3M^{2}_{PL}H^{2}-X\hat{\tilde{K}}_{X}-4X^{2}\hat{\tilde{K}}_{XX}-\frac{4X^{3}}{3}X\hat{\tilde{K}}_{XXX}-2H\dot{\phi}
\left(10X\tilde{G}_{X}+11X^{2}\tilde{G}_{XX}+2X^{3}\tilde{G}_{XXX}\right) \\
~~~~~~~~~~~~~~~~~~~~~~~~~~~~~~~~\displaystyle~~~~~~~~~~~~~~~~~~~~~~~~~~~~~~~~~~~~~~~~
~~~~~~~~~~~~~~~~~~+2X\tilde{G}_{\phi}+\frac{14X^{2}}{3}\tilde{G}_{\phi X}+\frac{4X^{3}}{3}\tilde{G}_{\phi XX},\\
\displaystyle a_{3}=-3a_{4}=-3\left[2M^{2}_{PL}H-2\dot{\phi}\left(2X\tilde{G}_{X}+X^{2}\tilde{G}_{XX}\right)\right],
\displaystyle a_{9}=-\frac{2}{3}a_{12}=-2M^{2}_{PL}.
\end{array}\ee
It is important to mention here that for scalar and tensor modes {\it ghosts} and {\it Laplacian} instabilities
can be avoided iff $c^{2}_{s}>0, Y_{s}>0$. Throughout the paper we use the required parameters from \cite{sayan1}
to compute the bispectrum and trispectrum.

Now following the prescription of {\it in-in formalism} in the interacting picture the {\it three point correlation function} for quasi-exponential limit,
after some trivial algebra, look:
\be\begin{array}{llll}\label{threept}
\displaystyle\langle\zeta(\vec{k_{1}})\zeta(\vec{k_{2}})\zeta(\vec{k_{3}})\rangle
=-i\sum^{8}_{j=1}\int^{0}_{-\infty}d\eta~ a ~\langle 0|\left[\zeta(\vec{k_{1}})\zeta(\vec{k_{2}})\zeta(\vec{k_{3}}),\left(H^{(j)}_{int}(\eta)\right)_{\zeta\zeta\zeta}\right]|0\rangle\\
~~~~~~~~~~~~~~~~~~~~~~~\displaystyle=(2\pi)^{3}\delta^{(3)}
(\vec{k_{1}}+\vec{k_{2}}+\vec{k_{3}}){\cal B}_{\zeta\zeta\zeta}(\vec{k_{1}},\vec{k_{2}},\vec{k_{3}}),\end{array}\ee
where the total Hamiltonian in the interaction picture can be expressed in terms of the third order Lagrangian density as
$\left(H_{int}(\eta)\right)_{\zeta\zeta\zeta}=\sum^{8}_{j=1}\left(H^{(j)}_{int}(\eta)\right)_{\zeta\zeta\zeta}
=-\int d^{3}x \left({\cal L}_{3}\right)_{\zeta\zeta\zeta}.$
 Throughout this article we use the Bunch-Davies mode function as
\be\begin{array}{llll}\label{ra2}\displaystyle  u_{m}(\eta,k)=\frac{\sqrt{-k\eta c_{m}}}{a\sqrt{2Y_{m}}}{\cal H}^{(1)}_{\nu_{m}}(-k\eta c_{m})
\rightarrow\frac{\left(-kc_{m}\eta\right)^{\frac{1}{2}-\nu_{m}}
\exp\left(i[\nu_{m}-\frac{1}{2}]\frac{\pi}{2}\right)2^{\nu_{m}
-\frac{3}{2}}}{2a\sqrt{Y_{m}c_{m}k}}\left(\frac{\Gamma(\nu_{m})}{\Gamma(\frac{3}{2})}\right)\end{array}\ee with $m=(S[scalar],T[tensor])$.
 Moreover following the momentum dependent ansatz given in \cite{felice},\cite{dan}
the {\it bispectrum} ${\cal B}_{\zeta\zeta\zeta}(\vec{k_{1}},\vec{k_{2}},\vec{k_{3}})$ is defined as:
\be\label{bispec}
{\cal B}_{\zeta\zeta\zeta}(\vec{k_{1}},\vec{k_{2}},\vec{k_{3}})=\frac{(2\pi)^{4}{\cal P}^{2}_{\zeta}}{\prod^{3}_{i=1}k^{3}_{i}}
{\cal A}_{\zeta\zeta\zeta}(\vec{k_{1}},\vec{k_{2}},\vec{k_{3}})=\frac{6}{5}f^{}_{NL;1}{ P}^{2}_{\zeta}\ee
where the symbol~ $;1$ is used for three scalar correlation.
Here ${\cal A}_{\zeta\zeta\zeta}(\vec{k_{1}},\vec{k_{2}},\vec{k_{3}})$ is the {\it shape function} for bispectrum and ${ P}^{2}_{\zeta}$ is used for 
normalization of E-mode polarization
expressed in terms of the new combination of the cyclic permutations of two-point correlation functions given by  
\be\label{bipow} { P}^{2}_{\zeta}={ P}_{\zeta}({k_{1}}){ P}_{\zeta}({k_{2}})+{ P}_{\zeta}({k_{2}})
{P}_{\zeta}({k_{3}})+{P}_{\zeta}({k_{3}}){P}_{\zeta}({k_{1}}).\ee
The {\it Power spectra} for scalar (${ P}_{\zeta}(k)$) and tensor modes (${ P}_{T}(k)$) at the horizon crossing can be written as:
\be\begin{array}{lllll}\label{ght1}
\displaystyle { P}_{\zeta}(k)=
\left(2^{2\nu_{s}-3}\left|\frac{\Gamma(\nu_{s})}{\Gamma(\frac{3}{2})}\right|^{2}\frac{\left(1-\epsilon_{V}-s^{S}_{V}\right)^{2}\sqrt{V(\phi)}}{8\pi^{2}Y_{S}
c^{3}_{s}\sqrt{\tilde{g}_{1}}M_{PL}}\right),
~~~~ \displaystyle { P}_{T}(k)=
 \left(2^{2\nu_{T}-3}\left|\frac{\Gamma(\nu_{T})}{\Gamma(\frac{3}{2})}\right|^{2}\frac{\left(1-\epsilon_{V}-s^{T}_{V}\right)^{2}\sqrt{V(\phi)}}{2\pi^{2}Y_{T}
c^{3}_{T}\sqrt{\tilde{g}_{1}}M_{PL}}\right).
          \end{array}
\ee
 Here for tensor modes we use 
$\left({P}_{T}(k)\right)_{ij;kl}=|u_{h}(\eta,k)|^{2}{\cal N}_{ij;kl},
 \displaystyle { P}_{T}(k)=\left({ P}_{T}(k)\right)_{ij;ij}$
 with the following helicity/spin dependent normalization factor:
${\cal N}_{ij;kl}=\sum_{\lambda}e^{\lambda}_{ij}(\vec{k})e^{\dagger(\lambda)}_{kl}(\vec{k}).$

 In this context $f^{}_{NL}$ represents 
the non linear parameter carrying the signature of primordial non-Gaussianities of the curvature perturbation in bispectrum. The explicit
 form of $f^{}_{NL}$ characterizing 
the bispectrum can be expressed as:

\be\begin{array}{llll}\label{shaefn1} \displaystyle f^{}_{NL;1}=\frac{10}{3\sum^{3}_{i=1}k^{3}_{i}}\left(\frac{k_{1}k_{2}k_{3}}{2K^{3}}\right)^{n_{\zeta}-1}
\left|\frac{\Gamma(\nu_{s})}{\Gamma(\frac{3}{2})}\right|^{2}\left\{\bar{C}_{1}\left[\frac{3}{4}{\cal I}_{1}(n_{\zeta}-1)
-\frac{3-\epsilon_{V}}{4c^{2}_{s}}\left(\frac{1+Y_{S}}{1+\epsilon_{V}}\right)^{2}{\cal I}_{1}(\tilde{\nu})\right]
\right.\\ \left.\displaystyle~~~~~~~~~~~~~~~~~+\frac{3(1-\epsilon_{V}-s^{S}_{V})}{2Y_{S}}\left[{\cal F}_{3}{\cal I}_{3}(n_{\zeta}-1)
+\frac{{\cal E}_{3}}{c^{2}_{s}}{\cal I}_{3}(\tilde{\nu})\right]+\frac{\bar{C}_{4}}{8}{\cal I}_{4}(\tilde{\nu})+\frac{\bar{C}_{5}Y_{S}}{4c^{2}_{s}}{\cal I}_{5}(\tilde{\nu})
\right.\\ \left.\displaystyle ~~~~~~~~~+\frac{3(1-\epsilon_{V}-s^{S}_{V})^{2}}{Y_{S}}\left[{\cal F}_{6}{\cal I}_{6}(n_{\zeta}-1)
+\frac{{\cal E}_{6}}{c^{2}_{s}}{\cal I}_{6}(\tilde{\nu})\right]+\frac{\bar{C}_{7}(1-\epsilon_{V}-s^{S}_{V})^{2}}{2Y_{S}c^{2}_{s}}{\cal I}_{7}(\tilde{\nu})
+\frac{\bar{C}_{8}(1-\epsilon_{V}-s^{S}_{V})}{8c^{2}_{s}}{\cal I}_{8}(\tilde{\nu})\right\}.
          \end{array}\ee
where the functional form of the momentum dependent functions ${\cal I}_{i}(x)\forall i$ are explicitly mentioned in the Appendix B.1.
From the coefficients of ${\cal I}_{i}(\tilde{\nu})$ with $i=1,3,5,7,8$ it seems that the non-Gaussian parameter $f^{}_{NL;1}$ is 
inverse proportional to the sound speed square for the scalar mode. But these co-efficients are not solely characterized by the
sound speed for scalar mode since they depend on other factors like (1) effective Gauss-Bonnet coupling ($\alpha_{(4)}$) and (2) 
higher order interaction between graviton and DBI Galileon in presence of quadratic correction of gravity in Einstein-Hilbert action.
Additionally in this context counter terms which appears as the coefficients of 
${\cal I}_{i}(n_{\zeta}-1)$ with $i=1,3,6$ and ${\cal I}_{4}(\tilde{\nu})$ 
originated from the effective Gauss-Bonnet coupling ($\alpha_{(4)}$) and
 higher order interaction between graviton (via Gauss-Bonnet correction) 
and DBI Galileon degrees of freedom in D3 brane in the background of
 four dimensional ${\cal N}$=1 SUGRA multiplet play a very crucial role in this context.
In $\alpha_{(4)}\neq~0$ limit such counter terms and dependence on the interaction between graviton and higher derivative DBI Galileon 
cannot be negligible in the slow-roll limit.
 Consequently, depending on the signature and the strength of the effective Gauss-Bonnet coupling three situation arises:
(1)the counter terms drives other terms, (2)the counter terms and other terms are tuned in such a way that the system is in equilibrium 
with respect to the sound speed and (3)the sound speed dominated terms win the war. Here the second situation is not physically interesting and the third
situation leads to the trivial feature of DBI Galileon. Only the non-trivial features comes from the first situation 
 in the context of single field DBI Galileon inflation.

In equation(\ref{shaefn1}) we have defined  $K=k_{1}+k_{2}+k_{3}$, $x=(n_{\zeta}-1,\tilde{\nu})$ and 
\be\begin{array}{llll}\label{constlab}\displaystyle\tilde{\nu}:=\left(\frac{s^{S}_{V}-2\epsilon_{V}}{1-\epsilon_{V}-s^{S}_{V}}\right),
\displaystyle {n}_{\zeta}-1=\left(3-2\nu_{s}\right)=-\left(\frac{2\epsilon_{V}+s^{S}_{V}+\delta_{V}}{1-\epsilon_{V}-s^{S}_{V}}\right)\\
\displaystyle{\cal F}_{3}:=-\frac{Y_{S}(1+Y_{S})}{1+\epsilon_{V}}\left[1+2\frac{Y_{S}-
\epsilon_{V}+(1+Y_{S})\rho_{3}}{1+\epsilon_{V}}+2{\cal T}_{3}\right],
\displaystyle\frac{\dot{\phi}X^{2}\tilde{G}_{XX}}{H}=\left(\rho_{3}+\frac{\rho_{4}}{c^{2}_{s}}\right),
\displaystyle\frac{\nu_{s}}{\Sigma_{G}}:=\left({\cal T}_{3}+\frac{{\cal T}_{4}}{c^{2}_{s}}\right),\\
\displaystyle {\cal E}_{3}:=-\frac{Y_{S}(1+Y_{S})}{1+\epsilon_{V}}\left[2{\cal T}_{4}-\frac{1+Y_{S}}{1+\epsilon_{V}}(1-2\rho_{4})\right],
 \displaystyle {\cal F}_{6}:=\frac{2(1+Y_{S})^{3}}{(1+\epsilon_{V})^{3}}\left[\frac{Y_{S}-\epsilon_{V}}{1+Y_{S}}+\rho_{3}\right],
\displaystyle {\cal E}_{6}:=\frac{2\rho_{4}(1+Y_{S})^{3}}{(1+\epsilon_{V})^{3}}\end{array}\ee
 with four new constants 
$\rho_{3},\rho_{4},{\cal T}_{3},{\cal T}_{4}$. In the present context $s^{S}_{V}=\frac{\dot{c}_{s}}{Hc_{s}}$ is an extra
 slow-roll parameter appearing due to the sound speed, $c_{s}\neq 1$ as defined in \cite{sayan1}.
For the numerical estimation we have further used the {\it equilateral configuration} $(k_{1}=k_{2}=k_{3}=k$ and $K=3k)$ in which 
the non-linear parameter $f_{NL}$ can be simplified to the following form as:

 \be\begin{array}{llll}\label{shapeas1}
\displaystyle  f^{equil}_{NL;1}=\frac{10}{9k^{3}}\left(\frac{1}{54}\right)^{n_{\zeta}-1}
\left|\frac{\Gamma(\nu_{s})}{\Gamma(\frac{3}{2})}\right|^{2}\left\{
\left(3\left(1-\frac{1}{c^{2}_{s}}\right)-\frac{Y_{S}\delta_{V}}{c^{2}_{s}}+\frac{Y^{2}_{S}}{c^{2}_{s}}
-\frac{2Y_{S}s^{S}_{V}}{c^{2}_{s}}\right)\right.\\\left. \displaystyle~~~~~~~~~~~~~~~~~~\times\left[\frac{3}{4}{\cal I}^{equil}_{1}(n_{\zeta}-1)
-\frac{3-\epsilon_{V}}{4c^{2}_{s}}\left(\frac{1+Y_{S}}{1+\epsilon_{V}}\right)^{2}{\cal I}^{equil}_{1}(\tilde{\nu})\right]
+\frac{3(1-\epsilon_{V}-s^{S}_{V})}{2Y_{S}}\left[{\cal F}_{3}{\cal I}^{equil}_{3}(n_{\zeta}-1)
+\frac{{\cal E}_{3}}{c^{2}_{s}}{\cal I}^{equil}_{3}(\tilde{\nu})\right]\right.\\ \left.
\displaystyle~~~~~~~~~~~~~~~~~~-\frac{1}{8}\left[\frac{Y_{S}}{2}+\frac{Y_{S}}{2}\left(3-Y_{S}\right)\right]{\cal I}^{equil}_{4}(\tilde{\nu})
+\frac{Y_{S}}{4c^{2}_{s}}\left(\frac{4\epsilon_{V}-Y_{S}(3-\epsilon_{V})}{4(1+\epsilon_{V})}\right){\cal I}^{equil}_{5}(\tilde{\nu})
\right.\\ \left.\displaystyle~~~~~~~~~~~~~~~~~~ +\frac{3(1-\epsilon_{V}-s^{S}_{V})^{2}}{Y_{S}}\left[{\cal F}_{6}{\cal I}^{equil}_{6}(n_{\zeta}-1)
+\frac{{\cal E}_{6}}{c^{2}_{s}}{\cal I}^{equil}_{6}(\tilde{\nu})\right]
-\frac{(1-\epsilon_{V}-s^{S}_{V})^{2}(1+Y_{S})^{2}(Y_{S}-\epsilon_{V})}{2Y_{S}c^{2}_{s}(1+\epsilon_{V})^{3}}{\cal I}^{equil}_{7}(\tilde{\nu})
\right.\\ \left.\displaystyle~~~~~~~~~~~~~~~~~~~~
+\frac{(1+Y_{S})(Y_{S}-\epsilon_{V})(1-\epsilon_{V}-s^{S}_{V})}{4c^{2}_{s}(1+\epsilon_{V})^{2}}{\cal I}^{equil}_{8}(\tilde{\nu})\right\}.\end{array}\ee

 Now using the tensor-to-scalar ratio at the pivot scale $k_{\ast}$:

 \be\begin{array}{lll}\label{ghtv3}\displaystyle {r}=\left(16. 2^{2(\nu_{T}-\nu_{s})}\left|\frac{\Gamma(\nu_{T})}{\Gamma(\nu_{s})}\right|^{2}\left(\frac{1-\epsilon_{V}-s^{T}_{V}}
{1-\epsilon_{V}-s^{S}_{V}}\right)^{2}c_{s}\epsilon_{s}\left[1-\frac{3}{2}{\cal O}(\epsilon^{2}_{T})\right]\right)_{\star}\end{array}\ee
the sound speed $c_{s}$ can be eliminated from the equation(\ref{shapeas1}) also.

Here $s^{T}_{V}=\frac{\dot{c}_{T}}{Hc_{T}}$ appearing due to the sound speed, $c_{T}\neq 1$. See \cite{sayan1} for the details.
The numerical value of $f^{equil}_{NL;1}$ in the equilateral limit
is obtained from our set up as $4<f^{equil}_{NL;1}<7$
within the window for tensor-to-scalar ratio $0.213<r<0.250$ \cite{sayan1}.  
This is extremely interesting result as it is different from other class of DBI models. The most impressing fact is that 
 the upper bound of $f^{equil}_{NL;1}$ in the quasi-exponential limit are in good agreement with combined constraint obtained
 from {\it Planck+WMAP9+high-L+BICEP2} \cite{Ade:2014xna,planck} data.

\subsection{\bf One scalar two tensor correlation}
After applying the gauge fixing condition to uniform gauge the one scalar and two tensor interaction  can be represented by the following third order action:
\be\begin{array}{lllll}\label{sstc}
    \displaystyle S_{\zeta h h}
=\int dtd^{3}x~a^{3}\left\{{\cal F}_{1}\zeta\dot{h_{ij}}^{2}+ +\frac{{\cal \tilde{F}}_{2}}{a^2}\zeta h_{ij,k}h_{ij,k}
+{\cal \tilde{F}}_{3}\psi_{,k}\dot h_{ij}h_{ij,k}+{\cal F}_{4}\dot\zeta\dot h_{ij}^2
+\frac{{\cal \tilde{F}}_{5}}{a^2}\partial^2\zeta \dot h_{ij}^2\right.\\ \left.\displaystyle~~~~~~~~~~~~~~~~~~~~~~~~~~~~~~~~~~~~~~~~~~~~~~~~~~~~~~~~~~~~~~~~~+{\cal \tilde{F}}_{6}\psi_{,ij}\dot h_{ik}\dot h_{jk}
+\frac{{\cal \tilde{F}}_{7}}{a^2}\zeta_{,ij}\dot h_{ik}\dot h_{jk}\right\}
   \end{array}\ee
where the dimensionful coefficients ${\cal F}_{i}(i=1,2.....7)$ are defined as:
\be\begin{array}{lllll}\label{ttwsha}
   \displaystyle {\cal \tilde{F}}_{1}=3Y_{T}\left[1-\frac{HL_{1}Y_{T}}{c^{2}_{T}}
+\frac{Y_{T}}{3}\frac{d}{dt}\left(\frac{L_{1}}{c^{2}_{T}}\right)\right],~~~{\cal \tilde{F}}_{2}=Y_{s}c^{2}_{s},~~~{\cal \tilde{F}}_{3}=-2Y_{s},\\ \displaystyle
{\cal \tilde{F}}_{4}=\frac{L_{1}}{c^{2}_{T}}\left(Y^{2}_{T}-\hat{\tilde{K}}_{XX}\right)
+2\sigma\left[\frac{Y_{s}}{Y_{T}}-1-\frac{HL_{1}Y_{T}}{c^{2}_{T}}\left(6+\frac{\dot{Y_{s}}}{HY_{s}}\right)\right]
+2Y^{2}_{T}\frac{d}{dt}\left(\frac{\sigma L_{1}}{Y_{T}c^{2}_{T}}\right),\\ \displaystyle
{\cal \tilde{F}}_{5}=2\sigma Y_{T}L_{1}\left(\frac{c^{2}_{s}}{c^{2}_{T}}-1\right),~~~{\cal \tilde{F}}_{6}=-\frac{4\sigma Y_{s}}{Y_{T}},~~~{\cal \tilde{F}}_{7}=4\sigma Y_{T}L_{1}
   \end{array}\ee
where we use $\sigma=\dot\phi XG_{5X}$. 
Now following the prescription of {\it in-in formalism} in the interaction picture {\it three point one scalar two tensor correlation function}
can be expressed in the following form:
\be\begin{array}{llll}\label{threept}
\displaystyle\langle\zeta(\vec{k}_1)h_{ij}(\vec{k}_2)h_{kl}(\vec{k}_3)\rangle
=-i\sum^{7}_{q=1}\int^{0}_{-\infty}d\eta~ a ~\langle 0|\left[\zeta(\vec{k}_1)h_{ij}(\vec{k}_2)h_{kl}(\vec{k}_3),\left(\left[H^{(q)}_{int}(\eta)\right]_{ij;kl}\right)_{\zeta hh}\right]|0\rangle\\
~~~~~~~~~~~~~~~~~~~~~~~~~~~~\displaystyle=(2\pi)^3\delta^{(3)}(\vec{k}_1+\vec{k}_2+\vec{k}_3)\left\{B_{\zeta hh}\right\}_{ij;kl}(
\vec{k}_1,\vec{k}_2,\vec{k}_3),\end{array}\ee
where the total Hamiltonian in the interaction picture can be expressed in terms of the third order Lagrangian density as
 $\left(\left[H_{int}(\eta)\right]_{ij;kl}\right)_{\zeta hh}=\sum^{7}_{q=1}\left(\left[H^{(q)}_{int}(\eta)\right]_{ij;kl}\right)_{\zeta hh}=-\int d^{3}x\left[ \left({\cal L}_{3}\right)_{\zeta hh}\right]_{ij;kl}$. 
 Moreover
the {\it cross bispectrum} $\left\{B_{\zeta hh}\right\}_{ij;kl}(
\vec{k}_1,\vec{k}_2,\vec{k}_3)$ is defined as:
\be\label{bispec}
\left\{B_{\zeta hh}\right\}_{ij;kl}(
\vec{k}_1,\vec{k}_2,\vec{k}_3)=\frac{(2\pi)^{4}{\cal P}^{2}_{u}}{\prod^{3}_{i=1}k^{3}_{i}}
\left({\cal A}_{\zeta hh}\right)_{ij;kl}(\vec{k_{1}},\vec{k_{2}},\vec{k_{3}})=\frac{6}{5}\left[f^{}_{NL;2}\right]^{u}_{ij;kl}{ P}^{2}_{u}\ee
 where the symbol  $;2$ stands for one scalar two tensor correlation.
Here $\left({\cal A}_{\zeta hh}\right)_{ij;kl}(\vec{k_{1}},\vec{k_{2}},\vec{k_{3}})$ is the {\it shape function} for bispectrum and 
the polarization indices are $u=1(E-mode),2(E\bigotimes B-mode),3(B-mode)$.
We adopt the following normalization depending on the polarization in which we are interested: 
\be\label{bipow} { P}^{2}_{u}=
\left\{
	\begin{array}{ll}
                    \displaystyle { P}_{\zeta}({k_{1}}){ P}_{\zeta}({k_{2}})+{ P}_{\zeta}({k_{2}})
{P}_{\zeta}({k_{3}})+{P}_{\zeta}({k_{3}}){P}_{\zeta}({k_{1}})& \mbox{ \it:{\cal u=1(E-mode)}}  \\
         \displaystyle  { P}_{\zeta}({k_{1}}){ P}_{h}({k_{2}})+{ P}_{\zeta}({k_{2}})
{P}_{h}({k_{3}})+{P}_{\zeta}({k_{3}}){P}_{h}({k_{1}})& ~\mbox{\it:{\cal u=2(E$\bigotimes$B mode)}}\\
{ P}_{h}({k_{1}}){ P}_{h}({k_{2}})+{ P}_{h}({k_{2}})
{P}_{h}({k_{3}})+{P}_{h}({k_{3}}){P}_{h}({k_{1}})& ~\mbox{\it:{\cal u=3(B-mode)}}.
          \end{array}
\right.
\ee
Consequently $\left[f^{}_{NL;2}\right]^{u}_{ij;kl}$ represents 
the non-linear parameter which carries the signature of primordial non-Gaussianities of the one scalar two tensor interaction. The explicit
 form of $\left[f^{}_{NL;2}\right]^{u}_{ij;kl}$ characterizing 
the bispectrum can be calculated as:
 \be\begin{array}{llll}\label{shapefnzxs}       
\displaystyle\left[f^{}_{NL;2}\right]^{\it u}_{ij;kl}=
\frac{10{\cal Q}^{POL}_{u}}{3\sum^{3}_{i=1}k^{3}_{i}}\frac{\left(\frac{3}{2}-\nu_{T}\right)^2
\underline{K}^{4\nu_T+2\nu_s-9}\left[Cos\left(\left[\nu_{s}-\frac{1}{2}\right]\frac{\pi}{2}\right)\right]^{\frac{1}{3}}
\left[Cos\left(\left[\nu_{T}-\frac{1}{2}\right]\frac{\pi}{2}\right)\right]^{\frac{2}{3}}}{c^{2\nu_s-3}_{s}c^{4\nu_T-6}_{T}\left(k_{1}\right)^{\nu_s}\left(k_{2}k_{3}\right)^{\nu_T}}\\
\displaystyle~~~~~~~~~~~~~~~~~~~~~~~~~~~ \times\left(2^{2\nu_{s}+4\nu_{T}-12}\left|\frac{\Gamma(\nu_{s})}{\Gamma(\frac{3}{2})}\right|^{2}\left|\frac{\Gamma(\nu_{T})}
{\Gamma(\frac{3}{2})}\right|^{4}\frac{\left(1-\epsilon_{V}-s^{S}_{V}\right)^{2}\left(1-\epsilon_{V}-s^{T}_{V}\right)^{4}{V^{\frac{3}{2}}(\phi)}}{Y_{S}Y^2_{T}
c^{4}_{T}c^{3}_{s}{\tilde{g}^{\frac{3}{2}}_{1}}M^3_{PL}}\right) \\~~~~~~~~~~~~~~~~~~~~~~~~~~~~~~
\displaystyle \times\left[32{\cal \tilde{F}}_{1}\left(\nabla_{1}\right)^u_{ij;kl}+4{\cal \tilde{F}}_{2}\left(\nabla_{2}\right)^u_{ij;kl}
+2\left({\cal \tilde{F}}_{3}\left(\nabla_{3}\right)^u_{ij;kl}\right.\right.\\ \left.\left.\displaystyle ~~~~~~~~~~~~~~~~~~~~~~~~~~~~~~~~~~~~~~
+{\cal \tilde{F}}_{4}\left(\nabla_{4}\right)^u_{ij;kl}
+{\cal \tilde{F}}_{5}\left(\nabla_{5}\right)^u_{ij;kl}+{\cal \tilde{F}}_{6}\left(\nabla_{6}\right)^u_{ij;kl}
+{\cal \tilde{F}}_{7}\left(\nabla_{7}\right)^u_{ij;kl}\right)\right]
          \end{array}\ee
with polarization index $u=1(E),2(E\bigotimes B),3(B)$.  
The functional form of the co-efficients $\left(\nabla_{i}\right)^u_{ij;kl}\forall i$ are explicitly mentioned in the Appendix B.2.
In this context we define $\underline{K}:=c_{s}k_{1}+c_{T}(k_{2}+k_{3})$.

The overall normalization factor for three types of polarization can be expressed as:
\be\label{normala}
{\cal Q}^{POL}_{u}=
\left\{
	\begin{array}{ll}
                    \displaystyle 8& \mbox{ \it:{\cal u=1(E-mode)}}  \\
         \displaystyle  128 & ~\mbox{\it:{\cal u=2(E$\bigotimes$B mode)}}\\
\displaystyle 2048 & ~\mbox{\it:{\cal u=3(B-mode)}}.
          \end{array}
\right.
\ee
Further, to make the computation simpler without loosing any essential information
 we reduce the complete set in terms of the two-polarization (helicity) mode instead of four complicated tensor indices. For this purpose let us define 
a reduced physical quantity:
\be\label{red} {\cal \bigoplus}^{\lambda}(\vec{k})=h_{ij}(\vec{k})e^{\dagger(\lambda)}_{ij}\ee
 in terms of which the one scalar two tensor correlation is defined as: 
\begin{equation}
  \langle\zeta(\vec{k}_1){\cal \bigoplus}^{\lambda_{2}}(\vec{k}_2){\cal \bigoplus}^{\lambda_{3}}(\vec{k}_3)\rangle
=(2\pi)^3\delta(\vec{k}_1+\vec{k}_2+\vec{k}_3)B^{(\lambda_{1};\lambda_{2})}_{(\zeta hh)}(
\vec{k}_1,\vec{k}_2,\vec{k}_3).
\end{equation}
where the {\it cross reduced bispectrum} is defined as:
\begin{eqnarray}
B^{(\lambda_{2};\lambda_{3})}_{(\zeta hh)}(
\vec{k}_1,\vec{k}_2,\vec{k}_3)=\frac{(2\pi)^4{\cal P}_{u}^2}{\prod^{3}_{i=1}k^{3}_{i}}
  {\cal A}^{(\lambda_{2};\lambda_{3})}_{(\zeta hh)}=\frac{6}{5}\left[f^{}_{NL;2}\right]^{u;(\lambda_{2};\lambda_{3})}{ P}^{2}_{u}.
\end{eqnarray}

Applying the basis transformation the explicit
 form of $\left[f^{}_{NL;2}\right]^{(\lambda_{2};\lambda_{3})}$ characterizing 
the crossed bispectrum can be written as:

\be\begin{array}{llll}\label{shapebnm1}\displaystyle
\left[f^{}_{NL;2}\right]^{\it u;(\lambda_{2};\lambda_{3})}=\frac{10{\cal Q}^{POL}_{u}}{3\sum^{3}_{i=1}k^{3}_{i}}\frac{\left(\frac{3}{2}-\nu_{T}\right)^2
\underline{K}^{4\nu_T+2\nu_s-9}\left[Cos\left(\left[\nu_{s}-\frac{1}{2}\right]\frac{\pi}{2}\right)\right]^{\frac{1}{3}}
\left[Cos\left(\left[\nu_{T}-\frac{1}{2}\right]\frac{\pi}{2}\right)\right]^{\frac{2}{3}}}{c^{2\nu_s-3}_{s}c^{4\nu_T-6}_{T}\left(k_{1}\right)^{\nu_s}\left(k_{2}k_{3}\right)^{\nu_T}}\\
\displaystyle ~~~~~~~~~~~~~~~~~~~~~~~~~~~~~~~~~~\times\left(2^{2\nu_{s}+4\nu_{T}-12}\left|\frac{\Gamma(\nu_{s})}{\Gamma(\frac{3}{2})}\right|^{2}\left|\frac{\Gamma(\nu_{T})}
{\Gamma(\frac{3}{2})}\right|^{4}\frac{\left(1-\epsilon_{V}-s^{S}_{V}\right)^{2}\left(1-\epsilon_{V}-s^{T}_{V}\right)^{4}{V^{\frac{3}{2}}(\phi)}}{Y_{S}Y^2_{T}
c^{4}_{T}c^{3}_{s}{\tilde{g}^{\frac{3}{2}}_{1}}M^3_{PL}}\right) \\
\displaystyle ~~~~~~~~~~~~~~~~~~~~~~~~~~~~~~~~
\times\left[32{\cal \tilde{F}}_{1}\left(\nabla_{1}\right)^{u;\lambda_{2};\lambda_{3}}+4{\cal \tilde{F}}_{2}\left(\nabla_{2}\right)^{u;\lambda_{2};\lambda_{3}}
+2\left({\cal \tilde{F}}_{3}\left(\nabla_{3}\right)^{u;\lambda_{2};\lambda_{3}}\right.\right.\\ \left.\left.\displaystyle ~~~~~~~~~~~~~~~~~~~~~~~~~~~~~~~~~~~~
+{\cal \tilde{F}}_{4}\left(\nabla_{4}\right)^{u;\lambda_{2};\lambda_{3}}
+{\cal \tilde{F}}_{5}\left(\nabla_{5}\right)^{u;\lambda_{2};\lambda_{3}}+{\cal \tilde{F}}_{6}\left(\nabla_{6}\right)^{u;\lambda_{2};\lambda_{3}}
+{\cal \tilde{F}}_{7}\left(\nabla_{7}\right)^{u;\lambda_{2};\lambda_{3}}\right)\right]
          \end{array}\ee
The functional form of the co-efficients $\left(\nabla_{i}\right)^{u;\lambda_{2};\lambda_{3}}\forall i$ after basis transformation 
are explicitly mentioned in the Appendix.
In the equilateral limit we have

\be\begin{array}{llll}\label{shapefn22z} 
\displaystyle \left[f^{equil}_{NL;2}\right]^{\bf u;(\lambda_{2};\lambda_{3})}=
\frac{10{\cal Q}^{POL}_{u}}{9k^{3}}\frac{\left(\frac{3}{2}-\nu_{T}\right)^2
((c_{s}+2c_{T})k)^{4\nu_T+2\nu_s-9}\left[Cos\left(\left[\nu_{s}-\frac{1}{2}\right]\frac{\pi}{2}\right)\right]^{\frac{1}{3}}
\left[Cos\left(\left[\nu_{T}-\frac{1}{2}\right]\frac{\pi}{2}\right)\right]^{\frac{2}{3}}}{c^{2\nu_s-3}_{s}c^{4\nu_T-6}_{T}k^{\nu_s+2\nu_T}}\\
\displaystyle~~~~~~~~~~~~~~~~~~~~~~~~~~~~~~~~~~~ \times\left(2^{2\nu_{s}+4\nu_{T}-12}\left|\frac{\Gamma(\nu_{s})}{\Gamma(\frac{3}{2})}\right|^{2}\left|\frac{\Gamma(\nu_{T})}
{\Gamma(\frac{3}{2})}\right|^{4}\frac{\left(1-\epsilon_{V}-s^{S}_{V}\right)^{2}\left(1-\epsilon_{V}-s^{T}_{V}\right)^{4}{V^{\frac{3}{2}}(\phi)}}{Y_{S}Y^2_{T}
c^{4}_{T}c^{3}_{s}{\tilde{g}^{\frac{3}{2}}_{1}}M^3_{PL}}\right) \\
\displaystyle~~~~~~~~~~~~~~~~~~~~~~~~~~~~~~~~~~~~~~~~ \left[32{\cal \tilde{F}}_{1}\left(\nabla_{1}\right)^{u;\lambda_{2};\lambda_{3}}_{equil}+4{\cal \tilde{F}}_{2}\left(\nabla_{2}\right)^{u;\lambda_{2};\lambda_{3}}_{equil}
+2\left({\cal \tilde{F}}_{3}\left(\nabla_{3}\right)^{u;\lambda_{2};\lambda_{3}}_{equil}\right.\right.\\ \left.\left.\displaystyle ~~~~~~~~~~~~~~~~~~~~~~~~~~~~~~~~~~~~~~~~~~~~~~~~~~~~
+{\cal \tilde{F}}_{4}\left(\nabla_{4}\right)^{u;\lambda_{2};\lambda_{3}}_{equil}
+{\cal \tilde{F}}_{5}\left(\nabla_{5}\right)^{u;\lambda_{2};\lambda_{3}}_{equil}+{\cal \tilde{F}}_{6}\left(\nabla_{6}\right)^{u;\lambda_{2};\lambda_{3}}_{equil}
+{\cal \tilde{F}}_{7}\left(\nabla_{7}\right)^{u;\lambda_{2};\lambda_{3}}_{equil}\right)\right]
          \end{array}
\ee
where each coefficients and functions 
are evaluated in equilateral limit.

\subsection{\bf Two scalar one tensor correlation}
After gauge fixing the interactions involving one tensor and two scalars
are given by the following third order action:
\be\begin{array}{lllll}\label{sst1}
    \displaystyle S_{\zeta\zeta h}
=\int dtd^{3}x~a^{3}\left\{\frac{{\cal Y}_1}{a^2}
h_{ij}\zeta_{,i}\zeta_{,j}
+\frac{{\cal Y}_2}{a^2}\dot h_{ij}\zeta_{,i}\zeta_{,j}
+{\cal Y}_3\dot h_{ij}\zeta_{,i}\psi_{,j}
+\frac{{\cal Y}_4}{a^2}\partial^2h_{ij}\zeta_{,i}\psi_{,j}
+\frac{{\cal Y}_5}{a^4}\partial^2 h_{ij}\zeta_{,i}\zeta_{,j}
\right.\\ \left.~~~~~~~~~~~~~~~~~~~~~~~~~~~~~~~~~~~~~~~~~~~~~~~~~~~~~~~~~~~~~~~~~~~~~~~~~~~~~~~~~~~
\displaystyle +{\cal Y}_6\partial^2 h_{ij}\psi_{,i}\psi_{,j}\right\}
   \end{array}\ee
where 
the dimensionful coefficients ${\cal Y}_{i}(i=1,2.....6)$ are defined as:

\be\begin{array}{llll}\label{uiycdr}
\displaystyle {\cal Y}_1=Y_s c^{2}_{s},
\\
\displaystyle{\cal Y}_2=\frac{L_{1}\hat{\tilde{K}}_{XX}}{4}\left(Y_s c^{2}_{s}-Y_Tc^{2}_{T}\right)
+L_{1}Y_T^2\left[
-\frac{1}{2}+\frac{HL_{1}\hat{\tilde{K}}_{XX}}{4}\left(3+\frac{\dot Y_T}{HY_T}\right)
-\frac{1}{4}\frac{d}{d t}\left(L_{1}{\hat{\tilde{K}}_{XX}}\right)
\right]\\~~~~~~~~~~~~~~~~~~~~~~~~~~~~~~~~~~~~~~~~~~~~~~~~~~~~~~~~~~~\displaystyle
+\frac{\sigma Y_s c^{2}_{s}}{Y_T}+{2HL_{1}Y_T\sigma}
-Y_T\frac{d}{d  t}\left({L_{1}\sigma}\right),
\\
\displaystyle{\cal Y}_3=Y_s\left[
\frac{3}{2}+\frac{d}{d t}\left(\frac{\hat{\tilde{K}}_{XX}L_{1}}{2}+\frac{\sigma}{Y_T}\right)
-\left(3H+\frac{\dot Y_T}{Y_T}\right)
\left(\frac{\hat{\tilde{K}}_{XX}L_{1}}{2}+\frac{\sigma}{Y_T}\right)
\right],
\\
\displaystyle{\cal Y}_4=Y_s\left[
-\frac{(Y_T-\hat{\tilde{K}}_{XX}c^{2}_{T})L_{1}}{2}
-{2H\sigma L_{1}}+\frac{d}{d t}\left({L_{1}\sigma}\right)
+\frac{\sigma}{Y_T^2}\left(Y_Tc^{2}_{T}-Y_s c^{2}_{s}\right)
\right],
\\
\displaystyle{\cal Y}_5=\frac{Y_T^2L_{1}}{2}\left[
\frac{(Y_T-\hat{\tilde{K}}_{XX}c^{2}_{T})}{2}
+{2HL_{1}\sigma}-\frac{d}{d t}\left({\sigma L_{1}}\right)
-\frac{\sigma}{Y_T^2}\left(3Y_Tc^{2}_{T}-Y_s c^{2}_{s}\right)
\right],
\\
\displaystyle{\cal Y}_6=\frac{Y_s^2}{4Y_T}\left[
1+\frac{6H\sigma}{Y_T}-2Y_T\frac{d}{d t}\left(\frac{\sigma}{Y_T^2}\right)
\right],
\end{array}\ee  

Following the prescription of {\it in-in formalism} 
in the interaction picture {\it three point two scalar one tensor correlation function} 
can be expressed in the following form:
\be\begin{array}{llll}\label{threept}
\displaystyle\langle\zeta(\vec{k}_1)\zeta(\vec{k}_2)h_{kl}(\vec{k}_3)\rangle
=-i\sum^{7}_{q=1}\int^{0}_{-\infty}d\eta~ a ~\langle 0|\left[\zeta(\vec{k}_1)\zeta(\vec{k}_2)h_{kl}(\vec{k}_3),
\left(\left[H^{(q)}_{int}(\eta)\right]_{kl}\right)_{\zeta \zeta h}\right]|0\rangle\\
~~~~~~~~~~~~~~~~~~~~~~~~~\displaystyle=(2\pi)^3\delta^{(3)}(\vec{k}_1+\vec{k}_2+\vec{k}_3)\left\{B_{\zeta\zeta h}\right\}_{kl}(
\vec{k}_1,\vec{k}_2,\vec{k}_3),\end{array}\ee
where the total Hamiltonian  can be expressed in terms of the third order Lagrangian density as
 $\left(\left[H_{int}(\eta)\right]_{kl}\right)_{\zeta \zeta h}
=\sum^{7}_{q=1}\left(\left[H^{(q)}_{int}(\eta)\right]_{kl}\right)_{\zeta \zeta h}=-\int d^{3}x\left[ \left({\cal L}_{3}\right)_{\zeta \zeta h}\right]_{kl}$.
Here the cross bispectrum $\left\{B_{\zeta \zeta h}\right\}_{kl}$ is defined as: 
\begin{equation}
\left\{B_{\zeta \zeta h}\right\}_{kl}=
  \frac{(2\pi)^4{\cal P}_{u}^2}{\prod^{3}_{i=1}k^{3}_{i}}
  \left({\cal A}_{\zeta \zeta h}\right)_{kl}=\frac{6}{5}\left[f^{}_{NL;3}\right]^{u}_{kl}P^{2}_{u}, \label{eq:Bzetazetah}
\end{equation}
where $\left({\cal A}_{\zeta \zeta h}\right)_{kl}$ is the two scalar one tensor correlation shape function and the symbol $;3$ represents two scalar one tensor correlation.
 Consequently the non-linear parameter
$\left[f^{}_{NL;3}\right]^{u}_{kl}$ can be expressed as:

\be\begin{array}{llll}\label{ujzxzatt}
         \displaystyle\left[f^{}_{NL;3}\right]^{u}_{kl}=\frac{10{ L}^{POL}_{u}{\cal N}_{ij;kl}}{3\sum^{3}_{i=1}k^{3}_{i}}
\frac{\underline{\underline{K}}^{4\nu_{s}+2\nu_{T}-9}\left[Cos\left(\left[\nu_{s}-\frac{1}{2}\right]\frac{\pi}{2}\right)\right]^{\frac{2}{3}}
\left[Cos\left(\left[\nu_{T}-\frac{1}{2}\right]\frac{\pi}{2}\right)\right]^{\frac{1}{3}}}{c^{4\nu_{s}-6}_{s}c^{2\nu_{T}-3}_{T}(k_{1}k_{2})^{\nu_{s}}k^{\nu_{T}}_{3}}
\\~~~~~~~~~~~~~~~~~~~~~~~~~~~~\displaystyle\times \left(2^{4\nu_{s}+2\nu_{T}-12}\left|\frac{\Gamma(\nu_{s})}
{\Gamma(\frac{3}{2})}\right|^{4}\left|\frac{\Gamma(\nu_{T})}{\Gamma(\frac{3}{2})}\right|^{2}\frac{\left(1-\epsilon_{V}-s^{S}_{V}\right)^{4}
\left(1-\epsilon_{V}-s^{T}_{V}\right)^{2}V^{\frac{3}{2}}(\phi)}{Y^{2}_{S}Y_{T}
c^{6}_{s}c^{3}_{T}\tilde{g}^{\frac{3}{2}}_{1}M^{3}_{PL}}\right)\left(\sum^{6}_{v=1}{\cal Y}_{v}\left(\hat{\nabla}_{v}\right)_{ij}\right)
          \end{array}
\ee

where the functional dependence of the co-efficients $\left(\hat{\nabla}_{v}\right)_{ij}\forall v$ are explicitly mentioned in the Appendix B.3.
In this context $\underline{\underline{K}}:=c_{s}(k_{1}+k_{2})+c_{T}k_{3}$.

For quasi-exponential limit the overall normalization factor for three types of polarization can be expressed as:
\be\label{normala}
{\cal L}^{POL}_{u}=
\left\{
	\begin{array}{ll}
                    \displaystyle 1& \mbox{ \it:{\cal u=1(E-mode)}}  \\
         \displaystyle  16 & ~\mbox{\it:{\cal u=2(E$\bigotimes$B mode)}}\\
\displaystyle 256 & ~\mbox{\it:{\cal u=3(B-mode)}}.
          \end{array}
\right.
\ee

As mentioned in the previous sub-section, performing basis transformation cross bispectrum for two scalars and one tensor can be expressed as:
\begin{equation}
  \langle\zeta(\vec{k}_1)\zeta(\vec{k}_2){\cal \bigoplus}^{\lambda}(\vec{k}_3)\rangle
=(2\pi)^3\delta^{(3)}(\vec{k}_1+\vec{k}_2+\vec{k}_3)B_{\lambda}^{(\zeta
\zeta h)}(\vec{k}_1,\vec{k}_2,\vec{k}_3).
\end{equation}

where we have used the following parameterization:
\begin{eqnarray}
B_{\lambda}^{(\zeta \zeta h)}=\frac{(2\pi)^4{\cal P}_{u}^2}{\prod^{3}_{i=1}k^{3}_{i}}
  {\cal A}^{\lambda}_{(\zeta \zeta h)}
= \frac{6}{5}\left[f^{}_{NL;3}\right]^{u;\lambda}P^{2}_{u}.
\end{eqnarray}
The polarized non-Gaussian parameter for two scalar and one tensor mode$\left[f^{}_{NL;3}\right]^{u;\lambda}$
can be rewritten as:

\be\begin{array}{llll}\label{shaad1}
\displaystyle \left[f^{}_{NL;3}\right]^{u}_{\lambda}
=\frac{20{ L}^{POL}_{u}\delta_{\lambda\lambda^{'}}}{3\sum^{3}_{i=1}k^{3}_{i}}
\frac{\underline{\underline{K}}^{4\nu_{s}+2\nu_{T}-9}\left[Cos\left(\left[\nu_{s}-\frac{1}{2}\right]\frac{\pi}{2}\right)\right]^{\frac{2}{3}}
\left[Cos\left(\left[\nu_{T}-\frac{1}{2}\right]\frac{\pi}{2}\right)\right]^{\frac{1}{3}}}{c^{4\nu_{s}-6}_{s}c^{2\nu_{T}-3}_{T}(k_{1}k_{2})^{\nu_{s}}k^{\nu_{T}}_{3}}
\\ \displaystyle~~~~~~~~~~~~~~~~~~~~~~~~~~~~\times \left(2^{4\nu_{s}+2\nu_{T}-12}\left|\frac{\Gamma(\nu_{s})}
{\Gamma(\frac{3}{2})}\right|^{4}\left|\frac{\Gamma(\nu_{T})}{\Gamma(\frac{3}{2})}\right|^{2}\frac{\left(1-\epsilon_{V}-s^{S}_{V}\right)^{4}
\left(1-\epsilon_{V}-s^{T}_{V}\right)^{2}V^{\frac{3}{2}}(\phi)}{Y^{2}_{S}Y_{T}
c^{6}_{s}c^{3}_{T}\tilde{g}^{\frac{3}{2}}_{1}M^{3}_{PL}}\right)\left(\sum^{6}_{v=1}{\cal Y}_{v}\left(\hat{\nabla}_{v}\right)_{\lambda^{'}}\right)
          \end{array}
\ee

where all the co-efficients $\left(\hat{\nabla}_{v}\right)_{\lambda^{'}}\forall v$ after basis transformation are explicitly written in the Appendix B.3.

In the equilateral limit the expression for the non-Gaussian parameter ($f_{NL}$) reduces to the following form:

  \be\begin{array}{lll}\label{shapefnasxc1}\displaystyle \left[f^{equil}_{NL;3}\right]^{u}_{\lambda}=
\frac{20{ L}^{POL}_{u}\delta_{\lambda\lambda^{'}}}{9k^{3}}
\frac{((2c_{s}+c_{T})k)^{4\nu_{s}+2\nu_{T}-9}\left[Cos\left(\left[\nu_{s}-\frac{1}{2}\right]\frac{\pi}{2}\right)\right]^{\frac{2}{3}}
\left[Cos\left(\left[\nu_{T}-\frac{1}{2}\right]\frac{\pi}{2}\right)\right]^{\frac{1}{3}}}{c^{4\nu_{s}-6}_{s}c^{2\nu_{T}-3}_{T}k^{2\nu_{s}+\nu_{T}}}
\\ \displaystyle~~~~~~~~~~~~~~~~~~~~~~~~~~~~~~~~\times \left(2^{4\nu_{s}+2\nu_{T}-12}\left|\frac{\Gamma(\nu_{s})}
{\Gamma(\frac{3}{2})}\right|^{4}\left|\frac{\Gamma(\nu_{T})}{\Gamma(\frac{3}{2})}\right|^{2}\frac{\left(1-\epsilon_{V}-s^{S}_{V}\right)^{4}
\left(1-\epsilon_{V}-s^{T}_{V}\right)^{2}V^{\frac{3}{2}}(\phi)}{Y^{2}_{S}Y_{T}
c^{6}_{s}c^{3}_{T}\tilde{g}^{\frac{3}{2}}_{1}M^{3}_{PL}}\right)\left(\sum^{6}_{v=1}{\cal Y}_{v}\left(\hat{\nabla}_{v}\right)^{equil}_{\lambda^{'}}\right)
          \end{array}
\ee
\subsection{\bf Three tensor correlation}

The interactions involving three tensors 
are given by the following third order action:
\be\begin{array}{lllll}\label{sst}
    \displaystyle S_{hhh}
=\int dtd^{3}x~a^{3}\left\{\frac{\sigma }{12} \dot h_{ij}\dot h_{jk}\dot h_{ki}
+\frac{Y_T}{4a^2c^2_T}\left(h_{ik}h_{jl}
-\frac{1}{2}h_{ij}h_{kl}\right)h_{ij,kl}\right\}
   \end{array}\ee
Now following the prescription of {\it in-in formalism} in the interaction picture {\it three point three tensor correlation function} 
can be expressed in the following form:
\be\begin{array}{llll}\label{threept}
\displaystyle\langle h_{i_1j_1}(\vec{k}_1)
h_{i_2j_2}(\vec{k}_2)
h_{i_3j_3}(\vec{k}_3)
\rangle
=-i\int^{0}_{-\infty}d\eta~ a ~\langle 0|\left[h_{i_1j_1}(\vec{k}_1)
h_{i_2j_2}(\vec{k}_2)
h_{i_3j_3}(\vec{k}_3),\left(\left[H_{int}(\eta)\right]_{i_1j_1i_2j_2i_3j_3}\right)_{hhh}\right]|0\rangle\\
~~~~~~~~~~~~~~~~~~~~~~~~~~~~~~~~~~\displaystyle=(2\pi)^3\delta^{(3)}(\vec{k}_1+\vec{k}_2+\vec{k}_3)\left\{B_{hhh}\right\}_{i_1j_1i_2j_2i_3j_3}(
\vec{k}_1,\vec{k}_2,\vec{k}_3),\end{array}\ee
where the total Hamiltonian is expressed in terms of the third order Lagrangian density as
 $\left(\left[H_{int}(\eta)\right]_{i_1j_1i_2j_2i_3j_3}\right)_{hhh}
=-\int d^{3}x\left[ \left({\cal L}_{3}\right)_{hhh}\right]_{i_1j_1i_2j_2i_3j_3}$. 
In this context the bispectrum for three tensor correlation can be expressed as:
\begin{eqnarray}
\left\{B_{hhh}\right\}_{i_1j_1i_2j_2i_3j_3}(
\vec{k}_1,\vec{k}_2,\vec{k}_3)&=&
\frac{(2\pi)^4{\cal P}_u^2}{\prod^{3}_{i=1}k^{3}_{i}}
{\cal A}^{hhh}_{i_1j_1i_2j_2i_3j_3}=\frac{6}{5}\left[f^{}_{NL;4}\right]P^{2}_{u}, \label{eq:Bhhh}
\end{eqnarray}
where the symbol $;4$ represents three tensor correlation. Also, the non-Gaussian parameter is given by:

 \be\begin{array}{llll}\label{shazaqw1} 
\displaystyle\left[f^{}_{NL;4}\right]_{i_{1}j_{1}i_{2}j_{2}i_{3}j_{3}}=\frac{10{\cal W}^{POL}_{u}}{3\sum^{3}_{i=1}k^{3}_{i}}\frac{K^{9-6\nu_{T}}
Cos\left(\left[\nu_{T}-\frac{1}{2}\right]\frac{\pi}{2}\right)}{(k_{1}k_{2}k_{3})^{2\nu_{T}}}
\displaystyle\left(2^{3(\nu_{s}+\nu_{T})-11}\left|\frac{\Gamma(\nu_{s})}{\Gamma(\frac{3}{2})}\right|^{3}\left|\frac{\Gamma(\nu_{T})}{\Gamma(\frac{3}{2})}\right|^{3}\right.\\ \left.
~~~~~~~~~~~~~~~~~~~~~~~~~~~~~~~~~~~~\displaystyle\frac{\left(1-\epsilon_{V}-s^{S}_{V}\right)^{3}\left(1-\epsilon_{V}-s^{T}_{V}\right)^{3}V^{\frac{3}{2}}(\phi)}{
Y^{\frac{3}{2}}_{S}Y^{\frac{3}{2}}_{T}
c^{\frac{9}{2}}_{s}c^{\frac{9}{2}}_{T}\tilde{g}^{\frac{3}{2}}_{1}M^{3}_{PL}}
\right)
\displaystyle\left(\sum^{3}_{p=1}\Delta^{(p)}_{i_{1}j_{1}i_{2}j_{2}i_{3}j_{3}}\right)
          \end{array}
\ee

where $K=k_1+k_2+k_3$ and the polarization index $u=1(E-mode),2(E\bigotimes B-mode),3(B-mode)$. 
The functional dependence of all the co-efficients $\Delta^{(p)}_{i_{1}j_{1}i_{2}j_{2}i_{3}j_{3}}\forall p$ are summarized in Appendix B.4.

For quasi-exponential limit the overall normalization factor for three types of polarization can be expressed as:
\be\label{normala}
{\cal W}^{POL}_{u}=
\left\{
	\begin{array}{ll}
                    \displaystyle 4 & \mbox{ \it:{\cal u=1(E-mode)}}  \\
         \displaystyle  64 & ~\mbox{\it:{\cal u=2(E$\bigotimes$B mode)}}\\
\displaystyle 1024 & ~\mbox{\it:{\cal u=3(B-mode)}}.
          \end{array}
\right.
\ee

After performing basis transformation the relevant three point correlation function for three tensor interaction can be expressed in terms 
of bispectrum as:
\begin{equation}
  \langle {\cal \bigoplus}^{\lambda_{1}}(\vec{k}_1)
{\cal \bigoplus}^{\lambda_{2}}(\vec{k}_2)
{\cal \bigoplus}^{\lambda_{3}}(\vec{k}_3)
\rangle
=(2\pi)^3\delta^{(3)}(\vec{k}_1+\vec{k}_2+\vec{k}_3)
B^{hhh}_{\lambda_1,\lambda_2,\lambda_3}.
\end{equation}
where
\begin{eqnarray}
 B^{hhh}_{\lambda_1,\lambda_2,\lambda_3}=\frac{(2\pi)^4{\cal P}_{u}^2}{\prod^{3}_{i=1}k^{3}_{i}}
  {\cal A}^{\lambda_{1},\lambda_{2},\lambda_{3}}_{(\zeta \zeta h)}
= \frac{6}{5}\left[f^{}_{NL;4}\right]^{u}_{\lambda_{1},\lambda_{2},\lambda_{3}}P^{2}_{u},
\end{eqnarray}

where the the non-linear parameter is given by:

\be\begin{array}{llll}\label{shapefnwqlook1}
         \displaystyle \left[f^{}_{NL;4}\right]^{u}_{\lambda_{1},\lambda_{2},\lambda_{3}}=
\frac{10{\cal W}^{POL}_{u}}{3\sum^{3}_{i=1}k^{3}_{i}}\frac{K^{9-6\nu_{T}}
Cos\left(\left[\nu_{T}-\frac{1}{2}\right]\frac{\pi}{2}\right)}{(k_{1}k_{2}k_{3})^{2\nu_{T}}}
\displaystyle\left(2^{3(\nu_{s}+\nu_{T})-11}\left|\frac{\Gamma(\nu_{s})}{\Gamma(\frac{3}{2})}\right|^{3}\left|\frac{\Gamma(\nu_{T})}{\Gamma(\frac{3}{2})}\right|^{3}\right.\\ \left.
\displaystyle~~~~~~~~~~~~~~~~~~~~~~~~~~~~~~~~~~~~~~~~~~~~~~~~~~~~~~~~~~~~~~\frac{\left(1-\epsilon_{V}-s^{S}_{V}\right)^{3}\left(1-\epsilon_{V}-s^{T}_{V}\right)^{3}V^{\frac{3}{2}}(\phi)}{
Y^{\frac{3}{2}}_{S}Y^{\frac{3}{2}}_{T}
c^{\frac{9}{2}}_{s}c^{\frac{9}{2}}_{T}\tilde{g}^{\frac{3}{2}}_{1}M^{3}_{PL}}
\right)
\displaystyle\left(\sum^{3}_{p=1}\Delta^{(p)}_{\lambda_{1}\lambda_{2}\lambda_{3}}\right)
          \end{array}
\ee

Once again, all the helicity dependent co-efficients $\Delta^{(p)}_{\lambda_{1}\lambda_{2}\lambda_{3}}\forall p$ after basis transformation
are explicitly mentioned in the Appendix B.4.

In the equilateral limit we have:

   \be\begin{array}{llll}\displaystyle\left[f^{equil}_{NL;4}\right]^{u}_{\lambda_{1},\lambda_{2},\lambda_{3}}=
\frac{10{\cal W}^{POL}_{u}}{9k^{3}}\frac{(3k)^{9-6\nu_{T}}
Cos\left(\left[\nu_{T}-\frac{1}{2}\right]\frac{\pi}{2}\right)}{k^{6\nu_{T}}}
\displaystyle\left(2^{3(\nu_{s}+\nu_{T})-11}\left|\frac{\Gamma(\nu_{s})}{\Gamma(\frac{3}{2})}\right|^{3}\left|\frac{\Gamma(\nu_{T})}{\Gamma(\frac{3}{2})}\right|^{3}\right.\\ \left.
~~~~~~~~~~~~~~~~~~~~~~~~~~~~~~~~~~~~~~~~~~~~~~~~~~~\displaystyle\frac{\left(1-\epsilon_{V}-s^{S}_{V}\right)^{3}\left(1-\epsilon_{V}-s^{T}_{V}\right)^{3}V^{\frac{3}{2}}(\phi)}{
Y^{\frac{3}{2}}_{S}Y^{\frac{3}{2}}_{T}
c^{\frac{9}{2}}_{s}c^{\frac{9}{2}}_{T}\tilde{g}^{\frac{3}{2}}_{1}M^{3}_{PL}}
\right)
\displaystyle\left(\sum^{3}_{p=1}\Delta^{(p);equil}_{\lambda_{1}\lambda_{2}\lambda_{3}}\right)
          \end{array}
\ee

Numerical values of all such non-Gaussian parameters from three point correlation for different polarizing modes are mentioned in the table(\ref{tab1}).
In this context {\bf PC} and {\bf PV} stands for the parity conserving and violating contribution for graviton degrees of freedom.

\section{\bf Tree Level Trispectrum Analysis from four scalar correlation}

To derive the expression for scalar trispectrum for D3 DBI Galileon let us start from fourth order action up to total derivatives.
Consequently the fourth order action in the uniform gauge can be expressed as:
   \be\begin{array}{llll}
       \displaystyle S_{\zeta\zeta\zeta\zeta}=S_{{\bf CI}}+S_{{\bf SE}}+S_{{\bf GE}}
      \end{array}\ee
where $S_{{\bf CI}}$, $S_{{\bf SE}}$ and $S_{{\bf GE}}$ represent the contribution from the {\it contact interaction}, {\it scalar exchange} and {\it graviton exchange}
appearing in the four point correlation. In the next subsections we will discuss the individual contributions separately.

\subsection{\bf Contact Interaction}
Taking into account the contribution coming from contact interaction of effective DBI Galileon in the fourth order action in uniform gauge we get: 
\be\begin{array}{lllll}\label{trispec}
    \displaystyle S_{{\bf CI}}
=\int dt d^{3}x \frac{a^{3}}{4}\left\{{\cal \bar{U}}_{1}\dot{\zeta}^{4}-\frac{(\partial\zeta)^{2}}{a^{2}}\dot{\zeta}^{2}{\cal \bar{U}}_{2}
+{\cal \bar{U}}_{3}\frac{(\partial\zeta)^{4}}{a^{4}}\right\},
   \end{array}\ee
where the co-efficients ${\cal \bar{U}}_{i}(i=1,2,3)$ for effective DBI Galileon are defined as:
\be\begin{array}{llllll}\label{cofe}
 \displaystyle   {\cal \bar{U}}_{1}=
\left(\frac{\dot{\phi}^{4}}{6}\left[\hat{\tilde{K}}_{4X}-\tilde{G}_{4X}\dot{\phi}^{2}\right]+
\dot{\phi}^{2}\left[\hat{\tilde{K}}_{XXX}-\tilde{G}_{XXX}\dot{\phi}^{2}\right]
+\frac{1}{2}\left[\hat{\tilde{K}}_{XX}-\tilde{G}_{XX}\dot{\phi}^{2}\right]\right),\\
\displaystyle{\cal \bar{U}}_{2}=\left(\dot{\phi}^{2}\left[\hat{\tilde{K}}_{XXX}-\tilde{G}_{XXX}
\dot{\phi}^{2}\right]+\hat{\tilde{K}}_{XX}-\tilde{G}_{XX}\dot{\phi}^{2}\right),
~~\displaystyle{\cal \bar{U}}_{3}=\frac{1}{2}\left[\hat{\tilde{K}}_{XX}-\tilde{G}_{XX}\dot{\phi}^{2}\right].
   \end{array}\ee
where $\hat{\tilde{K}}(\phi,X)$ and $\tilde{G}(\phi,X)$ are explicitly mentioned in equation(\ref{effcons}).
Using in-in procedure
the {\it four point correlation function} for quasi exponential situation can be expressed in the following form:
\be\begin{array}{llll}\label{threept}
\displaystyle\langle\zeta(\vec{k_{1}})\zeta(\vec{k_{2}})\zeta(\vec{k_{3}})\zeta(\vec{k_{4}})\rangle^{\bf CI}
=-i\sum^{3}_{j=1}\int^{0}_{-\infty}d\eta~ a ~\langle 0|\left[\zeta(\vec{k_{1}})\zeta(\vec{k_{2}})\zeta(\vec{k_{3}})\zeta(\vec{k_{4}}),\left(H^{(j)}_{int}(\eta)\right)^{\bf CI}_{\zeta\zeta\zeta\zeta}\right]|0\rangle\\
~~~~~~~~~~~~~~~~~~~~~~~~~~~~~\displaystyle
=(2\pi)^{3}\delta^{(3)}
(\vec{k_{1}}+\vec{k_{2}}+\vec{k_{3}}+\vec{k_{4}}){\cal T}^{\bf CI}_{\zeta}(\vec{k_{1}},\vec{k_{2}},\vec{k_{3}},\vec{k_{4}}),\end{array}\ee
where in the interaction picture the Hamiltonian can be written as: 
$\left(H_{int}(\eta)\right)^{\bf CI}_{\zeta\zeta\zeta\zeta}=\sum^{3}_{j=1}\left(H^{(j)}_{int}(\eta)\right)^{\bf CI}_{\zeta\zeta\zeta\zeta}.$ 

Here following the ansatz used in \cite{wands1} the {\it trispectrum} ${\cal T}^{\bf CI}_{\zeta}(\vec{k_{1}},\vec{k_{2}},\vec{k_{3}},\vec{k_{4}})$ for contact interaction 
is defined as:
\be\begin{array}{llllll}\label{bispec}
\displaystyle {\cal T}^{\bf CI}_{\zeta}(\vec{k_{1}},\vec{k_{2}},\vec{k_{3}},\vec{k_{4}})
\displaystyle=\frac{1}{\prod^{4}_{i=1}k^{3}_{i}}\left[(k^{3}_{1}k^{3}_{2}+k^{3}_{3}k^{3}_{4})\left(k^{-3}_{13}+k^{-3}_{14}\right)+
(k^{3}_{1}k^{3}_{4}+k^{3}_{2}k^{3}_{3})\left(k^{-3}_{12}+k^{-3}_{13}\right)
\right.\\ \left.~~~~~~~~~~~~~~~~~~~~~~~~~~~~~~~~~~~~~~~~~~~~~~~~~~~~~~~~~~~~~~~~~~~~~~~~
+(k^{3}_{1}k^{3}_{3}+k^{3}_{2}k^{3}_{4})\left(k^{-3}_{12}+k^{-3}_{14}\right)\right]
\left\{\tau^{\bf CI}_{NL}{P}^{3}_{\zeta(1)}+\frac{54}{25}
g^{\bf CI}_{NL}{P}^{3}_{\zeta(2)}\right\},\end{array}\ee
where 
\be\begin{array}{llll}\label{ghy} 
\displaystyle {P}^{3}_{\zeta(1)}=\sum_{j<p,i\neq j,p}{P}_{\zeta}({k_{ij}}){P}_{\zeta}({k_{j}}){P}_{\zeta}({k_{p}}), 
~~\displaystyle{P}^{3}_{\zeta(2)}=\sum_{i<j<p}{P}_{\zeta}({k_{i}}){P}_{\zeta}({k_{j}}){P}_{\zeta}({k_{p}})
\end{array}\ee
such that
\be\begin{array}{llll}\label{ghyf} \displaystyle {P}^{3}_{\zeta}={P}^{3}_{\zeta(1)}
+\frac{3456}{25}\frac{1}{\bar{K}^{3}}\left[(k^{3}_{1}k^{3}_{2}+k^{3}_{3}k^{3}_{4})\left(k^{-3}_{13}+k^{-3}_{14}\right)\right.\\ \left.~~~~~~~~~~~~~~~~~~~~~~~~~~~~~~~~~~~~~~~~~~~~~~
\displaystyle +
(k^{3}_{1}k^{3}_{4}+k^{3}_{2}k^{3}_{3})\left(k^{-3}_{12}+k^{-3}_{13}\right)
+(k^{3}_{1}k^{3}_{3}+k^{3}_{2}k^{3}_{4})\left(k^{-3}_{12}+k^{-3}_{14}\right)\right]{P}^{3}_{\zeta(2)}\end{array}\ee
and $\tau^{\bf CI}_{NL}$ and $g^{\bf CI}_{NL}$ are 
the two non linear parameters which carry the signatures of primordial non-Gaussianities of the curvature perturbation in trispectrum analysis.
By knowing $\tau^{\bf CI}_{NL}$ the other parameter $g^{\bf CI}_{NL}$ can be calculated by making use of the following relation \cite{mizko}:
\be\begin{array}{llll}\label{gnl}
\displaystyle g^{\bf CI}_{NL}=\frac{64}{\bar{K}^{3}}\left[(k^{3}_{1}k^{3}_{2}+k^{3}_{3}k^{3}_{4})\left(k^{-3}_{13}+k^{-3}_{14}\right)+
(k^{3}_{1}k^{3}_{4}+k^{3}_{2}k^{3}_{3})\left(k^{-3}_{12}+k^{-3}_{13}\right)
+(k^{3}_{1}k^{3}_{3}
+k^{3}_{2}k^{3}_{4})\left(k^{-3}_{12}+k^{-3}_{14}\right)\right]\tau^{\bf CI}_{NL},\end{array}\ee
where $\bar{K}=k_{1}+k_{2}+k_{3}+k_{4}$. So, there is only one independent piece of information,
 namely $\tau^{\bf CI}_{NL}$, that carries information about trispectrum obtained from {\it contact interaction}.

To proceed further we denote the angle between  $\vec{k_{i}}$ and $\vec{k_{j}}$ (with $i\neq j$)
by $\Theta_{ij}$ then 
\be\begin{array}{llll}\label{qweer}\displaystyle Cos(\Theta_{12})=Cos(\Theta_{34}):=Cos(\Theta_{3}),~~ 
 \displaystyle Cos(\Theta_{23})=Cos(\Theta_{14}):=Cos(\Theta_{1}),~~
\displaystyle Cos(\Theta_{13})=Cos(\Theta_{24}):=Cos(\Theta_{2})\end{array}\ee
 subject to the constraint 
$Cos(\Theta_{1})+Cos(\Theta_{2})+Cos(\Theta_{3})=-1$
comes from the conservation of momentum. Additionally we have used
 \be\begin{array}{llll}\label{vcxbhgd}\displaystyle k_{14}=k_{23}=|\vec{k_{1}}+\vec{k_{4}}|=|\vec{k_{2}}+\vec{k_{3}}|
=\sqrt{k^{2}_{1}+k^{2}_{4}+2k_{1}k_{4}Cos(\Theta_{1})}=\sqrt{k^{2}_{2}+k^{2}_{3}+2k_{2}k_{3}Cos(\Theta_{1})}, \\
\displaystyle k_{24}=k_{13}=|\vec{k_{2}}+\vec{k_{4}}|=|\vec{k_{1}}+\vec{k_{3}}|
=\sqrt{k^{2}_{2}+k^{2}_{4}+2k_{2}k_{4}Cos(\Theta_{2})}=\sqrt{k^{2}_{1}+k^{2}_{3}+2k_{1}k_{3}Cos(\Theta_{2})},
\\ \displaystyle  k_{34}=k_{12}=|\vec{k_{3}}+\vec{k_{4}}|=|\vec{k_{1}}+\vec{k_{2}}|
 =\sqrt{k^{2}_{3}
+k^{2}_{4}+2k_{3}k_{4}Cos(\Theta_{3})}=\sqrt{k^{2}_{1}+k^{2}_{2}+2k_{1}k_{2}Cos(\Theta_{3})}.\end{array}\ee

The explicit form of $\tau^{\bf CI}_{NL}$ characterizing the trispectrum obtained from {\it contact interaction} can be expressed for our model
as:

\be\begin{array}{lllll}\label{gfn11}
\displaystyle \tau^{\bf CI}_{NL}
=\frac{2^{8\nu_{s}-6}\pi^{6}Cos\left(\left[\nu_{s}-\frac{1}{2}\right]\frac{\pi}{2}\right)}{\left[(k^{3}_{1}k^{3}_{2}+k^{3}_{3}k^{3}_{4})\left(k^{-3}_{13}+k^{-3}_{14}\right)+
(k^{3}_{1}k^{3}_{4}+k^{3}_{2}k^{3}_{3})\left(k^{-3}_{12}+k^{-3}_{13}\right)
+(k^{3}_{1}k^{3}_{3}+k^{3}_{2}k^{3}_{4})\left(k^{-3}_{12}+k^{-3}_{14}\right)\right]}\\~~~~~~~~~~~~~~~~~~~
\times\displaystyle  \left|\frac{\Gamma(\nu_{s})}{\Gamma(\frac{3}{2})}\right|^{8}
\frac{(1-\epsilon_{V}-s^{S}_{V})^{8}H^{8}}
{Y^{4}_{S}c^{12}_{s}(k_{1}k_{2}k_{3}k_{4})^{2\nu_{s}}}
\displaystyle\left\{\frac{8\bar{\cal U}_{1}\bar{K}^{8\nu_{s}-5}}{13}\left[\Gamma(17-8\nu_{s})\bar{K}^{8}G_{1}
-i\Gamma(16-8\nu_{s})\bar{K}^{7}G_{2}\right.\right.\\ \left.\left.\displaystyle~~~~~~~~~~~~~~~~~~~
+\Gamma(15-8\nu_{s})\bar{K}^{6}G_{3}-i\Gamma(14-8\nu_{s})\bar{K}^{5}G_{4}+
\Gamma(13-8\nu_{s})\bar{K}^{4}G_{5}-
i\Gamma(12-8\nu_{s})\bar{K}^{3}G_{6}\right.\right.\\ \left.\left.\displaystyle~~~~~~~~~~~~~~~~~~~ +\Gamma(11-8\nu_{s})\bar{K}^{2}G_{7}
-i\Gamma(10-8\nu_{s})\bar{K}G_{8}+G_{9}\right]
\displaystyle +\frac{\bar{\cal U}_{2}\bar{K}^{8\nu_{s}-3}}{32}
\left[(\vec{k_{3}}.\vec{k_{4}})\bar{\cal I}(3,4;1,2)+(\vec{k_{2}}.\vec{k_{4}})\bar{\cal I}(2,4;1,3)\right.\right.\\ \left.\left.\displaystyle~~~~~~~~~~~~~~~~~~~
+(\vec{k_{2}}.\vec{k_{3}})\bar{\cal I}(2,3;1,4)+
(\vec{k_{1}}.\vec{k_{4}})\bar{\cal I}(1,4;2,3)
+(\vec{k_{1}}.\vec{k_{2}})\bar{\cal I}(1,2;3,4)+
(\vec{k_{1}}.\vec{k_{3}})\bar{\cal I}(1,3;2,4)\right]\right.\\ \left.
\displaystyle~~~~~~~~~~~~~~~~~~~ +\frac{\bar{\cal U}_{3}\bar{K}^{8\nu_{s}+12}}{8}
\left[(\vec{k_{1}}.\vec{k_{2}})(\vec{k_{3}}.\vec{k_{4}})+
(\vec{k_{1}}.\vec{k_{3}})(\vec{k_{2}}.\vec{k_{4}})+(\vec{k_{1}}.\vec{k_{4}})(\vec{k_{2}}.\vec{k_{3}})
\right]\displaystyle\left(\frac{\bar{\cal Z}_{1}\Gamma(13-8\nu_{s})}{(\bar{K})^{13}}\right.\right.\\ \left.\left.\displaystyle~~~~~~~~~~~~~~~~~~~
+\frac{\bar{\cal Z}_{2}\Gamma(14-8\nu_{s})}{(\bar{K})^{14}}-
\frac{\bar{\cal Z}_{3}\Gamma(15-8\nu_{s})}{(\bar{K})^{15}}
~\displaystyle-\frac{\bar{\cal Z}_{4}\Gamma(16-8\nu_{s})}{(\bar{K})^{16}}+
\frac{\bar{\cal Z}_{5}\Gamma(17-8\nu_{s})}{(\bar{K})^{17}}
\right)
\right\}\end{array}
\ee
where the functional dependence of the momentum dependent functions $G_{i}\forall i$, ${\cal Z}_{q}\forall q$ and $ \bar{\cal I}(i,j;m,n)$
are given in Appendix B.5.A. It is important to mention here that the 4D effective coupling and the interaction between 
the higher order graviton and DBI Galileon plays a significant role in the slow-role regime. From equation(\ref{gfn11}) it evident that 
the non-Gaussian parameter $\tau^{\bf CI}_{NL}$ obtained from the contact interaction is inversely proportional
to the 12th power of the sound  speed for scalar mode. But depending on the signature and strength of the Gauss-Bonnet coupling the behavior of the 
$\tau^{\bf CI}_{NL}$ changes.

Further, using the {\it equilateral configuration} $(k_{1}=k_{2}=k_{3}=k_{4}=k$ and $\bar{K}=4k)$ 
and incorporating the contribution from the maximum shape of the trispectrum
 ($Cos(\Theta_{1})=Cos(\Theta_{2})=Cos(\Theta_{3})=-\frac{1}{3}$ and $k_{ij}(for~ i<j)=\frac{2k}{\sqrt{3}}$) the non linear parameter 
can be expressed as:

\be\begin{array}{llll}\label{gfneqas2}\displaystyle\tau^{equil;\bf CI}_{NL}=
\frac{2^{24\nu_{s}-5}\pi^{6}Cos\left(\left[\nu_{s}-\frac{1}{2}\right]\frac{\pi}{2}\right)}{9\sqrt{3}k^{3}}
 \left|\frac{\Gamma(\nu_{s})}{\Gamma(\frac{3}{2})}\right|^{8}
\frac{(1-\epsilon_{V}-s^{S}_{V})^{8}H^{8}}
{Y^{4}_{S}c^{12}_{s}}\left\{\frac{8\bar{\cal U}_{1}}{13312k^{5}}\left[65536\Gamma(17-8\nu_{s})k^{8}G^{equil}_{1}\right.\right.\\ \left.\left.\displaystyle~~~~~~~~~~~~~~~~~~~~~~~~~~~~~~
-16384i\Gamma(16-8\nu_{s})k^{7}G^{equil}_{2}
+4096\Gamma(15-8\nu_{s})k^{6}G^{equil}_{3}-1024i\Gamma(14-8\nu_{s})k^{5}G^{equil}_{4}\right.\right.\\ \left.\left.\displaystyle~~~~~~~~~~~~~~~~~~~~~~~~~~~~~~~~~~~ +
256\Gamma(13-8\nu_{s})k^{4}G^{equil}_{5}-
64i\Gamma(12-8\nu_{s})k^{3}G^{equil}_{6}+16\Gamma(11-8\nu_{s})k^{2}G^{equil}_{7}
\right.\right.\\ \left.\left.\displaystyle~~~~~~~~~~~~~~~~~~~~~~~~~~~~~~~~-4i\Gamma(10-8\nu_{s})k G^{equil}_{8}+G^{equil}_{9}\right]
\displaystyle -\frac{\bar{\cal U}_{2}}{1024k}\bar{\cal I}^{equil}+\frac{838861\bar{\cal U}_{3}}{12}
\displaystyle\left(\frac{\bar{\cal Z}^{equil}_{1}k^{3}\Gamma(13-8\nu_{s})}{67108864}\right.\right.\\ \left.\left.\displaystyle~~~~~~~~~~~~~~~~~~~~~~~~~~~~~~~
+\frac{\bar{\cal Z}^{equil}_{2}k^{2}\Gamma(14-8\nu_{s})}{268435456}-
\frac{\bar{\cal Z}^{equil}_{3}k\Gamma(15-8\nu_{s})}{1073741824}
-\frac{\bar{\cal Z}^{equil}_{4}\Gamma(16-8\nu_{s})}{4294967296}+
\frac{\bar{\cal Z}^{equil}_{5}\Gamma(17-8\nu_{s})}{17179869184}
\right)
\right\}.
          \end{array}
\ee

\subsection{\bf Scalar Exchange}
Within in-in picture formalism, to calculate the four-point correlation function resulting from a correlation established 
via the scalar exchange mode of effective DBI Galileon we start with the following action in the uniform gauge as:

\be\begin{array}{lllll}\label{trispec2}
    \displaystyle S_{{\bf SE}}
=\int dt d^{3}x~a^{3}\left\{ {\bf A}\dot{\zeta}^{3}-\frac{(\partial\zeta)^{2}}{a^{2}}\dot{\zeta}~{\bf B}\right\},
   \end{array}\ee
where the co-efficients $(\bf A, B)$ are defined as:
\be\begin{array}{llllll}\label{cofe}
 \displaystyle   {\bf A}=\left(\frac{\dot{\phi}}{2}\left[\hat{\tilde{K}}_{XX}-
\tilde{G}_{XX}\right]+\frac{\dot{\phi}^{3}}{6}\left[\hat{\tilde{K}}_{XXX}-\tilde{G}_{XXX}\dot{\phi}^{2}\right]\right),~~
\displaystyle {\bf B}=-\frac{\dot{\phi}}{2}\left[\hat{\tilde{K}}_{XX}-\tilde{G}_{XX}\dot{\phi}^{2}\right].
   \end{array}\ee
Using in-in procedure
the {\it four point correlation function} for quasi-exponential limit can be expressed in the following form:
\be\begin{array}{llll}\label{threept2}
\displaystyle\langle\zeta(\vec{k_{1}})\zeta(\vec{k_{2}})\zeta(\vec{k_{3}})\zeta(\vec{k_{4}})\rangle^{\bf SE}
=-i\sum^{2}_{j=1}\sum^{2}_{p=1}\int^{0}_{-\infty}d\eta~\int^{\eta}_{-\infty}d\tilde{\eta} ~\langle 0|\left[\left[\zeta(\vec{k_{1}})\zeta(\vec{k_{2}})
\zeta(\vec{k_{3}})\zeta(\vec{k_{4}}),\left(H^{(j)}_{int}(\eta)\right)^{\bf SE}_{\zeta\zeta\zeta}\right],\left(H^{(p)}_{int}(\tilde{\eta})\right)^{\bf SE}_{\zeta\zeta\zeta}\right]|0\rangle\\
~~~~~~~~~~~~~~~~~~~~~~~~~~~~~~~~~\displaystyle
=(2\pi)^{3}\delta^{(3)}
(\vec{k_{1}}+\vec{k_{2}}+\vec{k_{3}}+\vec{k_{4}}){\cal T}^{\bf SE}_{\zeta}(\vec{k_{1}},\vec{k_{2}},\vec{k_{3}},\vec{k_{4}}),\end{array}\ee
where in the interaction picture the Hamiltonian can be written in terms of the third order Lagrangian density as:
$\left(H_{int}(\eta)\right)^{\bf SE}_{\zeta\zeta\zeta}=\sum^{2}_{j=1}\left(H^{(j)}_{int}(\eta)\right)^{\bf SE}_{\zeta\zeta\zeta}=-\int d^{3}x {\cal L}^{\bf SE}_{3}$.
 Hence following the ansatz used in \cite{wands1} the {\it trispectrum} ${\cal T}^{\bf SE}_{\zeta}(\vec{k_{1}},\vec{k_{2}},\vec{k_{3}},\vec{k_{4}})$ is defined as:
\be\begin{array}{llllll}\label{bispec2}
\displaystyle {\cal T}^{\bf SE}_{\zeta}(\vec{k_{1}},\vec{k_{2}},\vec{k_{3}},\vec{k_{4}})
=\frac{1}{\prod^{4}_{i=1}k^{3}_{i}}\left[(k^{3}_{1}k^{3}_{2}+k^{3}_{3}k^{3}_{4})\left(k^{-3}_{13}+k^{-3}_{14}\right)+
(k^{3}_{1}k^{3}_{4}+k^{3}_{2}k^{3}_{3})\left(k^{-3}_{12}+k^{-3}_{13}\right)
\right.\\ \left.~~~~~~~~~~~~~~~~~~~~~~~~~~~~~~~~~~~~~~~~~~~~~~~~~~~~~~~~~~~~~~~~~~~~~~~~
+(k^{3}_{1}k^{3}_{3}+k^{3}_{2}k^{3}_{4})\left(k^{-3}_{12}+k^{-3}_{14}\right)\right]
\left\{\tau^{\bf SE}_{NL}{P}^{3}_{\zeta(1)}+\frac{54}{25}
g^{\bf SE}_{NL}{P}^{3}_{\zeta(2)}\right\},\end{array}\ee
where 
$\tau^{\bf SE}_{NL}$ and $g^{\bf SE}_{NL}$ are 
the two non linear parameters which carry the signatures of primordial non-Gaussianities of the curvature perturbation obtained from scalar exchange contribution 
in trispectrum analysis.
By knowing $\tau^{\bf SE}_{NL}$ the other parameter $g^{\bf SE}_{NL}$ can be calculated by making use of the following relation \cite{mizko}:
\be\begin{array}{llll}\label{gnl2}
\displaystyle g^{\bf SE}_{NL}=\frac{64}{\bar{K}^{3}}\left[(k^{3}_{1}k^{3}_{2}+k^{3}_{3}k^{3}_{4})\left(k^{-3}_{13}+k^{-3}_{14}\right)+
(k^{3}_{1}k^{3}_{4}+k^{3}_{2}k^{3}_{3})\left(k^{-3}_{12}+k^{-3}_{13}\right)
+(k^{3}_{1}k^{3}_{3}
+k^{3}_{2}k^{3}_{4})\left(k^{-3}_{12}+k^{-3}_{14}\right)\right]\tau^{\bf SE}_{NL},\end{array}\ee
where $\bar{K}=k_{1}+k_{2}+k_{3}+k_{4}$. 
The explicit form of $\tau^{\bf SE}_{NL}$ characterizing the {\it scalar exchange} trispectrum can be expressed for our model
as:

\be\begin{array}{lllll}\label{gfn112}
\displaystyle \tau^{\bf SE}_{NL}
=\frac{2^{8\nu_{s}-14}{\bar K}^{10\nu_{s}-15}Cos\left(\left[\nu_{s}-\frac{1}{2}\right]\frac{\pi}{2}\right)}{\left[(k^{3}_{1}k^{3}_{2}+k^{3}_{3}k^{3}_{4})\left(k^{-3}_{13}+k^{-3}_{14}\right)+
(k^{3}_{1}k^{3}_{4}+k^{3}_{2}k^{3}_{3})\left(k^{-3}_{12}+k^{-3}_{13}\right)
+(k^{3}_{1}k^{3}_{3}+k^{3}_{2}k^{3}_{4})\left(k^{-3}_{12}+k^{-3}_{14}\right)\right]}\\~~~~~~~~~~
\times\displaystyle  \left|\frac{\Gamma(\nu_{s})}{\Gamma(\frac{3}{2})}\right|^{8}
\frac{(1-\epsilon_{V}-s^{S}_{V})^{8}H^{8}}
{Y^{5}_{S}c^{6}_{s}(k_{1}k_{2}k_{3}k_{4})^{2\nu_{s}}}
\displaystyle\left\{ 9 {\bf A}^{2}\left[{\bf\Xi}_{1}(-k_{1},-k_{2},-k_{12},k_{3},k_{4},k_{12})-{\bf\Xi}_{1}(k_{1},k_{2},-k_{12},k_{3},k_{4},k_{12})\right]
\right.\\ \left. \displaystyle~~~~~~~~~~+{\bf AB}\left[3(\vec{k_{3}}.\vec{k_{4}})
\left\{{\bf\Xi}_{3}(k_{1},k_{2},-k_{12},k_{12},k_{3},k_{4})-{\bf\Xi}_{3}(-k_{1},-k_{2},-k_{12},k_{12},k_{3},k_{4})\right\}
\right.\right.\\ \left.\left.~~~~~~~~~+6(\vec{k_{12}}.\vec{k_{4}})
\left\{{\bf\Xi}_{3}(k_{1},k_{2},-k_{12},k_{3},k_{4},k_{12})-{\bf\Xi}_{3}(-k_{1},-k_{2},-k_{12},k_{3},k_{4},k_{12})\right\}
\right.\right.\\ \left.\left.~~~~~~~~~+3(\vec{k_{1}}.\vec{k_{2}})
\left\{{\bf\Xi}_{4}(-k_{12},k_{1},k_{2},k_{3},k_{4},k_{12})-{\bf\Xi}_{4}(-k_{12},-k_{1},-k_{2},k_{3},k_{4},k_{12})\right\}
\right.\right.\\ \left.\left.~~~~~~~~~-6(\vec{k_{12}}.\vec{k_{2}})
\left\{{\bf\Xi}_{4}(k_{1},k_{2},-k_{12},k_{3},k_{4},k_{12})-{\bf\Xi}_{4}(-k_{1},-k_{2},-k_{12},k_{3},k_{4},k_{12})\right\}\right]
\right.\\ \left. \displaystyle~~~~~~~~~~-{\bf B}^{2}\left[(\vec{k_{1}}.\vec{k_{2}})(\vec{k_{3}}.\vec{k_{4}})
\left\{{\bf\Xi}_{2}(-k_{12},k_{1},k_{2},k_{12},k_{3},k_{4})-{\bf\Xi}_{2}(-k_{12},-k_{1},-k_{2},k_{12},k_{3},k_{4})\right\}
\right.\right.\\ \left.\left.~~~~~~~~~+2(\vec{k_{1}}.\vec{k_{2}})(\vec{k_{12}}.\vec{k_{4}})
\left\{{\bf\Xi}_{2}(-k_{12},k_{1},k_{2},k_{3},k_{4},k_{12})-{\bf\Xi}_{2}(-k_{12},-k_{1},-k_{2},k_{3},k_{4},k_{12})\right\}
\right.\right.\\ \left.\left.~~~~~~~~~-2(\vec{k_{3}}.\vec{k_{4}})(\vec{k_{12}}.\vec{k_{2}})
\left\{{\bf\Xi}_{2}(k_{1},k_{2},-k_{12},k_{12},k_{3},k_{4})-{\bf\Xi}_{2}(-k_{1},-k_{2},-k_{12},k_{12},k_{3},k_{4})\right\}
\right.\right.\\ \left.\left.~~~~~~~~~-4(\vec{k_{12}}.\vec{k_{4}})(\vec{k_{12}}.\vec{k_{2}})
\left\{{\bf\Xi}_{2}(k_{1},k_{2},-k_{12},k_{3},k_{4},k_{12})-{\bf\Xi}_{2}(-k_{1},-k_{2},-k_{12},k_{3},k_{4},k_{12})\right\}\right]
\right.\\ \left. \displaystyle~~~~~~~~~+{\bf 23~permutations~of~}
(k_{1},k_{2},k_{3},k_{4})\right\}\end{array}
\ee
where the momentum dependent functions  ${\bf\Xi}_{i}\forall i$
are mentioned in the Appendix. 
Further, using the {\it equilateral configuration}
 the non-Gaussian parameter from scalar exchange contribution 
can be expressed as:

\be\begin{array}{llll}\label{gfneqas22}\displaystyle\tau^{equil;\bf SE}_{NL}=
\frac{2^{20\nu_{s}-42}{ k}^{10\nu_{s}-18}Cos\left(\left[\nu_{s}-\frac{1}{2}\right]\frac{\pi}{2}\right)}
{9\sqrt{3}}\left|\frac{\Gamma(\nu_{s})}{\Gamma(\frac{3}{2})}\right|^{8}
\frac{(1-\epsilon_{V}-s^{S}_{V})^{8}H^{8}}
{Y^{5}_{S}c^{6}_{s}}\\~~~~~~~~~~~~~~~~~~
\times\displaystyle 
\displaystyle\left\{ 9 {\bf A}^{2}\left[{\bf\Xi}_{1}\left(-k,-k,-\frac{2k}{\sqrt{3}},k,k,\frac{2k}{\sqrt{3}}\right)-{\bf\Xi}_{1}\left(k,k,-\frac{2k}{\sqrt{3}},k,k,\frac{2k}{\sqrt{3}}\right)\right]
\right.\\ \left. \displaystyle~~~~~~~~~~~~~~~~~~~~~~~~+k^{2}{\bf AB}\left[3
\left\{{\bf\Xi}_{3}\left(k,k,-\frac{2k}{\sqrt{3}},\frac{2k}{\sqrt{3}},k,k\right)-{\bf\Xi}_{3}\left(-k,-k,-\frac{2k}{\sqrt{3}},\frac{2k}{\sqrt{3}},k,k\right)\right\}
\right.\right.\\ \left.\left.\displaystyle ~~~~~~~~~~~~~~~~~~~~~~~~+4\sqrt{3}
\left\{{\bf\Xi}_{3}\left(k,k,-\frac{2k}{\sqrt{3}},k,k,\frac{2k}{\sqrt{3}}\right)-{\bf\Xi}_{3}\left(-k,-k,-\frac{2k}{\sqrt{3}},k,k,\frac{2k}{\sqrt{3}}\right)\right\}
\right.\right.\\ \left.\left.\displaystyle ~~~~~~~~~~~~~~~~~~~~~~~~+3
\left\{{\bf\Xi}_{4}\left(-\frac{2k}{\sqrt{3}},k,k,k,k,\frac{2k}{\sqrt{3}}\right)-{\bf\Xi}_{4}\left(-\frac{2k}{\sqrt{3}},-k,-k,k,k,\frac{2k}{\sqrt{3}}\right)\right\}
\right.\right.\\ \left.\left.\displaystyle ~~~~~~~~~~~~~~~~~~~~~~~~-4\sqrt{3}
\left\{{\bf\Xi}_{4}\left(k,k,-\frac{2k}{\sqrt{3}},k,k,\frac{2k}{\sqrt{3}}\right)-{\bf\Xi}_{4}\left(-k,-k,-\frac{2k}{\sqrt{3}},k,k,\frac{2k}{\sqrt{3}}\right)\right\}\right]
\right.\\ \left. \displaystyle~~~~~~~~~~~~~~~~~~~~~~~~-k^{4}{\bf B}^{2}\left[
\left\{{\bf\Xi}_{2}\left(-\frac{2k}{\sqrt{3}},k,k,\frac{2k}{\sqrt{3}},k,k\right)-{\bf\Xi}_{2}\left(-\frac{2k}{\sqrt{3}},-k,-k,\frac{2k}{\sqrt{3}},k,k\right)\right\}
\right.\right.\\ \left.\left.\displaystyle ~~~~~~~~~~~~~~~~~~~~~~~~+\frac{4}{\sqrt{3}}
\left\{{\bf\Xi}_{2}\left(-\frac{2k}{\sqrt{3}},k,k,k,k,\frac{2k}{\sqrt{3}}\right)-{\bf\Xi}_{2}\left(-\frac{2k}{\sqrt{3}},-k,-k,k,k,\frac{2k}{\sqrt{3}}\right)\right\}
\right.\right.\\ \left.\left.\displaystyle ~~~~~~~~~~~~~~~~~~~~~~~~-\frac{4}{\sqrt{3}}
\left\{{\bf\Xi}_{2}\left(k,k,-\frac{2k}{\sqrt{3}},\frac{2k}{\sqrt{3}},k,k\right)-{\bf\Xi}_{2}\left(-k,-k,-\frac{2k}{\sqrt{3}},\frac{2k}{\sqrt{3}},k,k\right)\right\}
\right.\right.\\ \left.\left.\displaystyle ~~~~~~~~~~~~~~~~~~~~~~~~-\frac{16}{3}
\left\{{\bf\Xi}_{2}\left(k,k,-\frac{2k}{\sqrt{3}},k,k,\frac{2k}{\sqrt{3}}\right)-{\bf\Xi}_{2}\left(-k,-k,-\frac{2k}{\sqrt{3}},k,k,\frac{2k}{\sqrt{3}}\right)\right\}\right]
\right.\\ \left. \displaystyle~~~~~~~~~~~~~~~~~~~~~~~~~~~~~~~~~~~~~~~~~~~~~~~~~~~~~~~~~~~~~~~~~~~~~~~~~~~~~~~~~~~~~~~~~~~~~~~~~~~~+{\bf 23~permutations}\right\}.
          \end{array}
\ee
 
\subsection{\bf Graviton Exchange}
In this section we are interested to evaluate the contribution of four-point function of curvature perturbations from the exchange of graviton.
This process involves a third-order interaction among scalar fluctuations and tensor perturbations. To proceed, we need here
only the significant third order term in the action, which describes the graviton-scalar-scalar vertex in the uniform gauge  as:
\be\begin{array}{lllll}\label{trispec3}
    \displaystyle S_{{\bf GE}}
=\frac{1}{2}\int dt~ d^{3}x~a^{2}{\cal Y}_1~h_{ij}\zeta_{,i}\zeta_{,j},
   \end{array}\ee
where ${\cal Y}_1=Y_{S}c^{2}_{S}$.
Using in-in procedure
the {\it four point correlation function} both for quasi-exponential limit can be expressed in the following form:
\be\begin{array}{llll}\label{threept3}
\displaystyle\langle\zeta(\vec{k_{1}})\zeta(\vec{k_{2}})\zeta(\vec{k_{3}})\zeta(\vec{k_{4}})\rangle^{\bf GE}
=-i\lim_{\eta^{\star}\rightarrow 0}\int^{\eta_{\star}}_{-\infty}d\eta~\int^{\eta}_{-\infty}d\tilde{\eta} ~\langle 0|\left[\left[\zeta(\vec{k_{1}})\zeta(\vec{k_{2}})
\zeta(\vec{k_{3}})\zeta(\vec{k_{4}}),\left(H_{int}(\eta)\right)^{\bf GE}_{\zeta\zeta\zeta}\right],\left(H_{int}(\tilde{\eta})\right)^{\bf GE}_{\zeta\zeta\zeta}\right]|0\rangle\\
~~~~~~~~~~~~~~~~~~~~~~~~~~~~~~~~~~\displaystyle
=(2\pi)^{3}\delta^{(3)}
(\vec{k_{1}}+\vec{k_{2}}+\vec{k_{3}}+\vec{k_{4}}){\cal T}^{\bf GE}_{\zeta}(\vec{k_{1}},\vec{k_{2}},\vec{k_{3}},\vec{k_{4}}),\end{array}\ee
where in the interaction picture the Hamiltonian can be written in terms of the third order Lagrangian density as:
$\left(H_{int}(\eta)\right)^{\bf GE}_{\zeta\zeta h}=-\int d^{3}x~ {\cal L}^{\bf GE}_{3}.$
 
Here following the ansatz used in \cite{wands1} the {\it trispectrum} ${\cal T}^{\bf GE}_{\zeta}(\vec{k_{1}},\vec{k_{2}},\vec{k_{3}},\vec{k_{4}})$ obtained 
from the {\it graviton exchange} contribution is defined as:
\be\begin{array}{llllll}\label{bispec3}
\displaystyle {\cal T}^{\bf GE}_{\zeta}(\vec{k_{1}},\vec{k_{2}},\vec{k_{3}},\vec{k_{4}})
=\frac{1}{\prod^{4}_{i=1}k^{3}_{i}}\left[(k^{3}_{1}k^{3}_{2}+k^{3}_{3}k^{3}_{4})\left(k^{-3}_{13}+k^{-3}_{14}\right)+
(k^{3}_{1}k^{3}_{4}+k^{3}_{2}k^{3}_{3})\left(k^{-3}_{12}+k^{-3}_{13}\right)
\right.\\ \left.~~~~~~~~~~~~~~~~~~~~~~~~~~~~~~~~~~~~~~~~~~~~~~~~~~~~~~~~~~~~~~~~~~~~~~~~
+(k^{3}_{1}k^{3}_{3}+k^{3}_{2}k^{3}_{4})\left(k^{-3}_{12}+k^{-3}_{14}\right)\right]
\left\{\tau^{\bf GE}_{NL}{P}^{3}_{\zeta(1)}+\frac{54}{25}
g^{\bf GE}_{NL}{P}^{3}_{\zeta(2)}\right\},\end{array}\ee
where 
$\tau^{\bf GE}_{NL}$ and $g^{\bf GE}_{NL}$ are 
the two non linear parameters which carry the signatures of primordial non-Gaussianities of the curvature perturbation in trispectrum analysis.
By knowing $\tau^{\bf GE}_{NL}$ the other parameter $g^{\bf GE}_{NL}$ can be calculated by making use of the following relation \cite{mizko}:
\be\begin{array}{llll}\label{gnl3}
\displaystyle g^{\bf GE}_{NL}=\frac{64}{\bar{K}^{3}}\left[(k^{3}_{1}k^{3}_{2}+k^{3}_{3}k^{3}_{4})\left(k^{-3}_{13}+k^{-3}_{14}\right)+
(k^{3}_{1}k^{3}_{4}+k^{3}_{2}k^{3}_{3})\left(k^{-3}_{12}+k^{-3}_{13}\right)
+(k^{3}_{1}k^{3}_{3}
+k^{3}_{2}k^{3}_{4})\left(k^{-3}_{12}+k^{-3}_{14}\right)\right]\tau^{\bf GE}_{NL},\end{array}\ee
where $\bar{K}=k_{1}+k_{2}+k_{3}+k_{4}$. 
The explicit form of $\tau^{\bf GE}_{NL}$ characterizing the trispectrum obtained from the graviton exchange contribution can be expressed for our model
as:

\be\begin{array}{lllll}\label{gfn113}
\displaystyle \tau^{\bf GE}_{NL}
=\lim_{\eta^{\star}\rightarrow 0}\frac{2^{8\nu_{s}-31}\pi^{2}(1-\epsilon_{V}-s^{S}_{V})^{8}(1-\epsilon_{V}-s^{T}_{V})^{2}{\bar K}^{10\nu_{s}-15}Sin^{8}\left(\left[\nu_{s}-\frac{1}{2}\right]\frac{\pi}{2}\right)Sin^{2}\left(\left[\nu_{T}-\frac{1}{2}\right]\frac{\pi}{2}\right)}{\left[(k^{3}_{1}k^{3}_{2}+k^{3}_{3}k^{3}_{4})\left(k^{-3}_{13}+k^{-3}_{14}\right)+
(k^{3}_{1}k^{3}_{4}+k^{3}_{2}k^{3}_{3})\left(k^{-3}_{12}+k^{-3}_{13}\right)
+(k^{3}_{1}k^{3}_{3}+k^{3}_{2}k^{3}_{4})\left(k^{-3}_{12}+k^{-3}_{14}\right)\right]}\\~~~~~~~~~~
\times\displaystyle  \left|\frac{\Gamma(\nu_{s})}{\Gamma(\frac{3}{2})}\right|^{8}\left|\frac{\Gamma(\nu_{T})}{\Gamma(\frac{3}{2})}\right|^{2}
\frac{H^{10}}
{Y^{4}_{S}Y^{2}_{T}c^{12}_{s}c^{6}_{T}}\left\{
    \sum_{\substack{\lambda=+[+,-],\cr ~~~\times[+,-]}}~ \sum_{i,j,l,m} \sum_{\substack{a<b \cr c<d \cr {\bf 23 ~perms.}}}
    \epsilon_{ij}^{\lambda}(\vec{k}_{ab}) \epsilon_{lm}^{\lambda}(\vec{k}_{cd})
    \frac{k_a^i k_b^j k_c^l k_d^m}{k^{\nu_{T}+3}_{ab}(k_{a}k_{b}k_{c}k_{d})^{2\nu_{s}}}
    \cdot \mathcal{\vartheta}_{abcd}(\eta_{\star})\right\} \;,\end{array}
\ee
The momentum dependent functions $\mathcal{\vartheta}_{abcd}(\eta_{\star})$
are given in the Appendix. Here to write equation(\ref{gfn113}) we have used the fact that the exchange momentum dependent 
polarization tensor $\epsilon^{\lambda}_{ij}(\vec{k}_{ab})$
is a symmetric tensor and also the four-point correlator obtained from the graviton exchange is invariant under the exchange of the subscripts of the momenta
 $a\leftrightarrow b$
and $c\leftrightarrow d$. Additionally in equation(\ref{gfn113}) the sum is performed only over different indices $a,b,c,d$ and we have extracted an overall symmetry factor of 4
which takes care about the exchanges $a\leftrightarrow b$
and $c\leftrightarrow d$. Rewriting the sums appearing in equation(\ref{gfn113})  we get the following reduced formula for the non-Gaussian parameter:
\be\begin{array}{lllll}\label{gfn113xcs}
\displaystyle \tau^{\bf GE}_{NL}
=\frac{2^{8\nu_{s}-31}\pi^{2}(1-\epsilon_{V}-s^{S}_{V})^{8}(1-\epsilon_{V}-s^{T}_{V})^{2}{\bar K}^{10\nu_{s}-15}Sin^{8}\left(\left[\nu_{s}-\frac{1}{2}\right]\frac{\pi}{2}\right)Sin^{2}\left(\left[\nu_{T}-\frac{1}{2}\right]\frac{\pi}{2}\right)}{\left[(k^{3}_{1}k^{3}_{2}+k^{3}_{3}k^{3}_{4})\left(k^{-3}_{13}+k^{-3}_{14}\right)+
(k^{3}_{1}k^{3}_{4}+k^{3}_{2}k^{3}_{3})\left(k^{-3}_{12}+k^{-3}_{13}\right)
+(k^{3}_{1}k^{3}_{3}+k^{3}_{2}k^{3}_{4})\left(k^{-3}_{12}+k^{-3}_{14}\right)\right]}\\~~~~~~~~~~
\times\displaystyle  \left|\frac{\Gamma(\nu_{s})}{\Gamma(\frac{3}{2})}\right|^{8}\left|\frac{\Gamma(\nu_{T})}{\Gamma(\frac{3}{2})}\right|^{2}
\frac{H^{10}}
{Y^{4}_{S}Y^{2}_{T}c^{12}_{s}c^{6}_{T}}\left\{
    \sum_{\substack{\lambda=+[+,-],\cr ~~~\times[+,-]}}~ \sum_{i,j,l,m} \left[
    \epsilon_{ij}^{\lambda}(\vec{k}_{12}) \epsilon_{lm}^{\lambda}(\vec{k}_{34})
   \frac{ k_1^i k_2^j k_3^l k_4^m}{k^{\nu_{T}+3}_{12}(k_1 k_2 k_3 k_4)^{2\nu_{s}}}
    \cdot \left(\hat{\vartheta}_{1234}+\hat{\vartheta}_{3412}\right)\right.\right.\\ \left.\left.
\displaystyle ~~~~~~~~~~~~~~~~~~~~~~~~~~~~~~~~~~~~~~~~~~~~~~~~~~~~~~~~~~~~~~~~~~~+
    \epsilon_{ij}^{\lambda}(\vec{k}_{13}) \epsilon_{lm}^{\lambda}(\vec{k}_{24})
   \frac{ k_1^i k_3^j k_2^l k_4^m}{k^{\nu_{T}+3}_{13}(k_1 k_3 k_2 k_4)^{2\nu_{s}}}
    \cdot \left(\hat{\vartheta}_{1324}+\hat{\vartheta}_{2413}\right)\right.\right.\\ \left.\left.
\displaystyle ~~~~~~~~~~~~~~~~~~~~~~~~~~~~~~~~~~~~~~~~~~~~~~~~~~~~~~~~~~~~~~~~~~~+
    \epsilon_{ij}^{\lambda}(\vec{k}_{14}) \epsilon_{lm}^{\lambda}(\vec{k}_{23})
    \frac{k_1^i k_4^j k_2^l k_3^m}{k^{\nu_{T}+3}_{14}(k_1 k_4 k_2 k_3)^{2\nu_{s}}}
    \cdot \left(\hat{\vartheta}_{1423}+\hat{\vartheta}_{2314}\right)\right]\right\} \;,\end{array}
\ee
where we define $\lim_{\eta_{\star}\rightarrow 0}{\vartheta}_{abcd}(\eta_{\star}):=\hat{\vartheta}_{abcd}$.
There are divergent contributions in the limit $\eta_\ast
\rightarrow 0$ appear with a logarithmic dependence on the momenta,
but the additive cumulative contribution of $\mathcal{I}_{abcd}$
and $\mathcal{I}_{cdab}$ give rise to a finite contribution at late
times.

To represent Eq~(\ref{gfn113xcs}) in a simpler form, let us  start with
 the polarization sum $\sum_{s} \epsilon_{ij}^{s}(\vec k_{12}) \epsilon_{lm}^s
(\vec k_{34}) k_1^i k_2^j k_3^l k_4^m$ in terms of the relative
angles between the $\vec k_a$ and $\vec k_{12}$. The
polarization tensors $\epsilon^s_{ij}$ can be rewritten as
\be\begin{array}{lll}\label{jnt1}
   \displaystyle \epsilon_{ij}^+=
    \vec e_i \otimes\vec e_j - \vec {\bar e}_i \otimes\vec {\bar
    e}_j\;,
 ~~ \displaystyle \epsilon_{ij}^\times= \vec e_i \otimes\vec {\bar e}_j + \vec
    {\bar e}_i \otimes\vec e_j\;,
\end{array}\ee
where $\vec e$ and $\vec {\bar e}$ are orthogonal unit vectors
perpendicular to exchange momentum vector $\vec k_{12}$. It is convenient to
write the momentum vector $\vec k_a$ in a spherical polar coordinate system having $\{
\vec e, \vec {\bar e}, \vec {\hat k}_{12} \equiv \vec k_{12} / k_{12} \}$ as basis. In this coordinate
system one can express the momentum vector as:
$\vec k_a = k_a(\sin \theta_{a} \cos \phi_{a},\sin \theta_{a}
    \sin    \phi_{a}, \cos \theta_a)\;,$
where $\cos \theta_a \equiv \vec {\hat k}_a \cdot \vec {\hat
k}_{12} $ and $cos \phi_a \equiv \vec {\hat k}_a \cdot \vec e
$. This implies
\begin{eqnarray}
    \epsilon_{ij}^+ k_1^i k_2^j &=&  k_1 k_2 \sin \theta_1 \sin \theta_2
    cos (\phi_{1} + \phi_{2}) \;, 
\epsilon_{ij}^\times k_1^i k_2^j =  k_1 k_2
    sin \theta_1 \sin \theta_2 \sin (\phi_{1}+\phi_2) \;,
\end{eqnarray}
with an identical relation holding for $\epsilon_{ij}^+ k_3^i k_4^j$
and $\epsilon_{ij}^\times k_3^i k_4^j$ which will contribute to the polarization sum also. Since the projections of the momentum vectors $\vec
k_1$ and $\vec k_2$ ( similarly for $\vec k_3$ and $\vec k_4$) on the plane
orthogonal to exchange momentum vector $\vec k_{12}$ ($\vec k_{34}$) have the same amplitude but opposite
directions. Consequently we have two additional sets of constraint relationships given by:
\be\begin{array}{llll}\label{mnbs1}\displaystyle k_2 \sin \theta_2 = k_1 \sin \theta_1~~~  and~~~
\phi_2=\phi_1+\pi,~~~~
\displaystyle k_4 \sin \theta_4 = k_3 \sin \theta_3~~~ and
~~~\phi_4=\phi_3+\pi.\end{array}\ee
 Using these relations we get:
\be\begin{array}{llll}\label{a1}
 \displaystyle   \sum_s \epsilon_{ij}^s(\vec k_{ab}) \epsilon_{lm}^s (\vec
    k_{cd}) k_a^i k_b^j k_c^l k_d^m = k_a^2 k_c^2 \sin^2 \theta_a
    \sin^2 \theta_c \cos 2 \Upsilon_{ab,cd} \;,
\end{array}\ee
where we define a new angular coordinate $\Upsilon_{ab,cd} \equiv \phi_{a} - \phi_{c}$ with $a=1$, $(b,c)=2,3,4$, $d=3,4$ and $b>a,d>c,a\neq b\neq c \neq d$, which physically represents the angle
between the projections of the two momentum vectors $\vec k_a$ and $\vec k_c$ on the
plane orthogonal to $\vec k_{12}$. Alternatively this can be interpreted as the angle between
the two planes formed by the pair of momentum vectors $\{\vec k_1, \vec k_2 \}$ and $\{\vec
k_3, \vec k_4\}$.  Thus,
the expression for the non-Gaussian parameter calculated from the graviton exchange contribution from the trispectrum can
be simplified to the following expression:
\be\begin{array}{lllll}\label{gfn113xcxxc}
\displaystyle \tau^{\bf GE}_{NL}
=\frac{2^{8\nu_{s}-31}\pi^{2}(1-\epsilon_{V}-s^{S}_{V})^{8}(1-\epsilon_{V}-s^{T}_{V})^{2}{\bar K}^{10\nu_{s}-15}Sin^{8}\left(\left[\nu_{s}-\frac{1}{2}\right]\frac{\pi}{2}\right)Sin^{2}\left(\left[\nu_{T}-\frac{1}{2}\right]\frac{\pi}{2}\right)}{\left[(k^{3}_{1}k^{3}_{2}+k^{3}_{3}k^{3}_{4})\left(k^{-3}_{13}+k^{-3}_{14}\right)+
(k^{3}_{1}k^{3}_{4}+k^{3}_{2}k^{3}_{3})\left(k^{-3}_{12}+k^{-3}_{13}\right)
+(k^{3}_{1}k^{3}_{3}+k^{3}_{2}k^{3}_{4})\left(k^{-3}_{12}+k^{-3}_{14}\right)\right]}\\~~~~~~~~~~
\times\displaystyle  \left|\frac{\Gamma(\nu_{s})}{\Gamma(\frac{3}{2})}\right|^{8}\left|\frac{\Gamma(\nu_{T})}{\Gamma(\frac{3}{2})}\right|^{2}
\frac{H^{10}}
{Y^{4}_{S}Y^{2}_{T}c^{12}_{s}c^{6}_{T}}\left\{
    \frac{ k_1^2 k_3^2 [1-(\hat {\vec k}_{1} \cdot \hat {\vec
k}_{12})^2] [1-(\hat {\vec k}_{3} \cdot \hat {\vec k}_{12})^2]}{k^{\nu_{T}+3}_{12}(k_1 k_2 k_3 k_4)^{2\nu_{s}}}
\cos 2\Upsilon_{12,34}
    \cdot \left(\hat{\vartheta}_{1234}+\hat{\vartheta}_{3412}\right)\right.\\ \left.
\displaystyle ~~~~~~~~~~~~~~~~~~~~~~~~~~~~~~~~~~~~~~~~~~~~~~~~~~~~~~~~~~~~+
   \frac{ k_1^2 k_2^2 [1-(\hat {\vec k}_{1} \cdot \hat {\vec
k}_{13})^2] [1-(\hat {\vec k}_{2} \cdot \hat {\vec k}_{13})^2]}{k^{\nu_{T}+3}_{13}(k_1 k_3 k_2 k_4)^{2\nu_{s}}}
\cos 2\Upsilon_{13,24} 
    \cdot \left(\hat{\vartheta}_{1324}+\hat{\vartheta}_{2413}\right)\right.\\ \left.
\displaystyle ~~~~~~~~~~~~~~~~~~~~~~~~~~~~~~~~~~~~~~~~~~~~~~~~~~~~~~~~~~~~+
    \frac{k_1^2 k_2^2 [1-(\hat {\vec k}_{1} \cdot \hat {\vec
k}_{14})^2] [1-(\hat {\vec k}_{2} \cdot \hat {\vec k}_{14})^2]}{k^{\nu_{T}+3}_{14}(k_1 k_4 k_2 k_3)^{2\nu_{s}}}
\cos 2\Upsilon_{14,23} 
    \cdot \left(\hat{\vartheta}_{1423}+\hat{\vartheta}_{2314}\right)\right\} \;,\end{array}
\ee

Further, incorporating the contribution from the maximum shape of the trispectrum one can show that the 
graviton exchange contribution does not contribute anything in the equilateral limit. Now summing up all the significant contributions of four-point four scalar correlation coming from {\it contact interaction}, {\it scalar exchange} and {\it graviton exchange}
{\it interaction} the numerical value of $\tau^{equil}_{NL}$ in the equilateral limit
is obtained from our set up as  $48<\tau^{equil}_{NL}<97$
in quasi-exponential limit within the window for tensor-to-scalar ratio $0.213<r<0.250$
 which is significantly large from other class of DBI models and consistent with combined constraint obtained from {\it Planck+WMAP9+high-L+BICEP2 }\cite{Ade:2014xna,planck} data. 
\section{\bf Four point consistency conditions and violation of Suyama-Yamaguchi relation}

 In the {\it counter-collinear limit} collecting the contribution from the {\it scalar exchange} diagram 
we derive the following expression for the four point consistency condition:
\be\begin{array}{llll}\label{sq15}
    \displaystyle  \langle\zeta(\vec{k_{1}})\zeta(\vec{k_{2}})
\zeta(\vec{k_{3}})\zeta(\vec{k_{4}})\rangle^{SE}\approx
\displaystyle(2\pi)^{3}\delta^{3}({\vec k}_{1}+{\vec k}_{2}+{\vec k}_{3}+{\vec k}_{4})\frac{ \left(n_{\zeta}-1\right)^{2}}{4}
P_{\zeta}(k_{12})P_{\zeta}(k_{1})
\left[P_{\zeta}(k_{3}
)+\cdots\right]
   \end{array}\ee
which can be interpreted as the {\it scalar exchange } contribution arising from the product of two back-to-back bispectra in the squeezed limit.
Additionally, we consider the contribution from the {\it graviton exchange} diagram from which we derive another expression for the four point consistency condition:
\be\begin{array}{llll}\label{sq19}
    \displaystyle  \langle\zeta(\vec{k_{1}})\zeta(\vec{k_{2}})
\zeta(\vec{k_{3}})\zeta(\vec{k_{4}})\rangle^{GE}\approx
\displaystyle 9c_{s}\epsilon_{s}(2\pi)^{3}\delta^{3}({\vec k}_{1}+{\vec k}_{2}+{\vec k}_{3}+{\vec k}_{4})
P_{\zeta}(k_{12})P_{\zeta}(k_{1})\\
\displaystyle~~~~~~~~~~~~~~~~~~~~~~~~~~~~~~~~~~~~~~~~~~~~\times\left[\sum_{\substack{\lambda=+[+,-],\cr ~~~\times[+,-]}}~
    \sum_{i,j,l,m}\epsilon_{ij}^{\lambda}(\vec{k}_{12}) \epsilon_{lm}^{\lambda}(\vec{k}_{34})
\frac{k^{i}_{1}k^{j}_{1}k^{l}_{3}k^{m}_{3}}{k^{2}_{1}k^{2}_{3}}P_{\zeta}(k_{3})+\cdots\right]
   \end{array}\ee
Here using $k_{12}\rightarrow 0$, $\theta_{1},\theta_{3}\rightarrow \pi$ the polarization sum appearing in Eq~(\ref{sq19}) can be simplified to the following 
expression as:
\be\begin{array}{llll}\label{sq20}
    \displaystyle \sum_{\substack{\lambda=+[+,-],\cr ~~~\times[+,-]}}~
    \sum_{i,j,l,m}\epsilon_{ij}^{\lambda}(\vec{k}_{12}) \epsilon_{lm}^{\lambda}(\vec{k}_{34})\frac{k^{i}_{1}k^{j}_{1}k^{l}_{3}k^{m}_{3}}{k^{2}_{1}k^{2}_{3}}=\cos 2\Upsilon_{12,34}.
   \end{array}\ee
 Further substituting Eq~(\ref{sq20}) in Eq~(\ref{sq19}) and using Eq~(\ref{ghtv3}) the four-point correlation function from the {\it graviton exchange}
contribution in the counter-collinear limit ($k_{12}<<k_{1}\approx k_{2},k_{3}\approx k_{4}$) reduces to the following expression:
\be\begin{array}{llll}\label{sq21}
    \displaystyle  \langle\zeta(\vec{k_{1}})\zeta(\vec{k_{2}})
\zeta(\vec{k_{3}})\zeta(\vec{k_{4}})\rangle^{GE}=9.2^{2(\nu_{s}-\nu_{T}-4)}r_{\star}. \left[1+\frac{3}{2}{\cal O}(\epsilon^{2}_{T})\right]_{\star}\left(\frac{1-\epsilon_{V}-s^{S}_{V}}
{1-\epsilon_{V}-s^{T}_{V}}\right)^{2}_{\star}.\left|\frac{\Gamma(\nu_{s})}{\Gamma(\nu_{T})}\right|^{2}\\
\displaystyle~~~~~~~~~~~~~~~~~~~~~~~~~~~~~~~~~~~~~~~~~~~~\times(2\pi)^{3}\delta^{3}({\vec k}_{1}+{\vec k}_{2}+{\vec k}_{3}+{\vec k}_{4})
P_{\zeta}(k_{12})P_{\zeta}(k_{1})
\left[\cos 2\Upsilon_{12,34} P_{\zeta}(k_{3})+\cdots\right]
\end{array}\ee

 To check the validity of well known {\it Suyama-Yamguchi} consistency relation we start with the in-in picture where the four-point correlator can be written as: 
\be\begin{array}{llll}\label{sq22}
    \displaystyle \langle \zeta^{2}({\vec x})\zeta^{2}(0)\rangle_{\vec k}=\sum_{n}|\langle n_{\vec k}|\zeta^{2}(0)\rangle|^{2}
   \end{array}\ee
where n is a label for individual states or particle number within the momentum eigen space. Here the sum is written over positive definite terms. On the other hand 
in this context one of the contributions is the square of the squeezed limit of the three-point correlation function of the scalar contribution. This implies:
\be\begin{array}{llll}\label{sq23}
    \displaystyle \langle \zeta^{2}({\vec x})\zeta^{2}(0)\rangle_{\vec k}=\frac{|\langle \zeta({\vec k})|\zeta^{2}(0)\rangle|^{2}}{P_{\zeta}(k)}+\sum_{\tilde n}|\langle {\tilde n}_{\vec k}|\zeta^{2}(0)\rangle|^{2}.
   \end{array}\ee
As the second term in Eq~(\ref{sq23}) is always positive definite we conclude that:
    $\displaystyle \langle \zeta^{2}({\vec x})\zeta^{2}(0)\rangle_{\vec k}\geq\frac{|\langle \zeta({\vec k})|\zeta^{2}(0)\rangle|^{2}}{P_{\zeta}(k)}.$
 Further using this result in quasi-exponential limit we get:

\be\begin{array}{lllll}\label{opiu1}
 \displaystyle \lim_{q\to 0}\int_{\vec{k_{1}}} \frac{d^{3}k_{1}}{(2\pi)^{3}} \int_{\vec{k_{3}}} \frac{d^{3}k_{3}}{(2\pi)^{3}}
\langle\zeta(\vec{k_{1}})\zeta(\underbrace{\vec{q}-\vec{k_{1}}}_{\vec{ k_{2}}})\zeta(\vec{k_{3}})\zeta(\underbrace{-\vec{q}-\vec{k_{3}}}_{\vec{ k_{4}}})\rangle
\geq \lim_{q\to 0}\frac{\left|\int_{\vec{k_{2}}} \frac{d^{3}k_{2}}{(2\pi)^{3}}
\langle\zeta(\underbrace{\vec{k}}_{\vec{k_{1}}})
\zeta(\vec{k_{2}})\zeta(\underbrace{-\vec{q}-\vec{k_{2}}}_{\vec{ k_{3}}})\rangle\right|^{2}}{P_{\zeta}(q)}
\end{array}\ee

Hence using Eq~(\ref{opiu1}) finally we get:
\be\begin{array}{lllll}\label{opiuwe1}
\displaystyle\lim_{q\to 0}\int_{\vec{k_{1}}} \frac{d^{3}k_{1}}{(2\pi)^{3}} \int_{\vec{k_{3}}} \frac{d^{3}k_{3}}{(2\pi)^{3}}
P_{\zeta}(k_{1})P_{\zeta}(k_{2})\left\{\tau_{NL}-\frac{36}{25}\left(f_{NL}\right)^{2}\right\}\geq 0
   \end{array}\ee
resulting in a generic outcome of DBI Galileon inflation, viz,
\be
\displaystyle{\large{\hat\tau}_{NL}\geq\frac{36}{25}\left({\hat f}_{NL}\right)^{2}}
\ee
where  ${\hat\tau}_{NL}$ and ${\hat f}_{NL}$ are used to represent soft limits of the three and four point correlation functions.
This relation directly confirms
 the partial violation of {\it standard Suyama-Yamaguchi relation} \cite{ken,futa,alme} 
$\label{yguchi} {\hat \tau}_{NL}=\frac{36}{25}\left({\hat f}_{NL}\right)^{2}$
.
 These non-trivial features allow us to go beyond the no-go theorem in the present context.
Some other
aspects of the violation of well known {\it consistency relations} in the context of single field inflation has 
been studied in \cite{miskis,Rodriguez:2013cj}.

Let us now investigate for some possible explanations of 
the partial violation of {\it standard Suyama-Yamaguchi relation}.
The standard relations and limits of Non-Gaussianity are usually derived under the following assumptions:
\begin{itemize}
 \item The background is Einsteinian gravity,
 \item Inflation is driven by a single scalar field,
  \item The scalar field action is canonical, 
  \item Perfect slow roll conditions hold good throughout,
  \item The vacuum is Bunch-Davies.
\end{itemize}

Of course, most of the results derived using these assumptions are true to a great extent, it is not obvious that
they will still hold good if one or more assumptions are relaxed. Only when one deals with a context where 
he/she has to relax one or more assumptions, one can investigate for  the consequences and conclude if the relations are still valid or not.
In the present scenario, a non-Einstein framework forms the background 
along with a non-canonical action appearing in the matter sector for DBI Galilon.
The contributions of them arise through the first two terms of Eq~(\ref{model1})
which will further effect Eq~(\ref{sq15}) and Eq~(\ref{sq21}). 
On top of that, we have  higher derivative contributions for DBI Galileon matter sector, 
for contact interaction, scalar and graviton exchange contributions are coupled with 
the higher curvature contributions through highly non-linear terms as appearing in the perturbative 
action as mentioned in Eq~(\ref{trispec},\ref{trispec2},\ref{trispec3}),
which  directly affects the interaction vertex factors as well as the propagators of the setup,
resulting in deviation from standard results.
We suspect that  these {\it non-standard} inputs might have reflected in the violation of the no-go theorem.

Having said this, we do admit that this can at best serve as a qualitative explanation of the violation.
A huge amount of work needs to be done before one can comment conclusively on deviation from which assumption still respects 
the relation and deviation from which one leads to violation, and, in case it does, to what extent. 
This is a highly non-trivial task which one can only hope to attempt in future.


\begin{sidewaystable}
   \large\begin{tabular}{|c|c|c|c|c|c|} 
   \hline ${\bf \left[f_{NL;A}\right]^{u;(\lambda_{1}\lambda_{2}\lambda_{3})}}$  &{\bf }  
&  ${\bf \left[f_{NL;A}\right]^{u;(\lambda_{1}\lambda_{2}\lambda_{3})}}$ &{\bf } 
&  ${\bf \left[f_{NL;A}\right]^{u;(\lambda_{1}\lambda_{2}\lambda_{3})}}$  &{\bf }  \\
 {\bf (E-~mode)} &$\times10^{-3}$ & {\bf (E$\bigotimes$B-~mode)} &$\times10^{-3}$ & {\bf (B-~mode)} & $\times10^{-4}$ \\
\hline $\left[f_{NL;1}\right]^{1;(000)}$ {\bf(PC) }&4000~-~7000  & $\left[f_{NL;1}\right]^{2;(000)}$ {\bf(PC) }
 &0 & $\left[f_{NL;1}\right]^{3;(000)}$ {\bf(PC) }& 0\\
   \hline
   $\left[f_{NL;2}\right]^{1;(0++)}$ {\bf(PV) }&3.2~-~6.7 & $\left[f_{NL;2}\right]^{2;(0++)}$ {\bf(PV) } &2.1~-~4.5
 & $\left[f_{NL;2}\right]^{3;(0++)}$ {\bf(PV) } &2.8~-~8.7   \\
    \hline
    $\left[f_{NL;2}\right]^{1;(0--)}$  {\bf(PV) } &1.4~-~5.7 & $\left[f_{NL;2}\right]^{2;(0--)}$ {\bf(PV) } 
&2.1~-~8.9 &$\left[f_{NL;2}\right]^{3;(0--)}$ {\bf(PV) } &2.7~-~7.2\\
   \hline
   $\left[f_{NL;2}\right]^{1;(0+-)}$  {\bf(PV) }   &2.6~-~9.6 & $\left[f_{NL;2}\right]^{2;(0+-)}$ {\bf(PV) } &2.9~-~11.0
 &$\left[f_{NL;2}\right]^{3;(0+-)}$ {\bf(PV) }& 2.7~-~8.4 \\
   \hline
$\left[f_{NL;2}\right]^{1;(0-+)}$   {\bf(PV) }  &1.7~-~6.9 &$\left[f_{NL;2}\right]^{2;(0-+)}$  {\bf(PV) } &3.5~-~7.4
 &$\left[f_{NL;2}\right]^{3;(0-+)}$ {\bf(PV) } &1.8~-~10.6 \\
   \hline
$\left[f_{NL;3}\right]^{1;(00+)}$  {\bf(PC) }  & 121~-~432 &$\left[f_{NL;3}\right]^{2;(00+)}$  {\bf(PC) } &78~-~349
 &$\left[f_{NL;3}\right]^{3;(00+)}$ {\bf(PC) } & 45~-~221\\
   \hline
$\left[f_{NL;3}\right]^{1;(00-)}$   {\bf(PC) } &549~-~878 &$\left[f_{NL;3}\right]^{2;(00-)}$  {\bf(PC) } &304~-~883
 &$\left[f_{NL;3}\right]^{3;(00-)}$ {\bf(PC) }&189~-~588  \\
   \hline
$\left[f_{NL;4}\right]^{1;(+++)}$   {\bf(PV) } &0.23~-~0.97 & $\left[f_{NL;4}\right]^{2;(+++)}$ {\bf(PV) }
 &0.08~-~0.32 &$\left[f_{NL;4}\right]^{3;(+++)}$ {\bf(PV) } &0.02~-~0.34 \\
   \hline
$\left[f_{NL;4}\right]^{1;(---)}$  {\bf(PV) } &0.06~-~0.41 &$\left[f_{NL;4}\right]^{2;(---)}$ {\bf(PV) }
 &0.09~-~0.67 &$\left[f_{NL;4}\right]^{3;(---)}$ {\bf(PV) }&0.23~-~1.7\\
   \hline
$\left[f_{NL;4}\right]^{1;(++-)}$  {\bf(PV) }  &0.23~-~0.93 &$\left[f_{NL;4}\right]^{2;(++-)}$  {\bf(PV) }  &0.18~-~0.67
 &$\left[f_{NL;4}\right]^{3;(++-)}$ {\bf(PV) } &0.03~-~0.53\\
   \hline
$\left[f_{NL;4}\right]^{1;(+--)}$  {\bf(PV) }   &0.01~-~0.35 &$\left[f_{NL;4}\right]^{2;(+--)}$  {\bf(PV) } &0.07~-~0.44
 & $\left[f_{NL;4}\right]^{3;(+--)}$ {\bf(PV) } &0.02~-~0.42 \\
   \hline
$\left[f_{NL;4}\right]^{1;(-+-)}$  {\bf(PV) }   &0.04~-~0.39 &$\left[f_{NL;4}\right]^{2;(-+-)}$ {\bf(PV) } &0.02~-~0.32
 &$\left[f_{NL;4}\right]^{3;(-+-)}$ {\bf(PV) } &0.09~-~0.51\\
   \hline
$\left[f_{NL;4}\right]^{1;(-++)}$  {\bf(PV) }   &0.03~-~0.56 &$\left[f_{NL;4}\right]^{2;(-++)}$ {\bf(PV) }  &0.1~-~0.43
 &$\left[f_{NL;4}\right]^{3;(-++)}$ {\bf(PV) } &0.17~-~0.63\\
\hline
 $\left[f_{NL;4}\right]^{1;(--+)}$  {\bf(PV) }   &0.09~-~0.34 &$\left[f_{NL;4}\right]^{2;(--+)}$ {\bf(PV) } &0.07~-~0.41
 &$\left[f_{NL;4}\right]^{3;(--+)}$ {\bf(PV) } &0.05~-~0.44\\
   \hline  
\end{tabular}
  \caption{\large Different non-Gaussian ($\left[f_{NL;A}\right]^{u;(\lambda_{1}\lambda_{2}\lambda_{3})}$) parameters related to the primordial bispectrum 
  for A=1 (three scalar), 2(one scalar and two tensor), 3(two scalar and one tensor), 4(three tensor) with polarization index $u=1(E-mode), 2(E\bigotimes B-mode), 3(B-mode)$
including all helicity degrees of freedom represented by $\lambda_{1},\lambda_{2}$ and $\lambda_{3}$ estimated from our model. 
In this context ``+'',``-'' stands for two projections of helicity for graviton degrees of freedom and ``0'' represents helicity for scalar mode.
 Here {\bf PC} and {\bf PV} stands for
 the parity conserving and violating contributions appearing in the tree level primordial bispectrum analysis.}
  \label{tab1}
\end{sidewaystable}

\section{Summary and outlook}

In this article we have explored  primordial non-Gaussian features of DBI Galileon inflation 
in D3 brane. We have derived the expressions for three and four point correlation functions 
in terms of the non-linear parameters $f_{NL}$ and $\tau_{NL}$ for equilateral 
type of non-Gaussian configurations in the nontrivial polarization modes.
This resulted in a significantly large value for non-Gaussianity from this setup.
We could also find a parameter space for both non-Gaussianity and tensor-to-scalar ratio ($r$) 
consistent with combined constraint obtained from {\it Planck+WMAP9+high-L+BICEP2} data.
The detectable features of primordial non-Gaussianity lead to the conclusion that this type of models can directly be verified 
by upcoming data.
Moreover, the calculations reveal 
some other interesting results like partial violation of the  {\it Suyama-Yamaguchi} four-point consistency relation.

Some issues which can  be addressed in the context of non-Gaussianity for DBI Galileon are studies of mass spectrum of
 primordial black hole formation \cite{green}, \cite{kmai} as a tool for constraining non-Gaussianity at small scales;
 effect of the presence of one
loop and two loop radiative corrections in the presence of all possible scalar and tensor mode fluctuations in the bispectrum and trispectrum;
 study of different shapes in equilateral, local, orthogonal, squeezed limit configuration for the tree, one and two
loop level of non-Gaussianity and calculation of other higher order n-point correlation functions to find out the proper consistency relations between all
higher order non-Gaussian parameters as well as the analysis of CMB bispectrum and trispectrum in the presence of Galileon in SUGRA background.
Given the promise the results of the present paper shows, these open issues worth exploring in future as they may give rise to interesting results.


\section*{Acknowledgments}

SC thanks Council of Scientific and
Industrial Research, India for financial support through Senior
Research Fellowship (Grant No. 09/093(0132)/2010).

\section*{Appendix}
In this section we mention all the momentum dependent functions appearing in the context of bispectrum and trispectrum analysis
coming from all scalar-tensor three point correlations and four point scalar correlation.\\

\subsection*{\bf A. Functions appearing in three scalar correlation}

The functions appearing in the context of three scalar correlation can be expressed as:
\be\begin{array}{llllll}\label{dp1}
  {\cal I}_{1}(x)=Cos\left(\left[x-\frac{1}{2}\right]\frac{\pi}{2}\right)\Gamma(1+x)\left[\frac{2+x}{K}\sum_{i>j}k^{2}_{i}k^{2}_{j}
-\frac{1+x}{K^{2}}\sum_{i\neq j}k^{2}_{i}k^{3}_{j}\right],\\
 {\cal I}_{2}(x)=Cos\left(\left[x-\frac{1}{2}\right]\frac{\pi}{2}\right)\Gamma(1+x)\left[\frac{K}{1-x}
-\frac{1}{K}\sum_{i>j}k_{i}k_{j}-\frac{1+x}{K^{2}}k_{1}k_{2}k_{3}\right],
\displaystyle {\cal I}_{3}(x)=\frac{(k_{1}k_{2}k_{3})^{3}}{K^{3}}
\frac{\Gamma(3+x)}{2}Cos\left(\left[x-\frac{1}{2}\right]\frac{\pi}{2}\right),\\
 
 {\cal I}_{4}(x)=Cos\left(\left[x-\frac{1}{2}\right]\frac{\pi}{2}\right)\left\{\frac{(\vec{k_{1}}.\vec{k_{2}})k^{2}_{3}}{K}
\left[(3+x)\Gamma(1+x)-\Gamma(2+x)\frac{k_{3}}{K}\right]+\frac{(\vec{k_{2}}.\vec{k_{3}})k^{2}_{1}}{K}
\left[(3+x)\Gamma(1+x)-\Gamma(2+x)\frac{k_{1}}{K}\right]\right.\\ \left.~~~~~~~~~~~~~~~~~~~~~~~~~~~~~~~~~~~~~~~~~~~~~~~~~~~~~~~~~~~~~~~~~~~~~~~~~~~~~~~~~~~~~~~ +\frac{(\vec{k_{3}}.\vec{k_{1}})k^{2}_{2}}{K}
\left[(3+x)\Gamma(1+x)-\Gamma(2+x)\frac{k_{2}}{K}\right]\right\},\\
 {\cal I}_{5}(x)=Cos\left(\left[x-\frac{1}{2}\right]\frac{\pi}{2}\right)\left\{\frac{(\vec{k_{1}}.\vec{k_{2}})k^{2}_{3}}{K}
\left[\Gamma(1+x)+\Gamma(2+x)\frac{k_{3}}{K}\right] 
+\frac{(\vec{k_{2}}.\vec{k_{3}})k^{2}_{1}}{K}
\left[\Gamma(1+x)+\Gamma(2+x)\frac{k_{1}}{K}\right]+\right.\\ \left.~~~~~~~~~~~~~~~~~~~~~~~~~~~~~~~~~~~~~~~~~~~~~~~~~~~~~~~~~~~~~~~~~~~~~~~~~~~~~~~~~~~~~~~~~~~~~~~~~\frac{(\vec{k_{3}}.\vec{k_{1}})k^{2}_{2}}{K}
\left[\Gamma(1+x)+\Gamma(2+x)\frac{k_{2}}{K}\right]\right\},\\
 {\cal I}_{6}(x)=\frac{(k_{1}k_{2}k_{3})^{2}}{K^{3}}
Cos\left(\left[x-\frac{1}{2}\right]\frac{\pi}{2}\right)\frac{(6+x)\Gamma(3+x)}{12},\\

 {\cal I}_{7}(x)=Cos\left(\left[x-\frac{1}{2}\right]\frac{\pi}{2}\right)\frac{2+x}{2}\left[\Gamma(1+x)+\Gamma(2+x)
\left(\frac{k_{1}k_{2}+k_{2}k_{3}+k_{3}k_{1}}{K^{2}}+(3+x)\frac{k_{1}k_{2}k_{3}}{K^{3}}\right)\right]
\left\{\frac{(\vec{k_{1}}.\vec{k_{2}})k^{2}_{3}}{K}
 +\frac{(\vec{k_{2}}.\vec{k_{3}})k^{2}_{1}}{K} +\frac{(\vec{k_{3}}.\vec{k_{1}})k^{2}_{1}}{K}\right\},\\
 {\cal I}_{8}(x)=Cos\left(\left[x-\frac{1}{2}\right]\frac{\pi}{2}\right)\left\{
\frac{(\vec{k_{1}}.\vec{k_{2}})k^{2}_{3}}{K}\left[(3+x)\Gamma(1+x)+(3+x)\Gamma(2+x)\frac{k_{3}}{K}-\Gamma(3+x)\frac{k^{2}_{3}}{K^{2}}\right]
 +\frac{(\vec{k_{2}}.\vec{k_{3}})k^{2}_{1}}{K}
\left[(3+x)\Gamma(1+x)\right.\right.\\ \left.\left.~~~~~~~~~ +(3+x)\Gamma(2+x)\frac{k_{1}}{K}-\Gamma(3+x)\frac{k^{2}_{1}}{K^{2}}\right]
+\frac{(\vec{k_{3}}.\vec{k_{1}})k^{2}_{2}}{K}
\left[(3+x)\Gamma(1+x)+(3+x)\Gamma(2+x)\frac{k_{1}}{K}-\Gamma(3+x)\frac{k^{2}_{1}}{K^{2}}\right]\right\}.
   \end{array}\ee
In the equilateral configuration these functions are related through the following expression:
\be\begin{array}{lll}\label{eqconsd}\displaystyle{\cal I}^{equil}_{1}(x)={\cal I}^{equil}_{5}(x)=\frac{{\cal I}^{equil}_{4}(x)}{2},
 \displaystyle{\cal I}^{equil}_{2}(x)=\frac{3{\cal I}^{equil}_{8}(x)}{2(1-x)},
 \displaystyle{\cal I}^{equil}_{6}(x)=\left(1+\frac{x}{2}\right){\cal I}^{equil}_{3}(x).\end{array}\ee

Additionally in the squeezed limit these functions are reduced to the following expressions:
\be\begin{array}{llllll}\label{dp12}
\displaystyle   {\cal I}^{sq}_{1}=Cos\left(\left[x-\frac{1}{2}\right]\frac{\pi}{2}\right)k^{3}_{1}\Gamma(1+x)\left[\frac{2+x}{2}
-\frac{(1+x)}{2}\right],
 \displaystyle{\cal I}^{sq}_{2}=Cos\left(\left[x-\frac{1}{2}\right]\frac{\pi}{2}\right)\Gamma(1+x)\left[\frac{2k_{1}}{1-x}
-\frac{k_{1}}{2}-\frac{1+x}{4}k_{3}\right],\\
\displaystyle{\cal I}^{sq}_{3}=\frac{(k_{1}k_{3})^{3}}{8}
\frac{\Gamma(3+x)}{2}Cos\left(\left[x-\frac{1}{2}\right]\frac{\pi}{2}\right),\displaystyle {\cal I}^{sq}_{6}=\frac{k_{1}k^{2}_{3}}{8}
Cos\left(\left[x-\frac{1}{2}\right]\frac{\pi}{2}\right)\frac{(6+x)\Gamma(3+x)}{12},\\
\displaystyle{\cal I}^{sq}_{4}=Cos\left(\left[x-\frac{1}{2}\right]\frac{\pi}{2}\right)\left\{\frac{k_{1}k^{2}_{3}}{2}
\left[(3+x)\Gamma(1+x)-\Gamma(2+x)\frac{k_{3}}{2k_{1}}\right]
+(\vec{k_{1}}.\vec{k_{3}})k_{1}
\left[(3+x)\Gamma(1+x)-\Gamma(2+x)\frac{1}{2}\right]\right\},\\
 \displaystyle{\cal I}^{sq}_{5}=Cos\left(\left[x-\frac{1}{2}\right]\frac{\pi}{2}\right)\left\{\frac{k_{1}k^{2}_{3}}{2}
\left[\Gamma(1+x)+\Gamma(2+x)\frac{k_{3}}{2k_{1}}\right]
+\frac{(\vec{k_{1}}.\vec{k_{3}})k^{2}_{1}}{k_{1}}
\left[\Gamma(1+x)+\Gamma(2+x)\frac{k_{1}}{K}\right]\right\},\\


 \displaystyle{\cal I}^{sq}_{7}=Cos\left(\left[x-\frac{1}{2}\right]\frac{\pi}{2}\right)\frac{2+x}{2}\left[\Gamma(1+x)+\Gamma(2+x)
\left(\frac{1}{4}+(3+x)\frac{k_{3}}{8k_{1}}\right)\right]
\left\{\frac{k_{1}k^{2}_{3}}{2}
 +(\vec{k_{1}}.\vec{k_{3}})k_{1}\right\},\\
\displaystyle {\cal I}^{sq}_{8}=Cos\left(\left[x-\frac{1}{2}\right]\frac{\pi}{2}\right)\left\{
\frac{k_{1}k^{2}_{3}}{2}\left[(3+x)\Gamma(1+x)+(3+x)\Gamma(2+x)\frac{k_{3}}{2k_{1}}-\Gamma(3+x)\frac{k^{2}_{3}}{4k^{2}_{1}}\right]
\right.\\ \left.~~~~~~~~~~~~~~~~~~~~~~~~~~~~~~~~~~~~~~ +(\vec{k_{1}}.\vec{k_{3}})k_{1}
\left[(3+x)\Gamma(1+x)+(3+x)\Gamma(2+x)\frac{1}{2}-\Gamma(3+x)\frac{1}{4}\right]
\right\}.
   \end{array}\ee
\\

\subsection*{\bf B. Functions appearing in one scalar two tensor correlation}

The functional dependence of the co-efficients appearing in the context of one scalar two tensor correlation can be expressed as:
\begin{eqnarray}
&&\left(\nabla_{1}\right)^u_{ij;kl}=\sum^{6}_{p=1}\left\{\frac{\left[{\cal J}_{p}(\vec{k_{1}},\vec{k_{2}},\vec{k_{3}})\right]^u_{ij;kl}}{k^{\nu_{s}}_{1}(k_{2}k_{3})^{\nu_{T}}}
+\frac{\left[{\cal J}_{p}(\vec{k_{2}},\vec{k_{1}},\vec{k_{3}})\right]^u_{ij;kl}}{k^{\nu_{s}}_{2}(k_{1}k_{3})^{\nu_{T}}}+\frac{\left[{\cal J}_{p}
(\vec{k_{3}},\vec{k_{2}},\vec{k_{1}})\right]^u_{ij;kl}}{k^{\nu_{s}}_{3}(k_{2}k_{1})^{\nu_{T}}}\right\}\frac{\Gamma(7+p-4\nu_{T}-
2\nu_{s})}{c^{2\nu_s-\frac{p}{3}-\frac{7}{3}}_{s}c^{4\nu_T-\frac{2p}{3}-\frac{14}{3}}_{T}\underline{K}^{7+p-4\nu_{T}-2\nu_{s}}},
\nonumber\\
&&\left(\nabla_{2}\right)^u_{ij;kl}=\frac{\left[\frac{(\vec{k_{1}}.\vec{k_{2}})}{k^{\nu_{s}}_{1}(k_{2}k_{3})^{\nu_{T}}}+\frac{(\vec{k_{2}}.\vec{k_{3}})}{k^{\nu_{s}}_{2}(k_{1}k_{3})^{\nu_{T}}}
+\frac{(\vec{k_{3}}.\vec{k_{1}})}{k^{\nu_{s}}_{3}(k_{2}k_{3})^{\nu_{T}}}\right]}{\left(\frac{3}{2}-\nu_T\right)^2c^2_T}\sum^{4}_{p=1}H_{p}
\frac{\Gamma(9+p-4\nu_{T}-2\nu_{s})}{c^{2\nu_s-\frac{p}{3}-3}_{s}c^{4\nu_T-\frac{2p}{3}-6}_{T}\underline{K}^{9+p-4\nu_{T}-2\nu_{s}}}{\cal N}^u_{ij,kl},
\nonumber\\
&&\left(\nabla_{3}\right)^u_{ij;kl}=\frac{\left[\left(\vec{k_{1}}.\vec{k_{2}}\right)Y_{123}
+\left(\vec{k_{1}}.\vec{k_{3}}\right)Y_{132}+\left(\vec{k_{2}}.\vec{k_{3}}\right)Y_{213}+\left(\vec{k_{2}}.\vec{k_{1}}\right)Y_{231}
+\left(\vec{k_{3}}.\vec{k_{2}}\right)Y_{312}+\left(\vec{k_{3}}.\vec{k_{1}}\right)Y_{321}\right]{\cal N}^u_{ij,kl}}{\left(\frac{3}{2}-\nu_T\right)^2c^2_T},
\nonumber\\
&&\left(\nabla_{4}\right)^u_{ij;kl}=\left(\frac{3}{2}-\nu_s\right)\left[{\cal \tilde{J}}_{123}+{\cal \tilde{J}}_{132}
+{\cal \tilde{J}}_{213}+{\cal \tilde{J}}_{231}+{\cal \tilde{J}}_{312}+{\cal \tilde{J}}_{321}\right]{\cal N}^u_{ij,kl},
\nonumber\\
&&\left(\nabla_{5}\right)^u_{ij;kl}=\left[{\cal \tilde{C}}_{123}+{\cal \tilde{C}}_{132}+{\cal \tilde{C}}_{213}+{\cal \tilde{C}}_{231}
+{\cal \tilde{C}}_{312}+{\cal \tilde{C}}_{321}\right]^u_{ij,kl},
\nonumber\\
&&\left(\nabla_{6}\right)^u_{ij;kl}=\left[{\cal \hat{W}}_{123}+{\cal \hat{W}}_{132}+{\cal \hat{W}}_{213}+{\cal \hat{W}}_{231}+{\cal \hat{W}}_{312}+{\cal \hat{W}}_{321}\right]{\cal N}^u_{ij,kl},
\nonumber\\
&&\left(\nabla_{7}\right)^u_{ij;kl}=\left[k_{1m}k_{1m^{'}}\left\{\bar{X}_{123}+\bar{X}_{132}\right\}+k_{2m}k_{2m^{'}}\left\{\bar{X}_{231}+\bar{X}_{213}\right\}
+k_{3m}k_{3m^{'}}\left\{\bar{X}_{312}+\bar{X}_{321}\right\}\right]{\cal N}^u_{ij,kl}{\cal N}^u_{mn,m^{'}n}.
\end{eqnarray}
with
\be\begin{array}{llll}\label{jcas}
\left[{\cal J}_{1}(\vec{k_{a}},\vec{k_{b}},\vec{k_{c}})\right]^u_{ij;kl}={\cal N}^u_{ij,kl}\left(\frac{3}{2}-\nu_T\right)^2,
\left[{\cal J}_{2}(\vec{k_{a}},\vec{k_{b}},\vec{k_{c}})\right]^u_{ij;kl}=\left[{\cal J}_{1}(\vec{k_{a}},\vec{k_{b}},\vec{k_{c}})\right]^u_{ij;kl}\underline{K}c^{-\frac{1}{3}}_{s}c^{-\frac{2}{3}}_{T},
\\
\left[{\cal J}_{3}(\vec{k_{a}},\vec{k_{b}},\vec{k_{c}})\right]^u_{ij;kl}
={\cal N}^u_{ij,kl}\left(\frac{3}{2}-\nu_T\right)\left[(k^2_a+k^2_b+k_a k_b)+\left(\frac{3}{2}-\nu_T\right)k_{a}(k_{b}+k_{c})\right]
\\
\left[{\cal J}_{4}(\vec{k_{a}},\vec{k_{b}},\vec{k_{c}})\right]^u_{ij;kl}={\cal N}^u_{ij,kl}\left(\frac{3}{2}-\nu_T\right)\left[k_{b}k_{c}(k_{b}
+k_{c})+k_{a}(k^2_{b}+k^{2}_{c}+k_{b}k_{c})\right],
\\
\left[{\cal J}_{5}(\vec{k_{a}},\vec{k_{b}},\vec{k_{c}})\right]^u_{ij;kl}={\cal N}^u_{ij,kl}\left[k^2_bk^2_c+k_ak_bk_c(k_b+k_c)\right],
\left[{\cal J}_{6}(\vec{k_{a}},\vec{k_{b}},\vec{k_{c}})\right]^u_{ij;kl}=i{\cal N}^u_{ij,kl}k_{a}k^2_{b}k^2_{c},\\
~~~~~~~~~~~~~~~~~~~~~~~~~~~~~~~~~~~~~~~~~~~~~~~~~~~~~~~~~~~~~~~~~~~~~~~~(includes~3~permutations~of~a,b,c),
 \end{array}\ee
\be\begin{array}{llll}\label{integrals11}
 \displaystyle Y_{abc}=\frac{1}{k^{\nu_{s}}_{a}(k_{b}k_{c})^{\nu_{T}}}\left\{\sum^{4}_{p=1}H_{p}\frac{\Gamma(9+p-4\nu_{T}-2\nu_{s})}{c^{2\nu_s-\frac{p}{3}-3}_{s}c^{4\nu_T-\frac{2p}{3}-6}_{T}\underline{K}^{9+p-4\nu_{T}-2\nu_{s}}}
+\sum^{5}_{q=1}{\cal A}^{abc}_{q}\frac{\Gamma(8+q-4\nu_{T}-2\nu_{s})}{c^{2\nu_s-\frac{q}{3}-\frac{8}{3}}_{s}c^{4\nu_T-\frac{2q}{3}-\frac{16}{3}}_{T}
\underline{K}^{8+q-4\nu_{T}-2\nu_{s}}}\right\}\\~~~~~~~~~~~~~~~~~~~~~~~~~~~~~~~~~~~~~~~~~~~~~~~~~~~~~~~~~~~~~~~~~~~~~~~~~~~~~~~~~~~~~~~~~~~~
\displaystyle~(with~~a,b,c=1,2,3~~~with~a\neq b\neq c),\\
\displaystyle H_{1}=H_{2}=L_{1}, H_{3}=\frac{H_{1}\left(k_{a}k_{b}+k_{b}k_{c}+k_{c}k_{a}\right)}{c^{-\frac{2}{3}}_{s}c^{-\frac{4}{3}}_{T}\underline{K}^2},
 H_{4}=-\frac{k_{a}k_{b}k_{c}}{c^{-1}_{s}c^{-2}_{T}\underline{K}^3},\\
\displaystyle {\cal A}^{abc}_{1}=\frac{a^2Y_s c^2_s}{t_{1}}\left(\frac{3}{2}-\nu_s\right)^2,
{\cal A}^{abc}_{2}=c^{-\frac{1}{3}}_{s}c^{-\frac{2}{3}}_{T}\underline{K}{\cal A}^{abc}_{1},{\cal A}^{abc}_{3}=-\left(k_{a}k_{b}+k_{b}k_{c}+k_{c}k_{a}+k^{2}_{a}\right){\cal A}^{abc}_{1},\\
\displaystyle {\cal A}^{abc}_{4}=-\frac{a^2Y_s c^2_s}{t_{1}}\left(\frac{3}{2}-\nu_s\right)\left[k_{a}k_{b}k_{c}\left(\frac{3}{2}-\nu_{s}\right)+k^2_{a}(k_{b}+k_{c})\right],
{\cal A}^{abc}_{5}=\frac{a^2Y_s c^2_s}{t_{1}}k^2_{a}k_{b}k_{c},\\

\displaystyle \left[{\cal \tilde{C}}_{abc}\right]^{u}_{ij;kl}=\sum^{6}_{p=1}\frac{\left[{\cal J}_{p}(\vec{k_{a}},\vec{k_{b}},
\vec{k_{c}})\right]^u_{ij;kl}}{k^{\nu_{s}}_{a}(k_{b}k_{c})^{\nu_{T}}}\frac{\Gamma(7+p-4\nu_{T}-2\nu_{s})}{c^{2\nu_s-\frac{p}{3}-\frac{7}{3}}_{s}c^{4\nu_T-\frac{2p}{3}-\frac{14}{3}}_{T}
\underline{K}^{7+p-4\nu_{T}-2\nu_{s}}}~~~~~~~~~(includes~~6~~permuations~of~a,b,c),
\\ \displaystyle {\cal \hat{W}}_{abc}=\frac{1}{k^{\nu_{s}}_{a}(k_{b}k_{c})^{\nu_{T}}}\left\{
k^2_{a}\frac{\Gamma(8-4\nu_{T}-2\nu_{s})}{c^{2\nu_s-\frac{8}{3}}_{s}c^{4\nu_T-\frac{16}{3}}_{T}
\underline{K}^{8-4\nu_{T}-2\nu_{s}}}+\frac{a^2Y_s c^2_s}{t_{1}}\sum^{7}_{p=1}{a}^{abc}_{p}
\frac{\Gamma(6+p-4\nu_{T}-2\nu_{s})}{c^{2\nu_s-\frac{p}{3}-2}_{s}c^{4\nu_T-\frac{2p}{3}-4}_{T}\underline{K}^{6+p-4\nu_{T}-2\nu_{s}}}\right\},\\
\displaystyle \bar{X}_{abc}=\sum^{7}_{p=1}\frac{{a}^{abc}_{p}}{k^{\nu_{s}}_{a}(k_{b}k_{c})^{\nu_{T}}}
\frac{\Gamma(7+p-4\nu_{T}-2\nu_{s})}{c^{2\nu_s-\frac{p}{3}-\frac{7}{3}}_{s}c^{4\nu_T-\frac{2p}{3}-\frac{14}{3}}_{T}\underline{K}^{7+p-4\nu_{T}-2\nu_{s}}},\\

 \displaystyle {\cal \tilde{J}}_{abc}=\sum^{7}_{p=1}\frac{{a}^{abc}_{p}}{k^{\nu_{s}}_{a}(k_{b}k_{c})^{\nu_{T}}}
\frac{\Gamma(6+p-4\nu_T-2\nu_s)}{c^{2\nu_s-\frac{p}{3}-2}_{s}c^{4\nu_T-\frac{2p}{3}-4}_{T}\underline{K}^{6+p-4\nu_T-2\nu_s}},
{a}^{abc}_{1}=\left(\frac{3}{2}-\nu_{T}\right)^2\left(\frac{3}{2}-\nu_{s}\right), 
{a}^{abc}_{2}={a}^{abc}_{1}c^{-\frac{1}{3}}_{s}c^{-\frac{2}{3}}_{T}\underline{K},
\\ \displaystyle {a}^{abc}_{3}=\left(\frac{3}{2}-\nu_{T}\right)\left(\frac{3}{2}-\nu_{s}\right)\left[k_{a}(k_{b}+k_{c})+k^2_{b}+k^2_{c}+k_{b}k_{c}\right]
+\left(\frac{3}{2}-\nu_{T}\right)^2k^2_{a},\\
\displaystyle {a}^{abc}_{4}=\left[k^2_{a}(k_{b}+k_{c})\left(\frac{3}{2}-\nu_{T}\right)+\left\{k_{a}(k^2_{b}+k^2_{c}+k_{b}k_{c})+k_{b}k_{c}(k_{b}+k_{c})\right\}
\left(\frac{3}{2}-\nu_{T}\right)\left(\frac{3}{2}-\nu_{s}\right)\right],
\\ \displaystyle {a}^{abc}_{5}=\left[\left(\frac{3}{2}-\nu_{T}\right)k^2_{a}(k^2_{b}+k^2_{c}+k_{b}k_{c})+\left(\frac{3}{2}-\nu_{s}\right)k^2_{b}k^2_{c}
+\left(\frac{3}{2}-\nu_{T}\right)\left(\frac{3}{2}-\nu_{s}\right)k_{a}k_{b}k_{c}(k_{b}+k_{c})\right],
\\ \displaystyle {a}^{abc}_{6}=k_{a}k_{b}k_{c}\left[\left(\frac{3}{2}-\nu_{T}\right)k_{b}k_{c}+\left(\frac{3}{2}-\nu_{s}\right)k_{a}(k_{b}+k_{c})\right],
{a}^{abc}_{7}=-k^2_{a}k^2_{b}k^2_{c},
\end{array}\ee

After
using the basis transformation mentioned in equation(\ref{red}) the reduced form of the above mentioned co-efficients can be expressed in the following form:

\be\begin{array}{llll}\label{hjey}
\displaystyle\left(\nabla_{1}\right)^{u;\lambda_{2};\lambda_{3}}=\sum^{6}_{p=1}\left\{\frac{\left[{\cal J}_{p}(\vec{k_{1}},\vec{k_{2}},\vec{k_{3}})\right]^{u;\lambda_{2};\lambda_{3}}}{k^{\nu_{s}}_{1}(k_{2}k_{3})^{\nu_{T}}}
+\frac{\left[{\cal J}_{p}(\vec{k_{2}},\vec{k_{1}},\vec{k_{3}})\right]^{u;\lambda_{2};\lambda_{3}}}{k^{\nu_{s}}_{2}(k_{1}k_{3})^{\nu_{T}}}+\frac{\left[{\cal J}_{p}
(\vec{k_{3}},\vec{k_{2}},\vec{k_{1}})\right]^{u;\lambda_{2};\lambda_{3}}}{k^{\nu_{s}}_{3}(k_{2}k_{1})^{\nu_{T}}}\right\}\\
~~~~~~~~~~~~~~~~~~~~~~~~~~~~~~~~~~~~~~~~~~~~~~~~~~~~~~~~~~~~~~~~~~~~~~~~~~~~~~~~~~~\displaystyle\times
\frac{\Gamma(7+p-4\nu_{T}-2\nu_{s})}{c^{2\nu_s-\frac{p}{3}-\frac{7}{3}}_{s}c^{4\nu_T-\frac{2p}{3}-\frac{14}{3}}_{T}\underline{K}^{7+p-4\nu_{T}-2\nu_{s}}},
\\
\displaystyle\left(\nabla_{2}\right)^{u;\lambda_{2};\lambda_{3}}=\frac{2\left[\frac{(\vec{k_{1}}.\vec{k_{2}})}{k^{\nu_{s}}_{1}(k_{2}k_{3})^{\nu_{T}}}+\frac{(\vec{k_{2}}.\vec{k_{3}})}
{k^{\nu_{s}}_{2}(k_{1}k_{3})^{\nu_{T}}}
+\frac{(\vec{k_{3}}.\vec{k_{1}})}{k^{\nu_{s}}_{3}(k_{1}k_{2})^{\nu_{T}}}\right]^{u;\lambda_{2};\lambda_{3}}}{\left(\frac{3}{2}-\nu_T\right)^2c^2_T}\sum^{4}_{p=1}
H_{p}\frac{\Gamma(9+p-4\nu_{T}-2\nu_{s})}{c^{2\nu_s-\frac{p}{3}-3}_{s}c^{4\nu_T-\frac{2p}{3}-6}_{T}\underline{K}^{9+p-4\nu_{T}-2\nu_{s}}},
\\
\displaystyle\left(\nabla_{3}\right)^{u;\lambda_{2};\lambda_{3}}=\frac{2\left[\left(\vec{k_{1}}.\vec{k_{2}}\right)Y_{123}
+\left(\vec{k_{1}}.\vec{k_{3}}\right)Y_{132}+\left(\vec{k_{2}}.\vec{k_{3}}\right)Y_{213}+\left(\vec{k_{2}}.\vec{k_{1}}\right)Y_{231}
+\left(\vec{k_{3}}.\vec{k_{2}}\right)Y_{312}+\left(\vec{k_{3}}.\vec{k_{1}}\right)Y_{321}\right]^{u;\lambda_{2};\lambda_{3}}}{\left(\frac{3}{2}-\nu_T\right)^2c^2_T},
\\
\displaystyle\left(\nabla_{4}\right)^{u;\lambda_{2};\lambda_{3}}=2\left(\frac{3}{2}-\nu_s\right)\left[{\cal \tilde{J}}_{123}+{\cal \tilde{J}}_{132}
+{\cal \tilde{J}}_{213}+{\cal \tilde{J}}_{231}+{\cal \tilde{J}}_{312}+{\cal \tilde{J}}_{321}\right]\delta^{\lambda_{2}\lambda_{3}},
\\
\displaystyle\left(\nabla_{5}\right)^{u;\lambda_{2};\lambda_{3}}=\left[{\cal \tilde{C}}_{123}+{\cal \tilde{C}}_{132}+{\cal \tilde{C}}_{213}+{\cal \tilde{C}}_{231}
+{\cal \tilde{C}}_{312}+{\cal \tilde{C}}_{321}\right]^{u;\lambda_{2};\lambda_{3}},
\\
\displaystyle\left(\nabla_{6}\right)^{u;\lambda_{2};\lambda_{3}}=2\left[{\cal \hat{W}}_{123}+{\cal \hat{W}}_{132}
+{\cal \hat{W}}_{213}+{\cal \hat{W}}_{231}+{\cal \hat{W}}_{312}+{\cal \hat{W}}_{321}\right]\delta^{\lambda_{2}\lambda_{3}},
\\
\displaystyle \left(\nabla_{7}\right)^{u;\lambda_{2};\lambda_{3}}=\left[Z^{u;\lambda_{2};\lambda_{3}}_1\left\{\bar{X}_{123}+\bar{X}_{132}\right\}+Z^{u;\lambda_{2};\lambda_{3}}_2\left\{\bar{X}_{231}+\bar{X}_{213}\right\}
+Z^{u;\lambda_{2};\lambda_{3}}_3\left\{\bar{X}_{312}+\bar{X}_{321}\right\}\right].
\end{array}\ee
with
\be\begin{array}{llll}\label{jcas}
\left[{\cal J}_{1}(\vec{k_{a}},\vec{k_{b}},\vec{k_{c}})\right]^{u;\lambda_{2};\lambda_{3}}=2\left(\frac{3}{2}-\nu_T\right)^2,
\left[{\cal J}_{2}(\vec{k_{a}},\vec{k_{b}},\vec{k_{c}})\right]^{u;\lambda_{2};\lambda_{3}}=\left[{\cal J}_{1}(\vec{k_{a}},
\vec{k_{b}},\vec{k_{c}})\right]^{u;\lambda_{2};\lambda_{3}}c^{-\frac{1}{3}}_{s}c^{-\frac{2}{3}}_{T}\underline{K},
\\
\left[{\cal J}_{3}(\vec{k_{a}},\vec{k_{b}},\vec{k_{c}})\right]^{u;\lambda_{2};\lambda_{3}}
=2\lambda_{2}\lambda_{3}\left(\frac{3}{2}-\nu_T\right)\left[(k^2_a+k^2_b+k_a k_b)+\left(\frac{3}{2}-\nu_T\right)k_{a}(k_{b}+k_{c})\right]
\\
\left[{\cal J}_{4}(\vec{k_{a}},\vec{k_{b}},\vec{k_{c}})\right]^{u;\lambda_{2};\lambda_{3}}=2\left(\frac{3}{2}-\nu_T\right)\left[\lambda^{3}_{2}k_{b}k_{c}(k_{b}
+k_{c})+\lambda^{3}_{3}k_{a}(k^2_{b}+k^{2}_{c}+k_{b}k_{c})\right],
\\
\left[{\cal J}_{5}(\vec{k_{a}},\vec{k_{b}},\vec{k_{c}})\right]^{u;\lambda_{2};\lambda_{3}}=2\lambda^{2}_{2}\lambda^{2}_{3}\left[k^2_bk^2_c+k_ak_bk_c(k_b+k_c)\right],
\left[{\cal J}_{6}(\vec{k_{a}},\vec{k_{b}},\vec{k_{c}})\right]^{u;\lambda_{2};\lambda_{3}}=2i\lambda^{2}_{3}\lambda^{2}_{2}k_{a}k^2_{b}k^2_{c},\\
~~~~~~~~~~~~~~~~~~~~~~~~~~~~~~~~~~~~~~~~~~~~~~~~~~~~~~~~~~~~~~~~~~~~~~~~(includes~3~permutations~of~a,b,c),\\
 
\displaystyle \left[{\cal \tilde{C}}_{abc}\right]^{u;\lambda_{2};\lambda_{3}}=\sum^{6}_{p=1}\frac{\left[{\cal J}_{p}(\vec{k_{a}},\vec{k_{b}},
\vec{k_{c}})\right]^{u;\lambda_{2};\lambda_{3}}}{k^{\nu_{s}}_{a}(k_{b}k_{c})^{\nu_{T}}}\frac{\Gamma(7+p-4\nu_{T}
-2\nu_{s})}{c^{2\nu_s-\frac{p}{3}-\frac{7}{3}}_{s}c^{4\nu_T-\frac{2p}{3}-\frac{14}{3}}_{T}\underline{K}^{7+p-4\nu_{T}-2\nu_{s}}}~~~~~~~~~(includes~~6~~permuations~of~a,b,c),
\\ \displaystyle Z^{u;\lambda_{2};\lambda_{3}}_{a}=  \frac{c^{-\frac{1}{3}}_{s}c^{-\frac{2}{3}}_{T}\underline{K}}{32 k_a^2 k_b^2 k_c^2}
    (k_a-k_b-k_c)(k_a+k_b-k_c)(k_a-k_b+k_c)
    \left[k_a^2-(\lambda_2 k_b+\lambda_3 k_c)^2 \right]^{u}.
\end{array}\ee
\\
\subsection*{\bf C. Functions appearing in two scalar one tensor correlation}
The functional dependence of the co-efficients appearing in the context of two scalar one tensor correlation can be expressed as:
\begin{eqnarray}
&&\left(\hat{\nabla}_{1}\right)_{ij}=\left[\frac{\left(k_{2i}k_{3j}+k_{3i}k_{2j}\right)}{k^{\nu_{T}}_{1}(k_{2}k_{3})^{\nu_{s}}}+
\frac{\left(k_{1i}k_{3j}+k_{3i}k_{1j}\right)}{k^{\nu_{T}}_{2}(k_{1}k_{3})^{\nu_{s}}}+\frac{\left(k_{1i}k_{2j}+k_{2i}k_{1j}\right)}{k^{\nu_{T}}_{3}(k_{1}k_{2})^{\nu_{s}}}
\right]\tilde{O},
\nonumber\\
&&\left(\hat{\nabla}_{2}\right)_{ij}=c_{s}\left(\frac{3}{2}-\nu_{T}\right)\left[\frac{\left(k_{2i}k_{3j}P_{123}+k_{3i}k_{2j}P_{132}\right)}{k^{\nu_{T}}_{1}(k_{2}k_{3})^{\nu_{s}}}+
\frac{\left(k_{1i}k_{3j}P_{213}+k_{3i}k_{1j}P_{231}\right)}{k^{\nu_{T}}_{2}(k_{1}k_{3})^{\nu_{s}}}+\frac{\left(k_{1i}k_{2j}P_{312}+k_{2i}k_{1j}P_{321}\right)}{k^{\nu_{T}}_{3}(k_{1}k_{2})^{\nu_{s}}}
\right],
\nonumber\\
&&\left(\hat{\nabla}_{3}\right)_{ij}=c_{s}\left[\frac{\left(k_{2i}k_{3j}R_{123}+k_{3i}k_{2j}R_{132}\right)}{k^{\nu_{T}}_{1}(k_{2}k_{3})^{\nu_{s}}}+
\frac{\left(k_{1i}k_{3j}R_{213}+k_{3i}k_{1j}R_{231}\right)}{k^{\nu_{T}}_{2}(k_{1}k_{3})^{\nu_{s}}}+\frac{\left(k_{1i}k_{2j}R_{312}+k_{2i}k_{1j}R_{321}\right)}{k^{\nu_{T}}_{3}(k_{1}k_{2})^{\nu_{s}}}
\right],,
\nonumber\\
&&\left(\hat{\nabla}_{4}\right)_{ij}=\left[k^{2}_{1}\frac{\left(k_{2i}k_{3j}\tilde{R}_{123}+k_{3i}k_{2j}\tilde{R}_{132}\right)}{k^{\nu_{T}}_{1}(k_{2}k_{3})^{\nu_{s}}}+
k^{2}_{2}\frac{\left(k_{1i}k_{3j}\tilde{R}_{213}+k_{3i}k_{1j}\tilde{R}_{231}\right)}{k^{\nu_{T}}_{2}(k_{1}k_{3})^{\nu_{s}}}+k^{2}_{3}\frac{\left(k_{1i}k_{2j}\tilde{R}_{312}+k_{2i}k_{1j}\tilde{R}_{321}\right)}{k^{\nu_{T}}_{3}(k_{1}k_{2})^{\nu_{s}}}
\right],
\nonumber\\
&&\left(\hat{\nabla}_{5}\right)_{ij}=\left[k^{2}_{1}\frac{\left(k_{2i}k_{3j}+k_{3i}k_{2j}\right)}{k^{\nu_{T}}_{1}(k_{2}k_{3})^{\nu_{s}}}+
k^{2}_{2}\frac{\left(k_{1i}k_{3j}+k_{3i}k_{1j}\right)}{k^{\nu_{T}}_{2}(k_{1}k_{3})^{\nu_{s}}}+k^{2}_{3}\frac{\left(k_{1i}k_{2j}+k_{2i}k_{1j}
\right)}{k^{\nu_{T}}_{3}(k_{1}k_{2})^{\nu_{s}}}
\right]\tilde{O},
\nonumber\\
&&\left(\hat{\nabla}_{6}\right)_{ij}=
\left[k^{2}_{1}\frac{\left(k_{2i}k_{3j}\tilde{L}_{123}+k_{3i}k_{2j}\tilde{L}_{132}\right)}{k^{\nu_{T}}_{1}(k_{2}k_{3})^{\nu_{s}}}+
k^{2}_{2}\frac{\left(k_{1i}k_{3j}\tilde{L}_{213}+k_{3i}k_{1j}\tilde{L}_{231}\right)}{k^{\nu_{T}}_{2}
(k_{1}k_{3})^{\nu_{s}}}+k^{2}_{3}\frac{\left(k_{1i}k_{2j}\tilde{L}_{312}+k_{2i}k_{1j}\tilde{L}_{321}\right)}{k^{\nu_{T}}_{3}(k_{1}k_{2})^{\nu_{s}}}
\right].
\end{eqnarray}
with 
\be\begin{array}{lllll}\label{pform}
\displaystyle\tilde{O}= \left\{\sum^{4}_{p=1}O_{p}\frac{\Gamma(9+p-4\nu_{s}-2\nu_{T})}{c^{4\nu_{s}
-\frac{2p}{3}-6}_{s}c^{2\nu_{T}-\frac{p}{3}-3}_{T}\underline{\underline{K}}^{9+p-4\nu_{s}-2\nu_{T}}}\right\},
O_{1}=1,O_{2}=ic^{-\frac{2}{3}}_{s}c^{-\frac{1}{3}}_{T}\underline{\underline{K}},O_{3}=-(k_{a}k_{b}+k_{b}k_{c}+k_{c}k_{a}),O_{4}=-ik_{a}k_{b}k_{c},\\

\displaystyle P_{abc}=\sum^{5}_{p=1}m^{abc}_{p}\frac{\Gamma(8+p-4\nu_{T}-2\nu_{s})}{c^{4\nu_{s}-\frac{2p}{3}-\frac{16}{3}}_{s}c^{2\nu_{T}-\frac{p}{3}-\frac{8}{3}}_{T}\underline{\underline{K}}^{8+p-4\nu_{s}-2\nu_{T}}},
\displaystyle m^{abc}_{1}=\left(\frac{3}{2}-\nu_{T}\right),m^{abc}_{2}=c^{-\frac{2}{3}}_{s}c^{-\frac{1}{3}}_{T}\underline{\underline{K}}m^{abc}_{1},\\
m^{abc}_{3}=\left[\left(\frac{3}{2}-\nu_{T}\right)(k_{a}k_{b}+k_{b}k_{c}+k_{c}k_{a})+k^2_{a}\right],
\displaystyle m^{abc}_{4}=\left[\left(\frac{3}{2}-\nu_{T}\right)k_{a}k_{b}k_{c}+k^{2}_{a}(k_{b}+k_{c})\right],m^{abc}_{5}=k^{2}_{a}k_{b}k_{c},
\end{array}\ee
\be\begin{array}{llll}\label{kform}
\displaystyle R_{abc}=L_{1}\left(\frac{3}{2}-\nu_{T}\right)\sum^{5}_{p=1}\bar{{\cal A}}^{abc}_{p}\frac{\Gamma(8+p-2\nu_{T}-4\nu_{s}
)}{c^{4\nu_{s}-\frac{2p}{3}-\frac{16}{3}}_{s}c^{2\nu_{T}-\frac{p}{3}-\frac{8}{3}}_{T}\underline{\underline{K}}^{8+p-2\nu_{T}-4\nu_{s}}}
+\left\{\frac{a^{2}Y_{s}}{t_{1}}\frac{\left(\frac{3}{2}-\nu_{T}\right)\left(\frac{3}{2}-\nu_{s}\right)}{k^{2}_{c}}
\right.\\ \left.
~~~~~~~~~~~~~~~~~~~~~~~~~~~~~~~~~~~~~~~~~~~~~~~~~~~~~~~~~~~~~~~~~~~~~~~~~~~\displaystyle\times\sum^{6}_{q=1}
\left[\hat{\cal J}_{q}(\vec{k_{a}},\vec{k_{b}},\vec{k_{c}})\right]^{u}\frac{\Gamma(7+p-2\nu_{T}-4\nu_{s})}{c^{4\nu_{s}-\frac{2p}{3}-\frac{14}{3}}_{s}c^{2\nu_{T}-\frac{p}{3}-\frac{7}{3}}_{T}
\underline{\underline{K}}^{7+p-2\nu_{T}-4\nu_{s}}}\right\},\end{array}\ee

\be\begin{array}{llll}\label{kformzx1}
\displaystyle\left[\hat{\cal J}_{1}(\vec{k_{a}},\vec{k_{b}},\vec{k_{c}})\right]^{u}=\left(\frac{3}{2}-\nu_{s}\right)\left(\frac{3}{2}-\nu_{T}\right),
\left[\hat{\cal J}_{2}(\vec{k_{a}},\vec{k_{b}},\vec{k_{c}})\right]^{u}=ic^{-\frac{2}{3}}_{s}c^{-\frac{1}{3}}_{T}\underline{\underline{K}}\left[\hat{\cal J}_{1}(\vec{k_{a}},\vec{k_{b}},\vec{k_{c}})\right]^{u},
\\
\displaystyle\left[\hat{\cal J}_{3}(\vec{k_{a}},\vec{k_{b}},\vec{k_{c}})\right]^{u}=-\left[\left(\frac{3}{2}-\nu_{s}\right)k^{2}_{a}
+\left(\frac{3}{2}-\nu_{T}\right)k^{2}_{b}+\left(\frac{3}{2}-\nu_{s}\right)\left(\frac{3}{2}-\nu_{T}\right)\left\{k_{a}k_{b}+k_{b}k_{c}+k_{c}k_{a}\right\}\right],
\\
\displaystyle\left[\hat{\cal J}_{4}(\vec{k_{a}},\vec{k_{b}},\vec{k_{c}})\right]^{u}=--i\left[\left(\frac{3}{2}-\nu_{s}\right)k^{2}_{a}k_{c}+
\left(\frac{3}{2}-\nu_{T}\right)k^{2}_{c}k_{a}+\left(\frac{3}{2}-\nu_{s}\right)k^{2}_{a}k_{b}\right.\\ \left.~~~~~~~~~~~~~~~~~~~~~~~~~~~~~~~~~~~~~~~
~~~~~~~~~~~~~~~~~~~~~~~~~~~~~~~~~~~~~~~~~~~~~\displaystyle +\left(\frac{3}{2}-\nu_{T}\right)k^{2}_{c}k_{b}
+k_{a}k_{b}k_{c}\left[\hat{\cal J}_{1}(\vec{k_{a}},\vec{k_{b}},\vec{k_{c}})\right]^{u}\right],
\\
\displaystyle\left[\hat{\cal J}_{5}(\vec{k_{a}},\vec{k_{b}},\vec{k_{c}})\right]^{u}=\left[k^{2}_{a}k^{2}_{c}+\left(\frac{3}{2}-\nu_{s}\right)k^{2}_{a}k_{b}k_{c}
+\left(\frac{3}{2}-\nu_{T}\right)k_{a}k_{b}k^{2}_{c}\right],
\left[\hat{\cal J}_{6}(\vec{k_{a}},\vec{k_{b}},\vec{k_{c}})\right]^{u}=ik^{2}_{a}k_{b}k^{2}_{c},\\

\displaystyle\bar{{\cal A}}^{abc}_{1}=\left(\frac{3}{2}-\nu_{T}\right),\bar{{\cal A}}^{abc}_{2}=ic^{-\frac{2}{3}}_{s}c^{-\frac{1}{3}}_{T}\underline{\underline{K}}\bar{{\cal A}}^{abc}_{1},
\bar{{\cal A}}^{abc}_{3}=-\left(\frac{3}{2}-\nu_{T}\right)\left[k_{a}k_{b}+k_{b}k_{c}+k_{c}k_{a}+k^{2}_{a}\right],
\\
\displaystyle\bar{{\cal A}}^{abc}_{4}=\left[k_{a}k_{b}k_{c}\left(\frac{3}{2}-\nu_{T}\right)+k^{2}_{a}(k_{b}+k_{c})\right],
\bar{{\cal A}}^{abc}_{5}=k^{2}_{a}k_{b}k_{c},\\

   \displaystyle \tilde{R}_{abc}=k^{2}_{a}L_{1}\tilde{O}+\frac{a^{2}Y_{s}}{t_{1}}\frac{k^{2}_{a}}{k^{2}_{b}}\left(\frac{3}{2}-\nu_{s}\right)P_{abc},
\\
\displaystyle
 L_{abc}=L^{2}_{1}\tilde{O}-\frac{L_{1}a^{2}Y_{s}\left(\frac{3}{2}-\nu_{s}\right)}{t_{1}}
\sum^{5}_{p=1}n^{abc}_{q}\frac{\Gamma(8+p-4\nu_{s}-2\nu_{T})}{c^{4\nu_{s}-\frac{2p}{3}-\frac{16}{3}}_{s}c^{2\nu_{T}-\frac{p}{3}-\frac{8}{3}}_{T}\underline{\underline{K}}^{8+p-4\nu_{s}-2\nu_{T}}}
\\~~~~~~~~~~~~~~~~~~~~~~~~~~~~~~~~~~~~~~~~~~~~~~~~~~~~~~~~~~~~~~~~~~~~\displaystyle+\frac{a^{4}Y^{2}_{s}
\left(\frac{3}{2}-\nu_{s}\right)^{2}}{t^{2}_{1}k^{2}_{b}k^{2}_{c}}\sum^{6}_{r=1}d^{abc}_{r}\frac{\Gamma(7+p-4\nu_{s}-2\nu_{T})}{
c^{4\nu_{s}-\frac{2p}{3}-\frac{14}{3}}_{s}c^{2\nu_{T}-\frac{p}{3}-\frac{7}{3}}_{T}\underline{\underline{K}}^{7+p-4\nu_{s}-2\nu_{T}}},
\\
\displaystyle n^{abc}_{1}=\left(\frac{3}{2}-\nu_{s}\right)\left(\frac{1}{k^{2}_{a}}+\frac{1}{k^{2}_{b}}\right),
n^{abc}_{2}=ic^{-\frac{2}{3}}_{s}c^{-\frac{1}{3}}_{T}\underline{\underline{K}}n^{abc}_{1},\\
\displaystyle n^{abc}_{3}=-\left[2+\left(\frac{3}{2}-\nu_{s}\right)\left(\frac{k_{c}}{k_{b}}+\frac{k_{b}}{k_{c}}\right)+
\left(\frac{3}{2}-\nu_{s}\right)k^{2}_{a}(k_{b}+k_{c})\left(\frac{1}{k^{2}_{a}}+\frac{1}{k^{2}_{b}}\right)\right],
\\
\displaystyle n^{abc}_{4}=-i\left\{(k_{c}+k_{b})+k_{a}\left[2+\left(\frac{3}{2}-\nu_{s}\right)\left(\frac{k_{c}}{k_{b}}+\frac{k_{b}}{k_{c}}\right)\right]\right\},
n^{abc}_{5}=k_{a}(k_{c}+k_{b}),
\\
\displaystyle d^{abc}_{1}=\left(\frac{3}{2}-\nu_{s}\right)^{2},d^{abc}_{2}=ic^{-\frac{2}{3}}_{s}c^{-\frac{1}{3}}_{T}
\underline{\underline{K}}d^{abc}_{1},d^{abc}_{3}=-(k^{2}_{b}+k^{2}_{c}+k_{b}k_{c}+k_{a}k_{b}+k_{a}k_{c}),d^{abc}_{6}=ik_{a}k^{2}_{b}k^{2}_{c}\\
\displaystyle d^{abc}_{4}=-i\left[k^{2}_{b}k^{4}_{c}+\left(\frac{3}{2}-\nu_{s}\right)\left\{k_{b}k^{2}_{c}+k_{a}(k^{2}_{b}+k^{2}_{c}+k_{b}k_{c})\right\}\right],
d^{abc}_{5}=\left[k^{2}_{b}k^{2}_{c}+k_{a}\left(k^{2}_{b}k_{c}+k_{b}k^{2}_{c}\left(\frac{3}{2}-\nu_{s}\right)\right)\right].
   \end{array}\ee

 After using the basis transformation mentioned in equation(\ref{red}) we get:
\be\begin{array}{llll}\label{arr1}
\displaystyle\left(\hat{\nabla}_{1}\right)_{\lambda^{'}}=\left[\frac{\left(k^{\lambda^{'}}_{2}k^{\lambda^{''}}_{3}+k^{\lambda^{'}}_{3}
k^{\lambda^{''}}_{2}\right)}{k^{\nu_{T}}_{1}(k_{2}k_{3})^{\nu_{s}}}+
\frac{\left(k^{\lambda^{'}}_{1}k^{\lambda^{''}}_{3}+k^{\lambda^{'}}_{3}k^{\lambda^{''}}_{1}\right)}{k^{\nu_{T}}_{2}(k_{1}k_{3})^{\nu_{s}}}
+\frac{\left(k^{\lambda^{'}}_{1}k^{\lambda^{''}}_{2}+k^{\lambda^{'}}_{2}k^{\lambda^{''}}_{1}\right)}{k^{\nu_{T}}_{3}(k_{1}k_{2})^{\nu_{s}}}
\right]\tilde{O}\delta_{\lambda_{'}\lambda_{''}},
\\
\displaystyle\frac{\left(\hat{\nabla}_{2}\right)_{\lambda^{'}}}{c_{s}\left(\frac{3}{2}-\nu_{T}\right)}=\left[\frac{\left(k^{\lambda^{'}}_{2}k^{\lambda^{''}}_{3}P_{123}+k^{\lambda^{'}}_{3}k^{\lambda^{''}}_{2}P_{132}\right)}{k^{\nu_{T}}_{1}(k_{2}k_{3})^{\nu_{s}}}+
\frac{\left(k^{\lambda^{'}}_{1}k^{\lambda^{''}}_{3}P_{213}+k^{\lambda^{'}}_{3}k^{\lambda^{''}}_{1}P_{231}\right)}{k^{\nu_{T}}_{2}(k_{1}k_{3})^{\nu_{s}}}
+\frac{\left(k^{\lambda^{'}}_{1}k^{\lambda^{''}}_{2}P_{312}+k^{\lambda^{'}}_{2}k^{\lambda^{''}}_{1}P_{321}\right)}{k^{\nu_{T}}_{3}(k_{1}k_{2})^{\nu_{s}}}
\right]\delta_{\lambda_{'}\lambda_{''}},
\\
\displaystyle\left(\hat{\nabla}_{3}\right)_{\lambda^{'}}=c_{s}\left[\frac{\left(k^{\lambda^{'}}_{2}k^{\lambda^{''}}_{3}R_{123}+k^{\lambda6{'}}_{3}k^{\lambda^{''}}_{2}R_{132}\right)}{k^{\nu_{T}}_{1}(k_{2}k_{3})^{\nu_{s}}}+
\frac{\left(k^{\lambda^{'}}_{1}k^{\lambda^{''}}_{3}R_{213}+k^{\lambda^{''}}_{3}k^{\lambda^{''}}_{1}R_{231}\right)}{k^{\nu_{T}}_{2}
(k_{1}k_{3})^{\nu_{s}}}+\frac{\left(k^{\lambda^{'}}_{1}k^{\lambda^{''}}_{2}R_{312}+k^{\lambda^{'}}_{2}k^{\lambda^{''}}_{1}R_{321}\right)}{k^{\nu_{T}}_{3}(k_{1}k_{2})^{\nu_{s}}}
\right]\delta_{\lambda_{'}\lambda_{''}},,
\\
\displaystyle\left(\hat{\nabla}_{4}\right)_{\lambda^{'}}=\left[k^{2}_{1}\frac{\left(k^{\lambda^{'}}_{2}k^{\lambda^{''}}_{3}\tilde{R}_{123}+k^{\lambda^{''}}_{3}k^{\lambda^{''}}_{2}\tilde{R}_{132}\right)}{k^{\nu_{T}}_{1}(k_{2}k_{3})^{\nu_{s}}}+
k^{2}_{2}\frac{\left(k^{\lambda^{'}}_{1}k^{\lambda^{''}}_{3}\tilde{R}_{213}+k^{\lambda^{'}}_{3}k^{\lambda^{''}}_{1}\tilde{R}_{231}\right)}{k^{\nu_{T}}_{2}(k_{1}k_{3})^{\nu_{s}}}
+k^{2}_{3}\frac{\left(k^{\lambda^{'}}_{1}k^{\lambda^{''}}_{2}\tilde{R}_{312}+k_{2}k^{\lambda^{''}}_{1}\tilde{R}_{321}\right)}{k^{\nu_{T}}_{3}(k_{1}k_{2})^{\nu_{s}}}
\right]\delta_{\lambda_{'}\lambda_{''}},
\\
\displaystyle\left(\hat{\nabla}_{5}\right)_{\lambda^{'}}=\left[k^{2}_{1}\frac{\left(k^{\lambda^{'}}_{2}k^{\lambda^{''}}_{3}+k^{\lambda^{''}}_{3}k^{\lambda^{''}}_{2}\right)}{k^{\nu_{T}}_{1}(k_{2}k_{3})^{\nu_{s}}}+
k^{2}_{2}\frac{\left(k^{\lambda^{'}}_{1}k^{\lambda^{''}}_{3}+k^{\lambda^{'}}_{3}k^{\lambda^{''}}_{1}\right)}{k^{\nu_{T}}_{2}(k_{1}k_{3})^{\nu_{s}}}+k^{2}_{3}
\frac{\left(k^{\lambda^{'}}_{1}k^{\lambda^{''}}_{2}+k^{\lambda^{'}}_{2}k^{\lambda^{''}}_{1}
\right)}{k^{\nu_{T}}_{3}(k_{1}k_{2})^{\nu_{s}}}
\right]\tilde{O}\delta_{\lambda_{'}\lambda_{''}},
\\
\displaystyle\left(\hat{\nabla}_{6}\right)_{\lambda^{'}}=
\left[k^{2}_{1}\frac{\left(k^{\lambda^{'}}_{2}k^{\lambda^{''}}_{3}\tilde{L}_{123}+k^{\lambda^{''}}_{3}k^{\lambda^{''}}_{2}\tilde{L}_{132}\right)}{k^{\nu_{T}}_{1}(k_{2}k_{3})^{\nu_{s}}}+
k^{2}_{2}\frac{\left(k^{\lambda^{'}}_{1}k^{\lambda^{''}}_{3}\tilde{L}_{213}+k^{\lambda^{'}}_{3}k^{\lambda^{''}}_{1}\tilde{L}_{231}\right)}{k^{\nu_{T}}_{2}
(k_{1}k_{3})^{\nu_{s}}}+k^{2}_{3}\frac{\left(k^{\lambda^{'}}_{1}k^{\lambda^{''}}_{2}\tilde{L}_{312}+k^{\lambda^{'}}_{2}k^{\lambda^{''}}_{1}\tilde{L}_{321}\right)}{k^{\nu_{T}}_{3}(k_{1}k_{2})^{\nu_{s}}}
\right]\delta_{\lambda_{'}\lambda_{''}},
\end{array}\ee
where $k^{\lambda}_{i}=k_{i}$ where $i=1,2,3$. Most surprisingly, the above coefficients are independent of $\lambda$ due to no parity
violation.
\\
\\
\subsection*{\bf D. Functions appearing in three tensor correlation}

The functional dependence of the co-efficients appearing in the context of three tensor correlation can be expressed as:
\begin{eqnarray}
 &&\Delta^{(1)}_{i_{1}j_{1}i_{2}j_{2}i_{3}j_{3}}=\frac{\sigma}{12}{\cal N}_{i_{1}j_{1};ij}{\cal N}_{i_{2}j_{2};jk}
{\cal N}_{i_{3}j_{3};ki}c^{3}_{s}\left(\frac{3}{2}-\nu_{T}\right)^{3}\left[M_{123}+M_{132}+M_{213}+M_{231}+M_{312}+M_{321}\right],
\nonumber\\
&&\Delta^{(2)}_{i_{1}j_{1}i_{2}j_{2}i_{3}j_{3}}=\frac{Y_{T}}{2c^{2}_{T}}{\cal N}_{i_{1}j_{1};ik}{\cal N}_{i_{2}j_{2};jl}{\cal N}_{i_{3}j_{3};ij}\left[k_{3k}k_{3l}+k_{2k}k_{2l}+k_{1k}k_{1l}\right]{\cal Q},
\nonumber\\
&&\Delta^{(3)}_{i_{1}j_{1}i_{2}j_{2}i_{3}j_{3}}=-\frac{Y_{T}}{2c^{2}_{T}}{\cal N}_{i_{1}j_{1};i_{3}j_{3}}{\cal N}_{i_{2}j_{2};kl}\left[k_{3k}k_{3l}+k_{2k}k_{2l}+k_{1k}k_{1l}\right]{\cal Q}
\end{eqnarray}
with \be\begin{array}{lll}\label{hgdf}\displaystyle {\cal Q}=\sum^{4}_{p=1}O_{p}\frac{\Gamma(9+p-6\nu_{T})}{K^{9+p-6\nu_{T}}},
 \displaystyle M_{abc}=\sum^{7}_{p=1}b^{abc}_{p}\frac{\Gamma(6+p-6\nu_{T})}{K^{6+p-6\nu_{T}}},
{b}^{abc}_{1}=\left(\frac{3}{2}-\nu_{T}\right)^3, 
{b}^{abc}_{2}={b}^{abc}_{1}K,
\\ \displaystyle {b}^{abc}_{3}=\left(\frac{3}{2}-\nu_{T}\right)^2\left[k_{a}(k_{b}+k_{c})+k^2_{b}+k^2_{c}+k_{b}k_{c}
+k^2_{a}\right],\\
\displaystyle {b}^{abc}_{4}=\left[k^2_{a}(k_{b}+k_{c})\left(\frac{3}{2}-\nu_{T}\right)+\left\{k_{a}(k^2_{b}+k^2_{c}+k_{b}k_{c})+k_{b}k_{c}(k_{b}+k_{c})\right\}
\left(\frac{3}{2}-\nu_{T}\right)^2\right],
\\ \displaystyle {b}^{abc}_{5}=\left[\left(\frac{3}{2}-\nu_{T}\right)\left\{k^2_{a}(k^2_{b}+k^2_{c}+k_{b}k_{c})+k^2_{b}k^2_{c}\right\}
+\left(\frac{3}{2}-\nu_{T}\right)^2 k_{a}k_{b}k_{c}(k_{b}+k_{c})\right],
\\ \displaystyle {b}^{abc}_{6}=\left(\frac{3}{2}-\nu_{T}\right)k_{a}k_{b}k_{c}\left[k_{b}k_{c}+k_{a}(k_{b}+k_{c})\right],
{b}^{abc}_{7}=-k^2_{a}k^2_{b}k^2_{c}.
\end{array}\ee
After using the basis transformation mentioned in equation(\ref{red}) 
the helicity dependent functions are given by:
\begin{eqnarray}
 &&\Delta^{(1)}_{\lambda_{1}\lambda_{2}\lambda_{3}}=\frac{\sigma}{12}\delta_{\lambda_{1}\lambda^{'}}\delta_{\lambda_{2}\lambda^{''}}\delta_{\lambda_{3}\lambda^{'''}}
c^{3}_{s}\left(\frac{3}{2}-\nu_{T}\right)^{3}\left[M^{\lambda^{'}\lambda^{''}\lambda^{'''}}_{123}+M^{\lambda^{'}\lambda^{''}\lambda^{'''}}_{132}
+M^{\lambda^{'}\lambda^{''}\lambda^{'''}}_{213}+M^{\lambda^{'}\lambda^{''}\lambda^{'''}}_{231}
+M^{\lambda^{'}\lambda^{''}\lambda^{'''}}_{312}+M^{\lambda^{'}\lambda^{''}\lambda^{'''}}_{321}\right],
\nonumber\\
&&\Delta^{(2)}_{\lambda_{1}\lambda_{2}\lambda_{3}}=\frac{Y_{T}}{2c^{2}_{T}}\delta_{\lambda_{1}\lambda^{'}}\delta_{\lambda_{2}\lambda^{''}}\delta_{\lambda_{3}\lambda^{'''}}
\left[k^{\lambda^{'}}_{3}k^{\lambda^{''}}_{3}+k^{\lambda^{'}}_{2}k^{\lambda^{''}}_{2}+k^{\lambda^{'}}_{1}k^{\lambda^{''}}_{1}\right]{\cal Q}^{\lambda^{'''}},
\nonumber\\
&&\Delta^{(3)}_{\lambda_{1}\lambda_{2}\lambda_{3}}=-\frac{Y_{T}}{2c^{2}_{T}}
\delta_{\lambda^{''}\lambda^{'}}\delta_{\lambda_{2}\lambda_{1}}\delta_{\lambda_{3}\lambda^{'''}}\left[k^{\lambda^{'''}}_{3}k^{\lambda^{''}}_{3}
+k^{\lambda^{'''}}_{2}k^{\lambda^{''}}_{2}+k^{\lambda^{'''}}_{1}k^{\lambda^{''}}_{1}\right]{\cal Q},
\end{eqnarray}
with
\be\begin{array}{lll}\label{disp}
 \displaystyle M^{\lambda^{'}\lambda^{''}\lambda^{'''}}_{abc}=\sum^{7}_{p=1}
\left(b^{abc}_{p}\right)^{\lambda^{'}\lambda^{''}\lambda^{'''}}\frac{\Gamma(6+p-6\nu_{T})}{K^{6+p-6\nu_{T}}},{\cal Q}^{\lambda^{'''}}=\sum^{4}_{p=1}O^{\lambda^{'''}}_{p}\frac{\Gamma(9+p-6\nu_{T})}{K^{9+p-6\nu_{T}}},
\\ \displaystyle
\left({b}^{abc}_{1}\right)^{\lambda^{'}\lambda^{''}\lambda^{'''}}=\left(\frac{3}{2}-\nu_{T}\right)^3, 
\left({b}^{abc}_{2}\right)^{\lambda^{'}\lambda^{''}\lambda^{'''}}=\left({b}^{abc}_{1}\right)^{\lambda^{'}\lambda^{''}\lambda^{'''}}K,
\\ \displaystyle \left({b}^{abc}_{3}\right)^{\lambda^{'}\lambda^{''}\lambda^{'''}}=\left(\frac{3}{2}-\nu_{T}\right)^2\left[k^{\lambda^{'}}_{a}(k^{\lambda^{''}}_{b}
+k^{\lambda^{'''}}_{c})+(k^{\lambda^{''}}_{b})^{2}+(k^{\lambda^{'''}}_{c})^{2}+k^{\lambda^{''}}_{b}k^{\lambda^{'''}}_{c}
+(k^{\lambda^{'}}_{a})^{2}\right],\\
\displaystyle \left({b}^{abc}_{4}\right)^{\lambda^{'}\lambda^{''}\lambda^{'''}}=\left[(k^{\lambda^{'}}_{a})^{2}(k^{\lambda^{''}}_{b}+k^{\lambda^{'''}}_{c})
\left(\frac{3}{2}-\nu_{T}\right)+\left\{k^{\lambda^{'}}_{a}((k^{\lambda^{''}}_{b})^{2}+(k^{\lambda^{'''}}_{c})^{2}+k^{\lambda^{''}}_{b}k^{\lambda^{'''}}_{c})
+k^{\lambda^{''}}_{b}k^{\lambda^{'''}}_{c}(k^{\lambda^{''}}_{b}+k^{\lambda^{'''}}_{c})\right\}
\left(\frac{3}{2}-\nu_{T}\right)^2\right],
\\ \displaystyle \left({b}^{abc}_{5}\right)^{\lambda^{'}\lambda^{''}\lambda^{'''}}=\left[\left(\frac{3}{2}-\nu_{T}\right)
\left\{(k^{\lambda^{'}}_{a})^{2}((k^{\lambda^{''}}_{b})^{2}+(k^{\lambda^{'''}}_{c})^{2}+k^{\lambda^{''}}_{b}k^{\lambda^{'''}}_{c})+(k^{\lambda^{''}}_{b}k^{\lambda^{'''}}_{c})^{2}\right\}
+\left(\frac{3}{2}-\nu_{T}\right)^2 k^{\lambda^{'}}_{a}k^{\lambda^{''}}_{b}k^{\lambda^{'''}}_{c}(k^{\lambda^{''}}_{b}+k^{\lambda^{'''}}_{c})\right],
\\ \displaystyle \left({b}^{abc}_{6}\right)^{\lambda^{'}\lambda^{''}\lambda^{'''}}=\left(\frac{3}{2}-\nu_{T}\right)k^{\lambda^{'}}_{a}k^{\lambda^{''}}_{b}k^{\lambda^{'''}}_{c}
\left[k^{\lambda^{''}}_{b}k^{\lambda^{'''}}_{c}+k^{\lambda^{'}}_{a}(k^{\lambda^{''}}_{b}+k^{\lambda^{'''}}_{c})\right],
\left({b}^{abc}_{7}\right)^{\lambda^{'}\lambda^{''}\lambda^{'''}}=-(k^{\lambda^{'}}_{a}k^{\lambda^{''}}_{b}k^{\lambda^{'''}}_{c})^{2},
\\ \displaystyle
O^{\lambda^{'''}}_{1}=1,O^{\lambda^{'''}}_{2}
=iK^{\lambda^{'''}},O^{\lambda^{'''}}_{3}=-(k^{\lambda^{'''}}_{a}k_{b}+k^{\lambda^{'''}}_{b}
k_{c}+k^{\lambda^{'''}}_{c}k_{a}),O^{\lambda^{'''}}_{4}=-ik^{\lambda^{'''}}_{a}k_{b}k_{c},
\end{array}\ee
where $k^{\lambda^{'}}_{1}=\lambda^{'}k_{1}$, $k^{\lambda^{''}}_{2}=\lambda^{''}k_{2}$ and $k^{\lambda^{'''}}_{3}=\lambda^{'''}k_{3}$.
\\
\\
%
\subsection*{\bf E. Functions appearing in four scalar correlator}
\subsubsection*{\bf 1. Contact interaction}
The functional dependence of the momentum dependent functions appearing in the context of contact interaction of four scalar correlation can be expressed as:
\be\begin{array}{llll}\label{opi}
\displaystyle G_{1}=\left(\frac{3}{2}-\nu_{s}\right)^{4}, ~G_{2}=i\bar{K}G_{1},~G_{3}=G^{\frac{3}{4}}_{1}\sum^{4}_{i=1}k^{2}_{i}
+G_{1}\sum^{4}_{i>j=1}k_{i}k_{j},
\displaystyle G_{4}=iG_{1}\sum^{4}_{i>j>m=1}k_{i}k_{j}k_{m}
+iG^{\frac{3}{4}}_{1}\sum^{4}_{i\neq j=1}k^{2}_{i}k_{j},\\
\displaystyle G_{5}=\sqrt{G_{1}}\sum^{4}_{i>j=1}k^{2}_{i}k^{2}_{j}
+G^{\frac{3}{4}}_{1}\sum^{4}_{i>j>m=1}k^{2}_{i}k_{j}k_{m}+G_{1}\prod^{4}_{i=1}k_{i},
\displaystyle G_{6}=i\sqrt{G_{1}}\sum^{4}_{i,j,m=1}k^{2}_{i}k^{2}_{j}k_{m}
+i\prod^{4}_{i\neq j>m>n=1}k^{2}_{i}k_{j}k_{m}k_{n},\\
\displaystyle G_{7}=G^{\frac{1}{4}}_{1}\prod^{4}_{i>j>m=1}k^{2}_{i}k^{2}_{j}k^{2}_{m}
+\sqrt{G_{1}}\prod^{4}_{i<j,m<n,i\neq m,j\neq n=1}k^{2}_{i}k^{2}_{j}k_{m}k_{n},\displaystyle G_{8}=iG^{\frac{3}{4}}_{1}\prod^{4}_{i>j>m\neq n=1}k^{2}_{i}k^{2}_{j}k^{2}_{m}k_{n},
\\
\displaystyle G_{9}=\prod^{4}_{i=1}k^{2}_{i},~~\bar{\cal Z}_{1}=1,
\displaystyle \bar{\cal Z}_{2}=i\bar{K},~~\bar{\cal Z}_{3}=\prod^{4}_{i>j=1}k_{i}k_{j},~~\bar{\cal Z}_{4}=\prod^{4}_{i>j>m=1}k_{i}k_{j}k_{m}, 
\bar{\cal Z}_{5}=\prod^{4}_{i=1}k_{i}\end{array}\ee
and 
\be\begin{array}{llll}\label{yui}
\displaystyle \bar{\cal I}(i,j;m,n)=\left[\bar{K}^{4}\sqrt{G_{1}}\Gamma(10-6\nu_{s})
+\bar{K}^{3}\sqrt{G_{1}}(k_{i}+k_{j}+k_{m}+k_{n})\Gamma(11-6\nu_{s})\right.\\ \left.\displaystyle
~~~~~~~~~~~~~~~~~-\bar{K}^{2}\left\{k^{2}_{m}k^{2}_{n}-iG^{\frac{1}{4}}_{1}k_{m}k_{n}(k_{m}+k_{n})-k_{m}k_{n}\sqrt{G_{1}}
-k_{i}k_{j}\sqrt{G_{1}}-(k_{i}+k_{j})(k_{m}+k_{n})\sqrt{G_{1}}\right\}\Gamma(12-6\nu_{s})\right.\\ \left.\displaystyle
~~~~~~~~~~~~~~~~~~+\bar{K}\left\{k_{i}k_{j}(k_{m}+k_{n})\sqrt{G_{1}}
-(k_{i}+k_{j})k_{m}k_{n}\sqrt{G_{1}}\right\}\Gamma(13-6\nu_{s})+k_{m}k_{n}\sqrt{G_{1}}\Gamma(14-6\nu_{s})\right].
\end{array}\ee
\subsubsection*{\bf 2. Scalar exchange}
The functional dependence of the momentum dependent functions appearing in the context of scalar exchange contribution of four scalar correlation can be expressed as:
\be\begin{array}{lllll}\label{ecq1}
    \displaystyle {\bf \Xi}_{1}(k_{1},k_{2},k_{3},k_{4},k_{5},k_{6}):= \frac{\left(\frac{3}{2}-\nu_{s}\right)^{6}}{(k_{5}k_{6})^{\nu_{s}}}\sum^{6}_{b=0}\sum^{6}_{p=0}{\bf S}_{b}(k_{1},k_{2},k_{3}){\bf S}_{p}(k_{4},k_{5},k_{6})
\frac{(-1)^{b+p-12\nu_{s}}(i(k_{4}+k_{5}+k_{6}))^{b+p-6\nu_{s}}}{(k_{4}+k_{5}+k_{6})^{4}}\\
\displaystyle~~~~~~~~~~~~~~~~~~~~~~~~~~~~~~~~~~\times\Gamma\left(\frac{3}{2}+b-3\nu_{s}\right)\Gamma(3+b+p-6\nu_{s})
\,_2F^{\bf REG}_1\left[3+b+p-6\nu_{s};\frac{3}{2}+b-3\nu_{s};\right.\\ \left.\displaystyle~~~~~~~~~~~~~~~~~~~~~~~~~~~~~~~~~~~~~~~~~~~~~~~~~~~~~~~~~~~~~~~~~~~~
~~~~~~~~~~~~~~~~~~~~~~~~~~~~~~~~~~~~~~\frac{5}{2}+b-3\nu_{s};
-\frac{(k_{1}+k_{2}+k_{3})}{(k_{4}+k_{5}+k_{6})}\right],
   \end{array}\ee
\be\begin{array}{lllll}\label{ecq2}
    \displaystyle {\bf \Xi}_{2}(k_{1},k_{2},k_{3},k_{4},k_{5},k_{6}):=\frac{\left(\frac{3}{2}-\nu_{s}\right)^{2}}{c_{S}(k_{5}k_{6})^{\nu_{s}}}\sum^{4}_{m=0}\sum^{4}_{n=0}{\bf E}_{m}(k_{1},k_{2},k_{3}){\bf E}_{n}(k_{4},k_{5},k_{6})
\frac{(-1)^{2(m+n)-9\nu_{s}}(i(k_{4}+k_{5}+k_{6}))^{1-m-n+6\nu_{s}}}{(7+2m-6\nu_{s})(k_{4}+k_{5}+k_{6})^{8}}\\
\displaystyle~~~~~~~~~~~~~~~~~~~~~~~~~~~~~~~~~~\times\Gamma(7+m+n-6\nu_{s})\,_2F_1\left[7+m+n-6\nu_{s};\frac{7}{2}+m-3\nu_{s};\frac{9}{2}+b-3\nu_{s};
-\frac{(k_{1}+k_{2}+k_{3})}{(k_{4}+k_{5}+k_{6})}\right],
   \end{array}\ee
\be\begin{array}{lllll}\label{ecq3}
    \displaystyle {\bf \Xi}_{3}(k_{1},k_{2},k_{3},k_{4},k_{5},k_{6}):=\frac{c_{S}\left(\frac{3}{2}-\nu_{s}\right)^{4}}{(k_{5}k_{6})^{\nu_{s}}}\sum^{6}_{p=0}\sum^{4}_{q=0}{\bf S}_{p}(k_{1},k_{2},k_{3}){\bf E}_{q}(k_{4},k_{5},k_{6})
\frac{(-1)^{2(p+q)-12\nu_{s}+3}(i(k_{4}+k_{5}+k_{6}))^{1-p-q+6\nu_{s}}}{(k_{4}+k_{5}+k_{6})^{6}}\\
\displaystyle~~~~~~~~~~~~~~~~~~~~~~~~~~~~~~~~~~\times\Gamma\left(\frac{3}{2}+p-3\nu_{s}\right)\Gamma(5+p+q-6\nu_{s})
\,_2F^{\bf REG}_1\left[5+p+q-6\nu_{s};\frac{3}{2}+p-3\nu_{s};\right.\\ \left.\displaystyle~~~~~~~~~~~~~~~~~~~~~~~~~~~~~~~~~~~~~~~~~~~~~~~~~~~~~~~~~~~~~~~~~~~~
~~~~~~~~~~~~~~~~~~~~~~~~~~~~~~~~~~~~~~\frac{5}{2}+p-3\nu_{s};
-\frac{(k_{1}+k_{2}+k_{3})}{(k_{4}+k_{5}+k_{6})}\right],
   \end{array}\ee
\be\begin{array}{lllll}\label{ecq4}
    \displaystyle {\bf \Xi}_{4}(k_{1},k_{2},k_{3},k_{4},k_{5},k_{6}):=\frac{c_{S}\left(\frac{3}{2}-\nu_{s}\right)^{3}}{(k_{5}k_{6})^{\nu_{s}}}
\sum^{4}_{t=0}\sum^{6}_{r=0}{\bf E}_{t}(k_{1},k_{2},k_{3}){\bf S}_{r}(k_{4},k_{5},k_{6})
\frac{(-1)^{2(t+r)-12\nu_{s}+3}(i(k_{4}+k_{5}+k_{6}))^{1-t-r+6\nu_{s}}}{(k_{4}+k_{5}+k_{6})^{6}}\\
\displaystyle~~~~~~~~~~~~~~~~~~~~~~~~~~~~~~~~~~\times\Gamma\left(\frac{3}{2}+t-3\nu_{s}\right)\Gamma(5+t+r-6\nu_{s})
\,_2F^{\bf REG}_1\left[5+t+r-6\nu_{s};\frac{3}{2}+t-3\nu_{s};\right.\\ \left.\displaystyle~~~~~~~~~~~~~~~~~~~~~~~~~~~~~~~~~~~~~~~~~~~~~~~~~~~~~~~~~~~~~~~~~~~~
~~~~~~~~~~~~~~~~~~~~~~~~~~~~~~~~~~~~~~\frac{5}{2}+t-3\nu_{s};
-\frac{(k_{1}+k_{2}+k_{3})}{(k_{4}+k_{5}+k_{6})}\right]
   \end{array}\ee

where we use {\it Regularized Hypergeometric function} defined as:
   $ \displaystyle _2F^{\bf REG}_1\left[a~;b~;c~;d\right]= \frac{_2F_1\left[a~;b~;c~;d\right]}{\Gamma\left[ c\right]}.$
Additionally here we define two new sets of momentum dependent functions given by:
\be\begin{array}{lllll}\label{mom1}
    \displaystyle {\bf S}_{0}(k_{a},k_{b},k_{c})=\left(\frac{3}{2}-\nu_{s}\right)^{3},
 \displaystyle {\bf S}_{1}(k_{a},k_{b},k_{c})=-i(k_{a}+k_{b}+k_{c})\left(\frac{3}{2}-\nu_{s}\right)^{3},\\
\displaystyle {\bf S}_{2}(k_{a},k_{b},k_{c})=-\left(\frac{3}{2}-\nu_{s}\right)^{2}\left[(k^{2}_{a}+k^{2}_{b})+\left(\frac{3}{2}-\nu_{s}\right)k_{a}k_{b}\right]
-k_{c}(k_{a}+k_{b})\left(\frac{3}{2}-\nu_{s}\right)^{3}-k^{2}_{c}\left(\frac{3}{2}-\nu_{s}\right)^{2},\\
\displaystyle {\bf S}_{3}(k_{a},k_{b},k_{c})=ik_{a}k_{b}(k_{a}+k_{b})\left(\frac{3}{2}-\nu_{s}\right)^{2}+ik_{c}\left(\frac{3}{2}-\nu_{s}\right)^{2}\left[(k^{2}_{a}+k^{2}_{b})+\left(\frac{3}{2}-\nu_{s}\right)k_{a}k_{b}\right]
+ik^{2}_{c}(k_{a}+k_{b})\left(\frac{3}{2}-\nu_{s}\right)^{2},\\
\displaystyle {\bf S}_{4}(k_{a},k_{b},k_{c})=k^{2}_{a}k^{2}_{b}\left(\frac{3}{2}-\nu_{s}\right)^{2}
+k_{a}k_{b}k_{c}(k_{a}+k_{b})\left(\frac{3}{2}-\nu_{s}\right)^{2}+\left(\frac{3}{2}-\nu_{s}\right)
k^{2}_{c}\left[(k^{2}_{a}+k^{2}_{b})+\left(\frac{3}{2}-\nu_{s}\right)k_{a}k_{b}\right],\\
\displaystyle {\bf S}_{5}(k_{a},k_{b},k_{c})=-ik^{2}_{a}k^{2}_{b}k_{c}\left(\frac{3}{2}-\nu_{s}\right)
-ik_{a}k_{b}k^{2}_{c}(k_{a}+k_{b})\left(\frac{3}{2}-\nu_{s}\right),
\displaystyle {\bf S}_{6}(k_{a},k_{b},k_{c})=-k^{2}_{a}k^{2}_{b}k^{2}_{c},\\
    \displaystyle {\bf E}_{0}(k_{a},k_{b},k_{c})=\left(\frac{3}{2}-\nu_{s}\right),
\displaystyle {\bf E}_{1}(k_{a},k_{b},k_{c})=-i(k_{a}+k_{b}+k_{c})\left(\frac{3}{2}-\nu_{s}\right),
\displaystyle {\bf E}_{2}(k_{a},k_{b},k_{c})=-k_{a}k_{b}\left(\frac{3}{2}-\nu_{s}\right)
\\-k_{a}(k_{a}+k_{b})\left(\frac{3}{2}-\nu_{s}\right)-k^{2}_{a},
\displaystyle {\bf E}_{3}(k_{a},k_{b},k_{c})=ik_{a}k_{b}k_{c}\left(\frac{3}{2}-\nu_{s}\right)
+i(k_{a}+k_{b})k^{2}_{a},
\displaystyle {\bf E}_{4}(k_{a},k_{b},k_{c})=k^{2}_{a}k_{b}k_{c},\\
   \end{array}\ee
where the superscript indices of momentum $a=(1,4)$, $b=(2,5)$ and $c=(3,6)$. \\
 
\subsubsection*{\bf 3. Graviton exchange}
In this context the divergence free contributions of the momentum dependent functions appearing in the context of graviton exchange can be written as:

,
\begin{eqnarray}
 &&   \hat{\vartheta}_{abcd}+\hat{\vartheta}_{cdab} =
\frac{k_a+k_b}{U_{cd}^2}\Bigg[\frac12
(U_{cd}+k_{ab})(U_{cd}^2-2D_{cd})
+k_{ab}^2 (k_c+k_d) \Bigg] + (a,b \leftrightarrow c,d)\nonumber \\
 &&  \!\!\!\!\!\!\!\!\!\!\!\!\!\!\!~~~~~~~~~~~~~~~~~~~~~~~~~~~~~~
 +\frac{k_a k_b}{{\bar K}} \Bigg[\frac{D_{cd}}{U_{cd}} -
k_{ab} + \frac{k_{ab}}{U_{ab}} \left(k_c k_d -k_{ab}
\frac{D_{cd}}{U_{cd}}\right) \left(\frac1{{\bar K}} +\frac1{U_{ab}}
\right)\Bigg]
+(a,b \leftrightarrow c,d) \nonumber \\
  &&  \!\!\!\!\!\!\!\!\!\!\!\!\!\!\! ~~~~~~~~~~~~~~~~~~~~~~~~~~~~~~ -
 \frac{ k_{ab}}{U_{ab} U_{cd} {\bar K}} \Bigg[ D_{ab} D_{cd} + 2
k_{ab}^2 \big( \prod_a k_a \big) \left(\frac{1} {{\bar K}^2}
+\frac{1}{U_{ab} U_{cd}} +\frac{k_{ab}}{{\bar K} U_{ab} U_{cd}} \right)
\Bigg]~, \label{eq:I-integral}
\end{eqnarray}
where we define
$U_{ab}\equiv k_a+k_b+k_{ab}\;, \qquad D_{ab} \equiv (k_a+k_b) k_{ab}
    +k_a k_b$.



\end{document}